\documentclass[sigconf]{acmart}
\renewcommand\footnotetextcopyrightpermission[1]{} % removes footnote with conference information in first column
\setcopyright{none}

\usepackage{amsmath}
\usepackage[normalem]{ulem}
\usepackage{flushend}
\usepackage{pbox}

\usepackage{rotating}

\usepackage{url}
\usepackage{breakurl}
\usepackage{subfig}
\usepackage{caption}

\usepackage{courier}
\usepackage{pgfplots}

\usepackage{enumitem}
\usepackage{booktabs}

\usepackage{color}
\definecolor{bluekeywords}{rgb}{0.13,0.13,1}
\definecolor{greencomments}{rgb}{0,0.5,0}
\definecolor{redstrings}{rgb}{0.9,0,0}
\definecolor{grey}{rgb}{0.4,0.4,0.4}

\usepackage{listings}

\newcommand{\maciej}[1]{}

\usepackage{multirow}

\usepackage{color}
\usepackage{listings}

\newcommand{\smalltt}[1]{{\small\texttt{#1}}}
\newcommand{\realsmallsf}[1]{{\small\textsf{#1}}}
\newcommand{\smallsf}[1]{{\tt{#1}}}
\newcommand{\scriptsf}[1]{{\scriptsize\textsf{#1}}}

\newcommand{\goal}[1]{}

\newlength{\verticalcompensationlength}
\newcounter{verticalcompensationrows}

\usepackage{mathrsfs}

\usepackage[utf8]{inputenc}
\usepackage{cleveref}
\crefname{section}{§}{§§}
\Crefname{section}{§}{§§}

\begin{document}

\setlength{\pdfpageheight}{\paperheight}
\setlength{\pdfpagewidth}{\paperwidth}

%
%\special{papersize=8.5in,11in}
%\setlength{\pdfpageheight}{\paperheight}
%\setlength{\pdfpagewidth}{\paperwidth}

%\conferenceinfo{CONF 'yy}{Month d--d, 20yy, City, ST, Country} 
%\copyrightyear{20yy} 
%\copyrightdata{978-1-nnnn-nnnn-n/yy/mm} 
%\doi{nnnnnnn.nnnnnnn}

% Uncomment one of the following two, if you are not going for the 
% traditional copyright transfer agreement.

%\exclusivelicense                % ACM gets exclusive license to publish, 
                                  % you retain copyright

%\permissiontopublish             % ACM gets nonexclusive license to publish
                                  % (paid open-access papers, 
                                  % short abstracts)

%\title{Atomic Active Messages: A Hardware-supported Mechanism for Accelerating Graph Analytics}
\title{Accelerating Irregular Computations\\ with Hardware Transactional Memory and Active Messages}

\author{Maciej Besta}
       \affiliation{Department of Computer Science\\
       ETH Zurich}
       \email{maciej.besta@inf.ethz.ch}
\author{Torsten Hoefler}
       \affiliation{Department of Computer Science\\
       ETH Zurich}
       \email{htor@inf.ethz.ch}

\begin{abstract}
\sloppy
We propose Atomic Active Messages (AAM), a mechanism 
that accelerates irregular graph computations on both shared- and distributed-memory
machines.
The key idea behind AAM is that hardware transactional memory (HTM)
can be used for simple and efficient processing of
irregular structures in highly parallel environments.
We illustrate techniques such as coarsening and coalescing
that enable hardware transactions to considerably accelerate
graph processing.
We conduct a detailed performance analysis of AAM on Intel Haswell and IBM Blue Gene/Q
and we illustrate various performance tradeoffs between different HTM parameters that impact
the efficiency of graph processing.
%
%AAM facilitates constructing efficient graph algorithms that fully
%utilize intra- and inter-node parallelism in highly parallel manycore environments and large distributed
%memory machines.
%
AAM can be used to implement abstractions offered by existing programming models
and to improve the performance of
irregular graph processing codes such as Graph500 or Galois.
\end{abstract}


\maketitle
\pagestyle{plain}

{\vspace{-0.5em}\noindent \textbf{This is an extended version of a paper published at\\ ACM HPDC'15 under the same title}}

{\vspace{1em}\small\noindent\textbf{Project website:}\\\url{https://spcl.inf.ethz.ch/Research/Parallel\_Programming/HTM_in_Graphs}}

%\vspace{-0.5em}
\section{Introduction}
%\vspace{-0.2em}

%% graphs are important and challenging
%Big graphs stand behind many computational problems
%such as social network analysis, machine-learning and data-mining, electrical circuit design, or computational
%science~\cite{Pingali:2011:TPA:1993498.1993501}. Thus, developing
%efficient parallel graph algorithms is
%becoming an increasingly important research problem for the parallel
%programming community. However, such algorithms are inherently difficult
%to design due to several properties of 
%graphs computations~\cite{DBLP:journals/ppl/LumsdaineGHB07}. First, they are usually
%completely \emph{data-driven} and the data is \emph{unstructured} and
%\emph{irregular}, making parallelism based on partitioning of data
%difficult to express. Moreover, such computations are usually
%\emph{fine-grained} and have \emph{poor locality}. Finally, the
%runtime of graph computations is almost always dominated by memory
%accesses due to \emph{high data access to computation ratio}~\cite{DBLP:journals/ppl/LumsdaineGHB07}.

%~\cite{Pingali:2011:TPA:1993498.1993501}

% graphs are important and challenging
Big graphs stand behind many computational problems
in social network analysis, machine-learning, computational
science, and others~\cite{DBLP:journals/ppl/LumsdaineGHB07}. Yet, designing
efficient parallel graph algorithms is challenging
due to intricate properties of 
graph computations. First, they are often
\emph{data-driven} and \emph{unstructured},
making parallelism based on partitioning of data~\cite{tate2014programming}
difficult to express. Second, they are usually
\emph{fine-grained} and have \emph{poor locality}. 
%Moreover, the
%runtime of graph computations is almost always dominated by memory
%accesses due to \emph{high data access to computation ratio}
%
Finally, implementing synchronization based on locks or atomics is tedious, 
error prone, and typically requires concurrency specialists~\cite{DBLP:journals/ppl/LumsdaineGHB07}.

% HTM becomes available and shows promising results (Intel SC13, Schulz SC12)
\sloppy
Recent implementations of
hardware transactional memory
(HTM)~\cite{Herlihy:1993:TMA:165123.165164} promise a faster
and simpler programming for parallel algorithms. The key functionality
is that complex instructions or instruction sequences execute in
\emph{isolation} and become visible to other threads \emph{atomically}.
Available HTM implementations show promising performance in scientific
codes and industrial benchmarks~\cite{tsx-sc,Wang:2012:EBG:2370816.2370836}.
In this work, we show that 
the ease of programming and performance benefits are even
more promising for fine-grained, irregular, and data-driven
graph computations. 
%However, 
%the ease of programming and potential performance benefits seem even
%more promising for fine-grained, irregular, and data-driven
%graph computations. 

%Concurrency overheads in graph computations may dominate the runtime of the algorithm~\cite{DBLP:journals/ppl/LumsdaineGHB07}. The missing
%structure often forces programmers to protect access to every single vertex,
%resulting in costly fine-grained synchronization. This is
%usually implemented with standard locks at each vertex, causing high
%memory and execution time overheads, even for non-contended locks. Some
%locks can be avoided by manually employing atomic
%operations. However, this manual process is 
%tedious, error prone, and typically requires concurrency specialists.
%HTM, on the other hand allows to wrap one or more vertex or edge operations in a
%single transaction, providing isolation and atomicity while being
%nearly free of overheads in an uncontended execution. Yet, the costs of
%spawning a transaction for each single vertex or edge operation can be
%prohibitively expensive.

%Another challenge of graph computations is the size of the input that 
%often requires distributed memory
%machines~\cite{Malewicz:2010:PSL:1807167.1807184}. Such machines generally contain manycore compute nodes
%that may support HTM (cf. IBM Blue Gene/Q~\cite{Wang:2012:EBG:2370816.2370836}). Still, it is unclear how to handle inter-node
%transactions that are spawned on a remote node (e.g., if an edge leaves
%the current machine) or transactions spanning multiple nodes (e.g., if
%local as well as remote vertices are part of the transaction).

Another challenge of graph analytics is the size of the input that 
often requires distributed memory
machines~\cite{Malewicz:2010:PSL:1807167.1807184}. Such machines generally contain manycore compute nodes
that may support HTM (cf. IBM Blue Gene/Q~\cite{Wang:2012:EBG:2370816.2370836}). Still, it is unclear how to handle
transactions accessing vertices on both local and remote nodes.

%Processing real-world graphs often requires distributed memory
%machines~\cite{Malewicz:2010:PSL:1807167.1807184}. Such machines generally contain multiple cores
%that may provide HTM support. Still, it is unclear how to handle
%transactions that are spawned on a remote node (e.g., if an edge leaves
%the current machine) or transactions spanning multiple nodes (e.g., if
%local as well as remote vertices are part of the transaction).

%In this paper we propose a convenient abstraction and a mechanism called
%\emph{Atomic Active Messages} (AAM) that enables significant speedups in graph computations
%thanks to combining the active messaging (AM) model with hardware transactions and atomic operations.
%One of key notions in AAM is that a unit of graph computation (e.g., marking a vertex as visited in BFS~\cite{cormen2001introduction})
%may be \emph{coarsened}, i.e., multiple vertex computations can be
%executed atomically. Our detailed performance
%analysis illustrates when and why such coarsening has advantages over fine-grained
%approaches. Figure~\ref{fig:motivate} motivates AAM by illustrating
%the time to perform each phase in synchronized BFS using traditional fine-grained mechanisms
%and an AAM scheme based on coarser transactions.

\begin{figure}[h!]
%\vspace{-0.2em}
\centering

% \subfloat{
%  \includegraphics[width=0.22\textwidth]{res/motivate_more-eps-converted-to.pdf}
%  \label{fig:motivate_1}
% }\hfill
% \subfloat{
%  \includegraphics[width=0.22\textwidth]{res/motivate_bigger_graph_labels_more-eps-converted-to.pdf}
%  \label{fig:motivate_2}
% }
\includegraphics[width=0.32\textwidth]{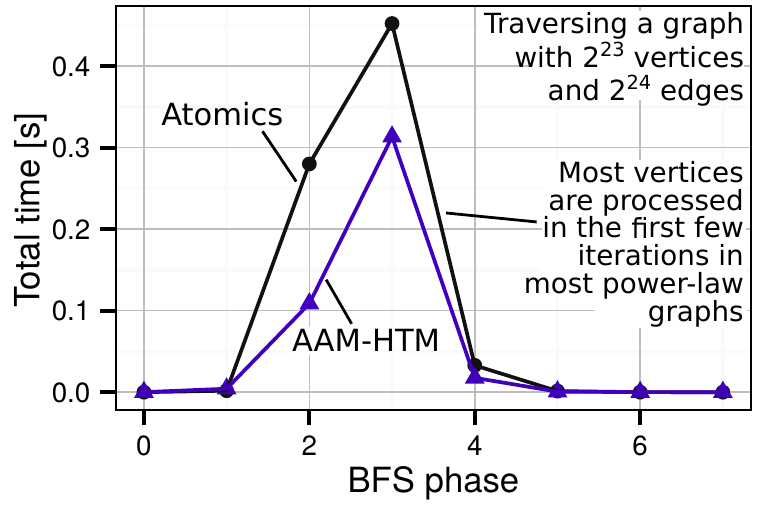}

 %\vspace{-0.9em}
 \caption{Comparison of the duration of an intra-node BFS traversal implemented with Blue Gene/Q fine-grained atomics and coarse hardware transactions (\smalltt{AAM-HTM}). One transaction modifies $2^{7}$ vertices. We use 64 threads and a Kronecker graph~\cite{Leskovec:2010:KGA:1756006.1756039} with a power-law vertex degree distribution.}
 
%with $2^{20}$ vertices and $2^{24}$ edges and 
 
%%\vspace{-0.3em}
\label{fig:motivate}
\end{figure}

In this paper we propose a mechanism called
\emph{Atomic Active Messages} (AAM) that accelerates graph analytics
by combining the active messaging (AM) model~\cite{von1992active} with HTM.
In AAM, fine units of graph computation (e.g., marking a vertex in BFS)
are \emph{coarsened} and executed as hardware transactions.
While software-based coarsening was proposed in the past~\cite{Kulkarni:2008:OPB:1346281.1346311}, in this paper we focus
on developing high performance hardware-supported techniques
to implement this mechanism on both shared- and distributed-memory machines,
on establishing principles and practice of the use of HTM for the processing of graphs,
and on illustrating various performance tradeoffs between different HTM parameters in the context of graph analytics.
%
%Our detailed performance
%analysis illustrates when and why such coarsening has advantages over fine-grained
%approaches.
Figure~\ref{fig:motivate} motivates AAM by showing
the time to perform each phase in a synchronized BFS traversal using traditional fine-grained atomics
and AAM based on coarser hardware transactions.

%\begin{figure}[h!]
%\vspace{-0.5em}
%\centering
% %\endminipage\hfill
% \subfloat[The traversed graph has $2^{20}$ vertices and $2^{24}$ edges (16 edges/vertex on average).]{
%  \includegraphics[width=0.22\textwidth]{res/motivate-eps-converted-to.pdf}
%  \label{fig:motivate_1}
% }\hfill
% \subfloat[The traversed graph has $2^{23}$ vertices and $2^{24}$ edges (2 edges/vertex on average).]{
%  \includegraphics[width=0.22\textwidth]{res/motivate_bigger_graph_labels-eps-converted-to.pdf}
%  \label{fig:motivate_2}
% }
% \vspace{-0.5em}
% \caption{Comparison of the duration of an intra-node BFS traversal implemented with Blue Gene/Q fine-grained atomics (\smalltt{Atomics}) and coarse hardware transactions (\smalltt{AAM-HTM}). A single transaction modifies $2^{7}$ vertices. We use 64 threads to traverse an R-MAT graph~\cite{Chakrabarti04r-mat:a} with a power-law vertex degree distribution~\cite{aiello2001random} and the R-MAT parameters $A=0.57, B=C=0.19$.}

%Another key insight of our work is that the proposed abstraction constitutes a hierarchy
%of atomic actives messages that can be used to express	
%graph computations on both shared- and distributed-memory machines.
%%
%We analyze this hierarchy in detail and
%conclude that AAM is a viable implementation abstraction for irregular graph
%computations\footnote{Link removed for double-blind purposes.}.
%%
%The key contributions of our work are:

Another key insight of our work is that AAM constitutes a hierarchy
of atomic actives messages that can be used to accelerate	
graph computations on both shared- and distributed-memory machines.
We analyze this hierarchy in detail and
conclude that AAM can be used to improve the performance of
generic graph analytics tools such as Galois or Graph500.
The key contributions of our work are:

%%\vspace{-0.5em}
\begin{itemize}[leftmargin=1em]
  \item We design the generic AAM mechanism that uses state-of-the-art HTM implementations to accelerate both shared- and distributed-memory graph computations.
  %\vspace{-0.6em}
  \item We establish the principles and practice of the use of HTM for graph computations. Specifically, we develop protocols for spawning
  remote/distributed hardware transactions.
%  \item We conduct a detailed performance analysis of
%  Intel Haswell HTM~\cite{tsx-sc}, Blue Gene/Q HTM~\cite{Wang:2012:EBG:2370816.2370836}, and hardware atomics~\cite{pami} to illustrate
%  which of these mechanisms are most advantageous for implementing AAM
%  and graph computations in general. Specifically, we find an optimum transaction size
%  for \smallsf{x86} and PowerPC architectures that provides highest performance in several graph analytics 
%  problems.
  %\vspace{-0.6em}
  \item We introduce a performance model and we conduct a detailed performance analysis of AAM based on
  Intel Haswell HTM~\cite{tsx-sc} and IBM Blue Gene/Q HTM~\cite{Wang:2012:EBG:2370816.2370836} to illustrate
  various performance tradeoffs between different HTM parameters in the context of graph analytics.
  Specifically, we find optimum transaction sizes
  for \smallsf{x86} and PowerPC machines that accelerate Graph500~\cite{murphy2010introducing} BFS
    code by $>$100\%.
      %\vspace{-1.8em}
      \item We show that AAM accelerates the processing of various synthetic and real-world graphs.
%  \item We demonstrate usability of AAM and automated coarsening for
%    real graph applications and we achieve a speedup of $>$100\%
%    over several graph analytics libraries.
\end{itemize}

\section{Background}
\label{sec:background}
%\vspace{-0.3em}

\goal{Introduce the section}

%We first present active messages and how they are used in distributed graph computations. We then discuss transactional memory and atomic operations.

We now describe active messages, atomics, transactional memory, and how they are used in graph computations.

%\vspace{-0.2em}
\subsection{Active Messages}
\label{sec:background_am}
%\vspace{-0.1em}

\goal{+Explain what AM are}

In the active messaging (AM) model~\cite{von1992active} processes exchange messages that carry: the address of a user-level handler function,
handler parameters, and optional
payload. When a message arrives, the parameters and the payload are
extracted from the network and the related handler runs at the
receiver~\cite{willcock-amplusplus}. Thus, AMs are conceptually similar to lightweight Remote
Procedure Calls (RPCs).

\goal{+Discuss where AM are used}

\sloppy
Active messages are often used to implement low-level performance-centric libraries that serve as a basis for developing higher-level libraries and runtime systems. Example libraries are Myrinet Express (MX), IBM's Deep Computing Messaging Framework (DCMF) for BlueGene/P, IBM's Parallel Active Message Interface (PAMI) for BlueGene/Q, GASNet~\cite{bonachea2002gasnet}, and AM++~\cite{willcock-amplusplus}.

\subsection{Active Messages in Graph Computations}
\label{sec:background_graphs}

A challenging part of designing a distributed graph algorithm is managing its data flow.
One way is to use a distributed data structure (e.g., a distributed queue) that spans all of its intra-node instances. Such structures are often hard to construct and debug~\cite{Edmonds:2013:EGA:2464996.2465441}. A BFS algorithm that uses a distributed queue is presented in Listing~\ref{lst:bfs_queue}.

%\vspace{-0.7em}
\begin{lstlisting}[float=h,caption=Distributed BFS using a
distributed queue~\cite{Gregor05theparallel} (\cref{sec:background_graphs}),label=lst:bfs_queue, mathescape=true]
if (source is local) Q.push(source);
while (!Q.empty()) {
	for (Vertex v : Q)
		if (v.visited == false) {
			v.visited = true;
			for (Vertex w : v.neighbors()) {Q.add(w); } } }
\end{lstlisting}
%\vspace{-1.5em}

%\vspace{-0.5em}
%\begin{lstlisting}[float=h,caption=Distributed BFS using a
%distributed queue~\cite{Gregor05theparallel} (\cref{sec:background_graphs}),label=lst:bfs_queue, mathescape=true]
%if (source is local) Q.push(source);
%while (!Q.empty()) {
%	for (Vertex v : Q)
%		if (v.visited == false) {
%			v.visited = true;
%			for (Vertex w : v.neighbors()) {Q2.push(w);}
%		}
%	Q.clear(); Q.swap(Q2); }
%\end{lstlisting}
%%\vspace{-1.5em}

Another approach uses active messages to express the data flow
of the program dynamically. When a process schedules 
computation for a vertex, it first checks whether it is the \emph{owner} of this
vertex. If yes, it performs the 
computation. Otherwise, the computation is
sent in an active message to a different node for
processing in a remote
handler~\cite{active-pebbles}. Thus,
no distributed data structures have to be used. We illustrate BFS using this approach in 
Listing~\ref{lst:bfs_am}.

%\vspace{-0.5em}
\begin{lstlisting}[float=h,caption=Distributed BFS using active messages~\cite{active-pebbles} (\cref{sec:background_graphs}),label=lst:bfs_am, mathescape=true]
struct bfs_AM_handler {
	bool operator()(const pair<Vertex, int>& x) {
		if (x.second < x.first.distance) {
			x.first.distance = x.second;
			send_active_message(x.first, x.second + 1); } } };
\end{lstlisting}
\subsection{Atomic Operations}
%\vspace{-0.1em}

Atomic operations appear to the rest of the system as if they occur
instantaneously. Atomics are used in lock-free graph computations to
perform fine-grained
updates~\cite{Gregor05theparallel,murphy2010introducing}. Yet, they are
limited to a single word and thus require complex protocols for
protecting operations involving multiple words~\cite{schweizer2015evaluating}.
We now present relevant atomics:

% \vspace{-0.3em}
\begin{description}[leftmargin=0.5em]
\item[\realsmallsf{Accumulate(*target, arg, op) (ACC)}:] it applies an
  operation \smallsf{op} (e.g., sum) to \smallsf{*target} using an argument \smallsf{arg}.
%  \vspace{-0.5em}
\item[\realsmallsf{Fetch-and-Op(*target, arg, op) (FAO)}:] similar to \smallsf{Accumulate} but
 it also returns the previous value
  of \smallsf{*target}.
%  \vspace{-0.5em}
  \sloppy
\item[\realsmallsf{Compare-and-Swap(*target, compare, value, *result) (CAS)}:] if \smallsf{*target == compare} then \smallsf{value} is written into \smallsf{*target} and the function sets \smallsf{*result} to \smallsf{true}, otherwise it does not change \smallsf{*target} and sets \smallsf{*result} to \smallsf{false}.   
  
%\item[\realsmallsf{Compare-and-Swap(*target, compare, value, *result) (CAS)}] compares \smallsf{*target} to a specified value \smallsf{compare}. If they are the same, a new value \smallsf{value} is written into \smallsf{*target} and the function writes \smallsf{true} to \smallsf{*result}, otherwise it does not change \smallsf{*target} and writes \smallsf{false} to \smallsf{*result}.
\end{description}

%\subsubsection{Atomic Operations in Graph Computations}
%\label{sec:atomics_in_graphs}

%\vspace{-1.0em}
\subsection{Transactional Memory}
\label{sec:background_tm}
%\vspace{-0.1em}

Transactional Memory (TM)~\cite{Herlihy:1993:TMA:165123.165164} is a
technique in which portions of code (\emph{transactions}) are executed 
in isolation and their memory effects become visible atomically. Thus,
such code portions are linearizable and easy to reason about. The
underlying TM mechanism records all modifications to specific memory
locations and commits them atomically. It also detects dependencies between
transactions accessing the same memory locations and solves potential \emph{conflicts} between such accesses by rolling back any changes to the data. TM can be based on software
emulation~\cite{Shavit:1995:STM:224964.224987} (software TM; STM) or native hardware
support~\cite{Herlihy:1993:TMA:165123.165164} (HTM).
%Existing hardware-based implementations of TM are present in, e.g., Intel Haswell~\cite{intel_tsx} and IBM BlueGene/Q~\cite{Wang:2012:EBG:2370816.2370836}.

%\subsubsection{Hardware Transactional Memory}

%Several vendors introduced HTM implementations: IBM, Sun, and Azul added HTM to Blue Gene/Q (BG/Q)
%machines~\cite{Wang:2012:EBG:2370816.2370836}, the Rock processor~\cite{Chaudhry:2009:RHS:1550399.1550516}, and the Vega CPUs~\cite{azul}, respectively. AMD has released the specification of its future HTM~\cite{Chung:2010:AAE:1934902.1935010}.
%Finally, Intel implemented two HTM instruction sets in the Haswell processor:
%Hardware Lock Elision (HLE) and Restricted Transactional Memory (RTM)
%that together constitute Transactional
%Synchronization Extensions (TSX)~\cite{tsx}. HLE allows for fast and simple porting of
%legacy lock-based code into code that uses TM.
%RTM enables programmers to define
%transactional regions in a more flexible manner than that possible with
%HLE~\cite{tsx-sc}.

Several vendors introduced HTM implementations: IBM, Sun, and Azul added HTM to Blue Gene/Q (BG/Q)
machines~\cite{Wang:2012:EBG:2370816.2370836}, the Rock processor~\cite{Chaudhry:2009:RHS:1550399.1550516}, and the Vega CPUs~\cite{azul}, respectively.
Intel implemented two HTM instruction sets in the Haswell processor:
Hardware Lock Elision (HLE) and Restricted Transactional Memory (RTM)
that together constitute Transactional
Synchronization Extensions (TSX)~\cite{tsx-sc}. HLE allows for fast and simple porting of
legacy lock-based code into code that uses TM.
RTM enables programmers to define
transactional regions in a more flexible manner than that possible with
HLE~\cite{tsx-sc}.

%\subsubsection{Transactional Memory in Graph Computations}
%\label{sec:tm_in_graphs}

There are few existing studies on STM in graph computations~\cite{Kang:2009:ETM:1504176.1504182}. Using HTM in graph processing has been largely unaddressed and
only a few initial works exist~\cite{Dice:2009:EEC:1508244.1508263,weigert2013dream}.

%Very early designs that provided transaction-like semantics were implemented in Transmeta Crusoe~\cite{crusoe} and Efficeon~\cite{efficeon} processors; we skip them in the paper because they did not address the problem of parallel execution of a critical section~\cite{crusoe}.

%\vspace{-0.4em}
\section{Atomic Active Messages}
\label{sec:properties_of_graph_algs}
%\vspace{-0.1em}

%Synchronization in graph computations has traditionally been done 
%with locks or atomics. 
%%
%The first approach-
%usually scales poorly and makes deadlock-freedom difficult to achieve~\cite{}.
%The second approach may also be inefficient because atomics
%are costly operations~\cite{} that are inherently fine-grained and thus
%have to be issued in significant numbers when processing big graphs.

Atomic Active Messages (AAM) is a mechanism
motivated by recent advances in deploying transactional memory in hardware.
An atomic active message is a message that, upon its arrival, executes a user-specified handler called an \emph{operator}.
A \emph{spawner} is a process (or a thread within this process, depending on the context) that issues atomic active messages.
An \emph{activity} is the computation that takes place as a result of executing an operator.
Activities run speculatively, isolated from one another, and they either \emph{commit} atomically or do not commit
at all.
We distinguish between operators and the activities to keep our discussion generic.

%AAM consists of three key components: the \emph{programming model} used by the developer to 
%design a graph algorithm, the \emph{runtime system} that executes algorithms and optimizes the execution, and the actual \emph{implementation} of the algorithms.

To use AAM, the developer specifies the operator code that modifies elements (vertices or edges) of the graph.
We use single-element operators for easy and intuitive programming of graph algorithms. Still,
multiple-element \emph{coarse} operators can be specified by experienced users.
The developer also determines
the structure of a vertex or an edge and defines the \emph{failure handler}, an additional piece of code
executed in certain types of algorithms (explained in~\cref{sec:types-of-aam}).
%if an activity returns a specified value.

Our runtime system executes algorithms by exchanging messages, spawning activities to run the operator code, running failure handlers, and optimizing the execution.
An activity can be \emph{coarse}: it may execute
\emph{several} operators atomically. Note that operators are (optionally) coarsened by the developer while activities are
coarsened by the runtime.

The implementation determines how activities are isolated from one another.
An activity can execute as a critical section guarded by locks, or (if it modifies one element) as an atomic operation (e.g., CAS in BFS).
However, we argue that in many cases running activities as hardware transactions provides the highest speedup; we support this claim with a detailed
performance study in Sections~\ref{sec:microbenchmarks} and~\ref{sec:evaluation}.

%%%Second, atomic active messages also differ from protocols based on locks because they spawn activities that execute optimistically
%%%and rollback if a conflict between threads occurs. Speculative execution
%%%reduces the risk of deadlocks and is significantly more efficient than a lock-based approach~\cite{Wang:2012:EBG:2370816.2370836}.

%In this section we first introduce several definitions and notation, and
%then discuss the classification of graph algorithms using an operator
%formulation similar to Tao~\cite{Pingali:2011:TPA:1993498.1993501}.

%\vspace{-0.2em}
\subsection{Definitions and Notation}
\label{sec:definitions}
%\vspace{-0.1em}

%Assume there are $N$ processes $p_1, ..., p_N$ in the system.
%A process $p_i$ runs on a compute node $n_i, 1 \ge i \ge N$
%and it may contain up to $T$ threads.

Assume there are $N$ processes $p_1, ..., p_N$ in the system.
A process $p_i$ runs on a compute node $n_i, 1 \ge i \ge N$
and it may contain up to $T$ threads.
Then, we model the analyzed graph $G$ as a tuple $(V,E)$; $V$ is a set of
vertices and $E \subseteq V \times V$ is a set of edges between
vertices. Without loss of generality we assume that $G$ is
partitioned and distributed using a one-dimensional scheme~\cite{bulucc2012graph}:
$V$ is divided into $N$ subsets $V_i$ and every $V_i \subseteq
V$ is stored on node $n_i$.
%Directed edges are stored together with
%their source vertices.
%We model each undirected edge $(s,d)$ as a pair
%of directed ones, one of which points from vertex $s$ to vertex $d$, and
%the other one from $d$ to $s$.
We call process $p_i$ the \emph{owner}
of every vertex $v \in V_i$ and every edge $(v, w)$ such that $v \in
V_i, w \in V$. We denote the average degree in $G$ as $\bar{d}$.
\subsection{Types of Atomic Active Messages}
\label{sec:types-of-aam}
%\vspace{-0.1em}

%The concept of \emph{activity} applied to distributed graph algorithms
%immediately leads to several questions: what is the flow of data, how
%activities interact with one another and what are the connections
%between activities and nodes utilized for the computation. These
%questions are indeed the foundation of three aspects which allow us to
%categorize distributed graph algorithms: \emph{direction of data flow},
%\emph{activity conflicts}, and the \emph{span of activities}.

AAM accelerates
graph computations that run on a single ($N=1$) or multiple ($N > 1$) nodes.
If $N=1$ then messages only spawn intra-node activities. If $N > 1$ then
a message may also be sent over the network to execute a remote activity.
Now, we identify two further key criteria of categorizing messages:
\emph{direction of data flow} and \emph{activity commits}.
They enable four types of messages; each type
improves the performance of different graph algorithms.

%
%\emph{Direction of data flow} determines whether or not an operator has
%to return information to its spawner. \emph{Activity conflicts}
%describes whether the executing operators conflict with one another in
%a way that some of them may have to rollback. Finally, \emph{span of
%operators} determines where the operator is going to be executed and how
%many nodes it may span.

%We present the above-described hierarchy in
%Figure~\ref{fig:general_classification}. We now proceed to analyze the
%classification in more detail. We use this classification to distill
%several similarities between seemingly different graph algorithms and
%then we show (Section~\ref{sec:aam_protocols}) that each distributed
%graph algorithm can be implemented using one of four generic
%communication protocols.
%
%\begin{figure}[h!]
%\centering
%\includegraphics[width=0.44\textwidth]{general_classification-eps-converted-to.pdf}
% %\vspace{-0.8em}
%\caption{The classification of graph algorithms.}
%\label{fig:general_classification}
%%\vspace{-1.0em}
%\end{figure}

%\vspace{-0.1em}
\subsubsection{Direction of Data Flow}
%\vspace{-0.1em}

This criteria determines if an activity has to communicate some data
back to its spawner. In some graph algorithms the data flow is \emph{unidirectional}
and messages are \emph{Fire-and-Forget} (FF): they start activities that do not return any data.
Other algorithms require the activity to return some data to the spawner to run a failure handler.
We name a message that executes such an activity a
\emph{Fire-and-Return} (FR) message
(the flow of data is \emph{bidirectional}).

%First, there are algorithms in which the spawner of the
%activity does not have to wait for or bother about any results. We call
%the algorithms from this class \emph{FF (Fire-and-Forget)
%algorithms} and the data flow is always \emph{one-directional}. The
%second class is connected with the algorithms where the result of some
%computation performed by the activity has to be returned to the spawner
%which, depending on this result, may take different actions. We name the
%algorithms from this class \emph{FR (Fire-and-Return) algorithms}
%(the flow of data is \emph{two-directional}). 

%\vspace{-0.1em}
\subsubsection{Activity Commits}
%\vspace{-0.1em}

In some graph algorithms messages belong to the type \emph{Always-Succeed} (AS):
they spawn activities that have to successfully commit,
even if it requires multiple rollbacks
or serialized execution. An example such algorithm is PageRank~\cite{Brin:1998:ALH:297805.297827}
where each
vertex $v$ has a parameter \emph{rank} that is augmented with
the normalized ranks of $v$'s neighbors.
%
%Assume that an activity increments the rank of a vertex.
Now, if we implement activities with transactions, then such transactions
may conflict while
concurrently updating the rank of the same vertex $v$, but \emph{finally each
of them has to succeed} to add its normalized rank.
The other
type are \emph{May-Fail} (MF) messages that spawn
activities that may also fail ultimately and not re-execute after a rollback. An example is BFS in which two activities, which concurrently change the distance of the same vertex, conflict and only one of them succeeds.
%
%%%Note that we distinguish between rollbacks of activities at the algorithm level, and aborts of transactions caused by, e.g., context switches. In the latter case the transaction is reexecuted by the runtime to ensure correctness.
%
Note that we distinguish between rollbacks of activities at the algorithm level, and aborts of transactions due to cache eviction, context switches, and other reasons caused by hardware/OS. In the latter case the transaction is reexecuted by the runtime to ensure correctness.

Our criteria entail four message types:
FF\&AS, FF\&MF, FR\&AS, FR\&MF.
We now show examples on how each of these types can be used to program
graph algorithms.
%
%%We now show how the AAM programming model uses these different types of messages to express different graph algorithms.
%
%We now show that these different types are used by different graph algorithms
%and that they need different operators in the AAM programming model.

%\vspace{-0.2em}
%\section{Expressing Graph Algorithms with the AAM Programming Model}
%\section{The AAM Programming Model}
\subsection{Example Case Studies}
\label{sec:alg_case_studies}
%\vspace{-0.1em}

%In this work, we advocate for a new programming model for graph
%computations based on those properties. We argue in the following that
%the concept of atomic active messages naturally fits all investigated
%algorithms. While we do not focus on the language abstractions provided
%to the programmer, we believe that high-level language concepts like
%futures~\cite{Baker:1977:IGC:800228.806932} or more low-level atomic
%regions~\cite{bocchino2011safe} could be used to expose atomic active
%messages to a programmer. In the following, we focus on the most complex
%part of graph algorithms, the update operator that performs
%modifications of graph objects. Without prescribing a specific syntax,
%we show C-like code to implement the operator in isolation, assuming
%that it forms a transaction. Our implementation utilizes
%system-specific annotations to mark atomic regions in C.
%
%We now show that the AAM abstraction can be used to naturally express 
%various distributed graph computations.

%We now express different graph computations using the AAM programming model.
%
In AAM, a single graph algorithm uses only one type of atomic active messages. This type 
determines the form of the related operator and the existence of the failure handler.
Here, we focus on the operator as the most complex
part of graph algorithms. %Without prescribing a specific syntax,
We show C-like code to implement the operator in isolation. Our implementation utilizes
system annotations to mark atomic regions in C.
%
%In the following, we focus on the most complex
%part of algorithms: the operator that perform
%modifications of graph objects. Without prescribing a specific syntax,
%we show C-like code to implement the operator in isolation. Our implementation utilizes
%system annotations to mark atomic regions in C.
%%
We present the code of six single-element operators.
When necessary, we discuss the failure handlers.
We describe multiple-element operators at the end of this section.

%\subsection{Algorithm Case Studies}
%\label{sec:alg_case_studies}

%\maciej{transactions are almost always ``either local or remote''. What about a category 'only local'? all shared memory algs fits here. And 'only remote'? Seems as if DFS could be something into this...}

%In this section we illustrate several examples of distributed graph
%algorithms. For each, we show how to formulate it using the operator
%abstraction and how it can be categorized with the proposed classification. We will use these examples to show that seemingly different distributed graph algorithms can be implemented using the same generic protocol. For each case study we omit all the implementation details connected with the application control flow (e.g., termination detection), instead we focus on the operator.

%\vspace{-0.2em}
%\subsubsection{PageRank}
%\label{sec:pr_op}
\subsubsection{PageRank (FF \& AS)}
\label{sec:pr_op}
%\vspace{-0.1em}

PageRank (PR)~\cite{Brin:1998:ALH:297805.297827} is an iterative algorithm that
calculates the \emph{rank} of each vertex $v \in V$: $rank(v) = \frac{1 - d}{|V|} +
\sum_{w \in n(v)} (d \cdot \frac{rank(w)}{out\_deg(w)})$. $n(v)$ is the
set of $v$'s neighbors, $d$ is the \emph{dump factor}~\cite{Brin:1998:ALH:297805.297827} and
$out\_deg(w)$ is the number of links leaving $w$.
Depending on the operator design, PR may be either \emph{vertex-centric} and \emph{edge-centric}. 

The pseudocode of the vertex-centric variant is presented in
Listing~\ref{lst:pagerank}. The operator increases the ranks of $v$'s
neighbors with a factor $d \cdot
\frac{rank(v)}{out\_deg(v)}$. It also adds
$\frac{1-d}{|V|}$ to $rank(v)$. The copies of stale ranks
from a previous iteration are kept and used for calculating
new ranks.
%The runtime system ensures that each PR step begins with swapping
%old ranks and new ranks and setting new ranks to zero.
%%
Assuming that each vertex $v$ is processed by one activity, this PR variant uses AS messages: each activity has to successfully add the factors to the ranks of respective vertices (which may require serialization). Data flow is unidirectional (messages are FF) because activities do not have to communicate any results back to their spawners. 
Thus, the operator returns \smalltt{void}.
%Finally, activities may span multiple nodes.

%\vspace{-0.5em}
%\begin{lstlisting}[float=h,caption=The operator in the vertex-centric PageRank variant (\cref{sec:pr_op}),label=lst:pagerank, mathescape=true]
%void Operator(Vertex v) {
%	v.rank += (1 - d) / vertices_nr;
%	forall(Vertex n: v.neighbors) {
%		n.rank += d * v.old_rank / v.out_deg; } }
%\end{lstlisting}

%\vspace{-0.2em}
\begin{lstlisting}[float=h,caption=The operator in the vertex-centric PageRank variant (\cref{sec:pr_op}),label=lst:pagerank, mathescape=true]
void Operator(Vertex v) {
	v.rank += (1 - d) / vertices_nr;
	for(int i = 0; i < v.neighbors.length; i++) {
		v.neighbors[i].rank += d * v.old_rank/v.out_deg; } }
\end{lstlisting}

There exist other PR variants. Specifically, one can 
analyze incoming edges to dispose of conflicts.
We will later (Section~\ref{sec:evaluation}) show that a careful AAM design
outperforms such approaches used in various codes such as PBGL.

\subsubsection{Breadth First Search (FF \& MF)}
\label{sec:bfs_sync_op}
%\vspace{-0.1em}

Breadth First Search (BFS) uses
FF \& MF messages. Spawners do not have to wait
for any results, but some activities may fail when concurrently updating
vertices using different distance values. Such a
conflict is solved at the node owning the vertex and no
information has to be sent back to any of the spawners,
thus the operator returns \smalltt{void}.
We present the operator pseudocode in Listing~\ref{lst:sync_bfs}.

%\vspace{-0.5em}
\begin{lstlisting}[float=h,caption=BFS operator (\cref{sec:bfs_sync_op}),label=lst:sync_bfs, mathescape=true]
void Operator(Vertex v, int new_distance) {
	if(v.distance > new_dist) {v.distance = new_dist;} }
\end{lstlisting}

%			forall(Vertex n: v.neighbors) {
%				invoke Operator(n, distance+1);

%\begin{lstlisting}[float, caption=Asynchronous BFS operator,label=lst:async_bfs, mathescape=true]
%void Operator(Vertex v, int new_distance) {
%	if(v.distance > new_distance) {
%		v.distance = new_distance;
%		forall(Vertex n: v.neighbors) {
%			invoke Operator(n, distance+1);
%		}
%	}
%}
%\end{lstlisting}

\subsubsection{Boruvka Minimum Spanning Tree}
\label{sec:boruvka_op}

%During the algorithm execution one activity is spawned for a set of supervertices

Boruvka is an algorithm for finding the minimum spanning tree of a graph
with weighted edges. First, a \emph{supervertex} is created out of each vertex. The activity
finds the smallest-weight edge incident to each accessed
supervertex, merges two supervertices connected with this edge, and
appropriately modifies the remaining incident edges. Boruvka
uses May-Fail messages: if two concurrent activities conflict, then one of them will fail. Messages are
also Fire-and-Return as the activity has to
communicate back whether it succeeded or failed, so that
the spawner may choose to either retry or, e.g., backoff for some time. 
Listing~\ref{lst:boruvka} illustrates the operator pseudocode.

%Boruvka is an algorithm for finding the minimum spanning tree of a graph
%with weighted edges. The distributed variant of Boruvka
%first creates a \emph{supervertex} out of each vertex. During the
%algorithm execution one activity is spawned for a set of supervertices. The activity
%finds the smallest-weight edge incident to each accessed
%supervertex, merges two supervertices connected with this edge, and
%appropriately modifies the remaining incident edges. Distributed Boruvka
%uses May-Fail messages: if two concurrent activities conflict, then one of them will fail. Messages are
%also Fire-and-Return as the activity has to
%communicate back whether it succeeded or failed, so that
%the spawner may choose to either retry or, e.g., backoff for some time. 
%Listing~\ref{lst:boruvka} illustrates the operator pseudocode.

%(we discuss such implementation details in
%Section~\ref{sec:aam_protocols})
%Finally, for obvious reasons (the distribution of graph data) the activities are spanning.

\begin{lstlisting}[float=h,caption=Boruvka operator (\cref{sec:boruvka_op}),label=lst:boruvka, mathescape=true]
void Operator(Supervertex vertices[]) {
	forall(Supervertex v: vertices) {
		Edge edge = get_smallest_weight_edge(v);
		
		forall(Edge e: get_incident_edges(edge.dest)) {
			if(e.dest == v) continue;
			//set 'v' as a new destination for e
			e.change_edge_dest(v);	
		}
		edge.dest.delete();
} }
\end{lstlisting}
%	//'smallest_e.dest' points to the neighboring supervertex 

%\vspace{-0.6em}
\subsubsection{ST Connectivity (FR \& AS)}
\label{sec:st_op}
%\vspace{-0.3em}

ST connectivity~\cite{Reingold:2005:USL:1060590.1060647} determines if two given
vertices ($s$ and $t$) are connected. First, the algorithm marks each vertex as ``white''. Then, it starts
two concurrent BFS traversals from $s$ and $t$. Both traversals use different colors (``grey''
and ``green'') to mark vertices as visited. Each activity returns the
information on the colors of visited vertices. In case of ``white'' no
action is taken and the operator returns \smalltt{false}. If the found color is
used by the other BFS, then $s$ and $t$ are connected, the operator returns \smalltt{true}, and the runtime
executes a failure handler at the
spawner that terminates the algorithm. The operator is presented
in Listing~\ref{lst:cc}.

%\vspace{-0.7em}
\begin{lstlisting}[float=h,caption=ST Connectivity operator (\cref{sec:st_op}),label=lst:cc, mathescape=true]
bool Operator(Vertex v, Color new_col) {
	if(v.color != WHITE && v.color != new_col) return true;
	v.color = new_col; return false; }	
\end{lstlisting}

\subsubsection{Boman Graph Coloring (FR \& MF)}
\label{sec:boman_op}

Graph coloring proposed by Boman et al.~\cite{Boman05ascalable} is a heuristic algorithm that
minimizes the number of colors assigned to graph vertices. In this algorithm as expressed using AAM (see Listing~\ref{lst:boman}), an activity
changes the color or vertex $v$ to $X$.
Then, if any of $v$'s neighbors has color $X$, either $v$ or the neighbor has to change its color;
the choice is random.
Activities are spawned by MF \& FR messages because
multiple processes trying to
update $v$'s color may conflict and the
spawners have to be notified if they need to assign new colors to $v$'s neighbors in failure handlers.
%Thus, the operator returns an \smalltt{int} indicating  .

%\vspace{-0.7em}
\begin{lstlisting}[float=h,caption=Boman graph coloring operator (\cref{sec:boman_op}),label=lst:boman, mathescape=true]
int Operator(Vertex v, Color X) {
	v.Color = X;
	if(v.hasNeighborWithColor(X)) {
		//return the ID of a vertex to be recolored
		if(rand([0;1]) < 0.5) return v.neighborWithCol(X).ID;
		else return v.ID;
	} else {	//NO_VERTEX_ID means no vertex is recolored
		return NO_VERTEX_ID; }
\end{lstlisting}

\subsection{Discussion}
\label{sec:activities_atomics}
%\vspace{-0.3em}

The introduced AAM operators modify single vertices. Thus, they
enable intuitive developing and reasoning about
graph computations that are also fine-grained by nature.
Still, some users may want to specify coarser operators to use additional 
knowledge that they have about the graph structure for higher performance.
Here, the user determines the number of elements to be modified
in the operator and the policy of their selection
(e.g., the operator may choose each vertex randomly, or try to
modify elements stored in a contiguous block of memory to avoid HTM aborts).

Manual coarsening of operators may be challenging. Our runtime
system automatically coarsens activities
for easier AAM programming.
We now discuss the implementation details of coarsening and other optimizations.
While single-element operators can be implemented with atomics or fine-grained locks,
we argue that a more performant approach is based on {coarse} transactions.

\section{Implementing Activities}
%\vspace{-0.1em}

We now discuss the details of implementing activities;
we skip most of the issues related to the runtime as they
were properly addressed in other studies~\cite{willcock-amplusplus,Edmonds:2013:EGA:2464996.2465441,active-pebbles}.

%We now describe the implementation of
%activities and the related optimizations.
%%
%%%%Activities are executed by the runtime;
%%%%here, we skip most of the issues related to the runtime as they
%%%%were properly addressed in other studies~\cite{willcock-amplusplus,Edmonds:2013:EGA:2464996.2465441,active-pebbles}. We instead focus on the core AAM concepts: optimizing graph computations and implementing activities.
%
%%%The runtime system exchanges messages,
%%%optimizes the algorithm execution, and runs the code provided by the user.
%%%Here, we skip most of these issues as they
%%%were properly addressed in other studies~\cite{willcock-amplusplus,Edmonds:2013:EGA:2464996.2465441,Pingali:2011:TPA:1993498.1993501,active-pebbles}. We instead focus on the core AAM concepts: optimizing graph computations and implementing activities.

%We now discuss how to coarsening graph computations. We then show how to implement activities and 
%we finish with presenting the \emph{ownership protocol}
%for executing distributed transactions. In the whole section we assume in-memory computations.

%\vspace{-0.2em}
\subsection{Implementing Activities with HTM}
\label{sec:impl_act}
%\vspace{-0.1em}

%\noindent
In this paper we advocate for using HTM to implement activities.
However, locks and atomics would also match the activity semantics (atomics can implement fine activities that modify single words).
We thus compared the performance of all the three mechanisms to illustrate HTM's advantages.
Locks consistently entailed generally lower performance and we thus skip them.

Transactions can implement an activity of any size. We use Intel Haswell HLE and RTM ISAs\footnote{\scriptsize \textsf{We verify the correctness of all the results to ensure that the limitations of TSX~\cite{intel-errata} do not affect our evaluation and the conclusions drawn.}} and IBM BG/Q HTM.
%At the assembly level, 
RTM provides two key functions: \smallsf{XBEGIN}
that starts a transaction and \smallsf{XEND} that performs a commit.
Yet, it does not guarantee progress. Thus, we repeat aborted transactions
and we use exponential backoff to avoid livelock.
The HTM in BG/Q automatically retries aborted transactions and it serializes the execution when the number of retries is equal to a certain value; we use the default value (10).
HLE performs serialization after the first abort.
\subsection{Optimizing the Execution of Activities}
\label{sec:coarsening_a}
%\vspace{-0.1em}

Two most significant optimizations applied by the runtime
are \emph{coarsening} and \emph{coalescing} of activities.
First, in the intra-node computations, the runtime coarsens activities by atomically executing more than one operator; an example is
presented in Listing~\ref{lst:coarsened_a}. We denote activities that are not coarse as \emph{fine}.
Coarsening amortizes the overhead of starting and committing an activity;
it also reduces the amount of fine-grained synchronization.
Second, activities targeted at the same remote node
are sent in a single message, i.e., coalesced. This
reduces the overhead of sending and receiving an atomic active message
and saves bandwidth.
Finally, we also use various optimizations that attempt to reduce
the amount of synchronization even further. For example,
the runtime avoids executing the BFS operator
for each vertex by verifying if the vertex has already been visited.

%A graph computation based on AAM is coarsened if activities atomically execute more than one operator.
%An example of such a \emph{coarse activity} in the BFS algorithm is
%presented in Listing~\ref{lst:coarsened_a}. We denote activities that are not coarse as \emph{fine}.
%

%%An activity can be programmed with a loop. However, if an activity runs as a transaction, unrolling the loop may result in better performance because it would reduce the number of conditional expressions and memory accesses tracked by the transaction.
%
%A straightforward way to implement activities is based on iterating
%over all the elements of \smallsf{vertices[]} in a loop.
%However, if an activity is implemented as a hardware transaction, this may result
%in performance penalties due to a more complicated 

%\vspace{-0.4em}
\begin{lstlisting}[float=h,caption=A BFS coarse activity (\cref{sec:coarsening_a}),label=lst:coarsened_a, mathescape=true]
void Activity(Vertex vertices[], int new_distance) {
	forall(Vertex v: vertices) {
		//call the BFS operator from Listing 5
		Operator(v, new_distance); } }
\end{lstlisting}
\subsection{A Protocol for Distributed Activities}
%\vspace{-0.3em}

The \emph{ownership protocol} enables activities implemented as hardware transactions that access or modify data from remote nodes.
The basic idea behind the protocol is that a handler running such an activity has to first physically relocate all required vertices/edges to the memory of the node where the activity executes. 
This approach is dictated by the fact that a hardware transaction cannot simply send a message because it would not be able to rollback remote changes that this message caused. In addition, most HTM implementations prevent many types of operations (e.g., system calls) from being executed inside a transaction~\cite{Wang:2012:EBG:2370816.2370836}.

Our protocol assumes that each graph element
has an \emph{ownership marker} that can be
modified atomically by any process.
Each marker is initially set to a value $\perp$ different from any process id.
When a transaction from a node $n_i$ accesses a remote graph element, it aborts and 
the runtime uses CAS or a different mechanism (e.g., an active message) to set the marker of this element to the id of process $p_i$. 
If the CAS succeeds, the marked element is
transferred to node $n_i$ and the transaction restarts.
If the CAS fails, the handler sets all previously marked elements to $\perp$ and backs off for a random amount of time.
If a local transaction attempts to access a marked element, it aborts.
This mechanism is repeated until all remote elements are cached locally.
Finally, after the transaction succeeds, the elements are sent back to
their original nodes and their markers are set to $\perp$.

%\begin{lstlisting}[float, caption=The \emph{Ownership} Protocol,label=lst:ownership, mathescape=true]
%//message handler which is the entry point of the distributed transaction.
%//In this example, a transaction modifies each neighbor of the input vertex 'v'
%void message_handler(Vertex v) {
%	int my_id = get_node_id();	//we first get the id of the local node
%	
%	for(int i = 0; i < v.neighbors.size; i++) {
%		Vertex neighbor = v.neighbors[i];
%		//try to mark the neighbor atomically. OWNED_BY_NONE is a unique value which indicates that the vertex is currently owned by no node
%		//if 'neighbor.mark' == 'my_id' then set 'neighbor.mark' to 'my_id'. 
%		int previous_mark = compare_and_swap(neighbor.mark, my_id, OWNED_BY_NONE);
%
%		//in case of a failure clear previously marked vertices
%		if(previous_mark != OWNED_BY_NONE) {
%			for(int j = 0; j < i-1; j++) {
%				set(v.neighbors[j].mark, OWNED_BY_NONE);
%			}
%			//'backoff_fn' is an arbitrary random function which depends on 'my_id'
%			float backoff_time = backoff_fn(my_id);
%			sleep(backoff_time);
%			retry message_handler;
%		}
%	}
%	//at this moment, each vertex has been marked and can be safely relocated
%	for(int i = 0; i < v.neighbors.size; i++) {
%		relocate(v.neighbors[i]);	//implementation-dependent function
%	}
%	run_transaction();
%	//after the transaction, vertices can be optionally sent back to their original nodes
%}
%\end{lstlisting}

\begin{figure*}
%\vspace{-1.0em}
%\centering
 %\endminipage\hfill
 \subfloat[Has-C, RTM.]{
  \includegraphics[width=0.2\textwidth]{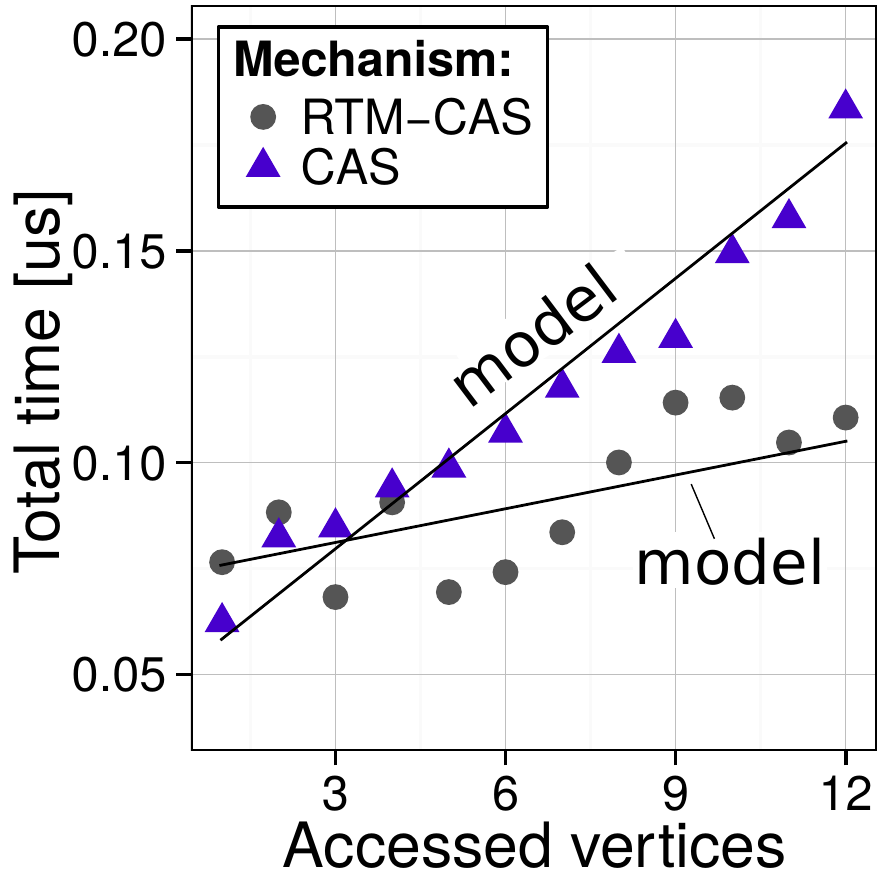}
  \label{fig:galilei-cas-rtm-model}
 }
  \subfloat[Has-C, HLE.]{
  \includegraphics[width=0.2\textwidth]{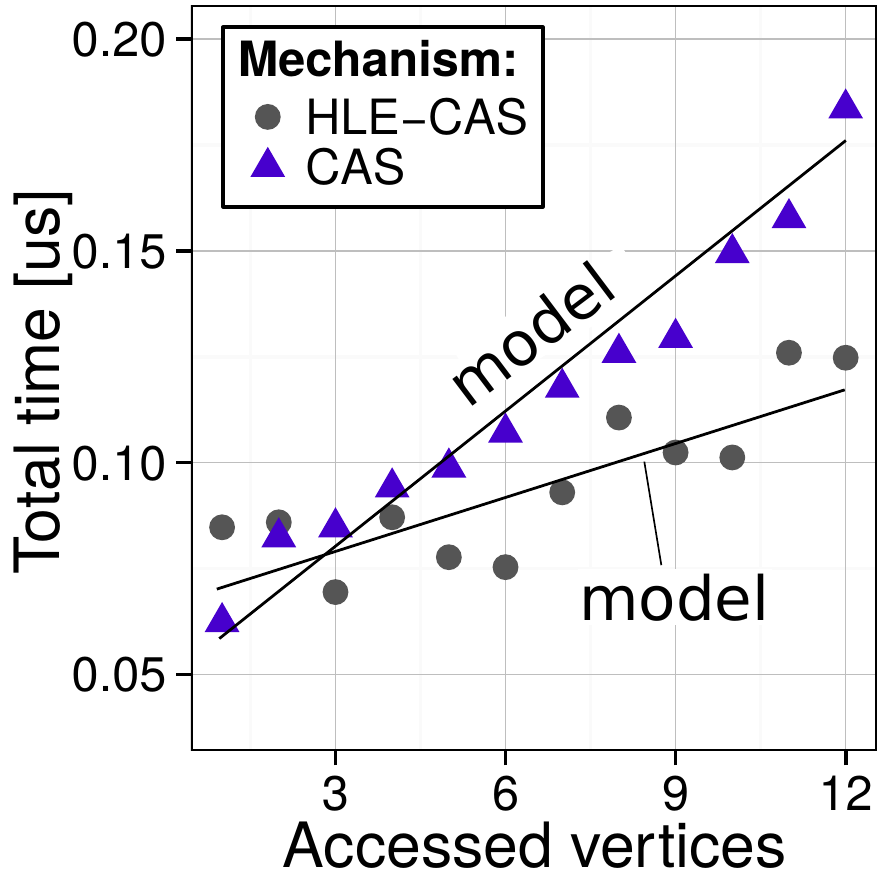}
  \label{fig:galilei-cas-hle-model}
 }
%  \subfloat[Has-P, RTM.]{
% %  \includegraphics[width=0.19\textwidth]{cas_offnode_coal_study_labels-eps-converted-to.pdf}
%   \includegraphics[width=0.155\textwidth]{greina-cas-rtm_e-eps-converted-to.pdf}
%   \label{fig:greina-cas-rtm-model}
%  }\hfill
%   \subfloat[Has-P, HLE.]{
%   \includegraphics[width=0.155\textwidth]{greina-cas-hle_e-eps-converted-to.pdf}
%   \label{fig:greina-cas-hle-model}
%  }\hfill
  \subfloat[BGQ, the short mode.]{
  \includegraphics[width=0.2\textwidth]{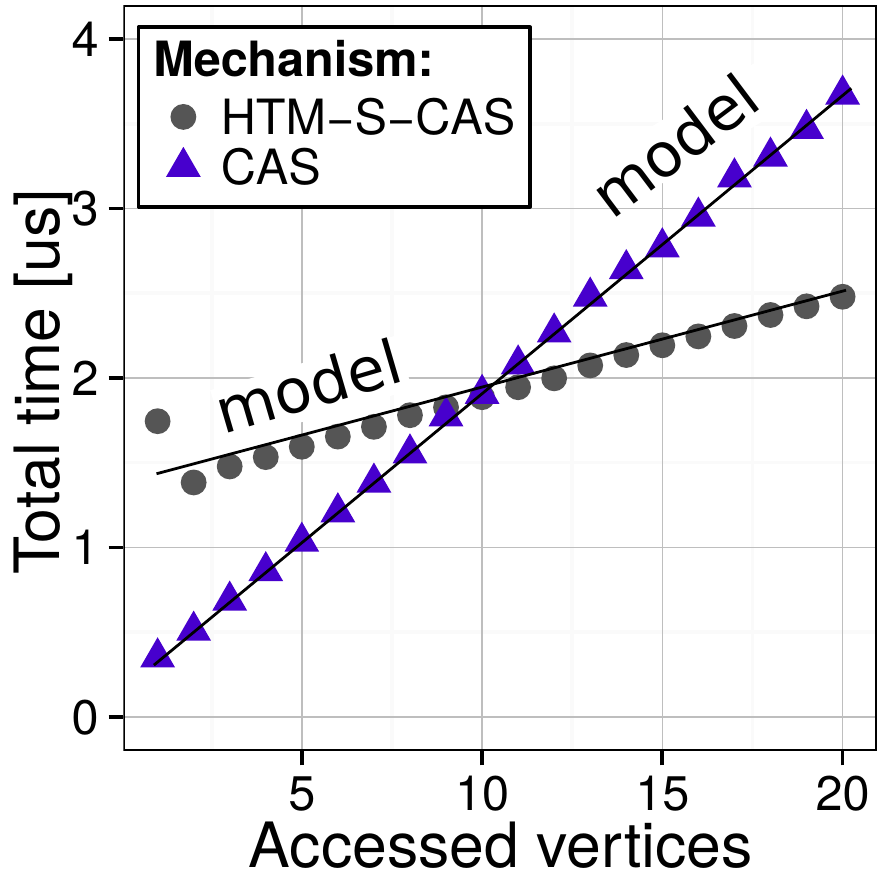}
  \label{fig:greina-cas-rtm-model}
 }
  \subfloat[BGQ, the long mode.]{
  \includegraphics[width=0.2\textwidth]{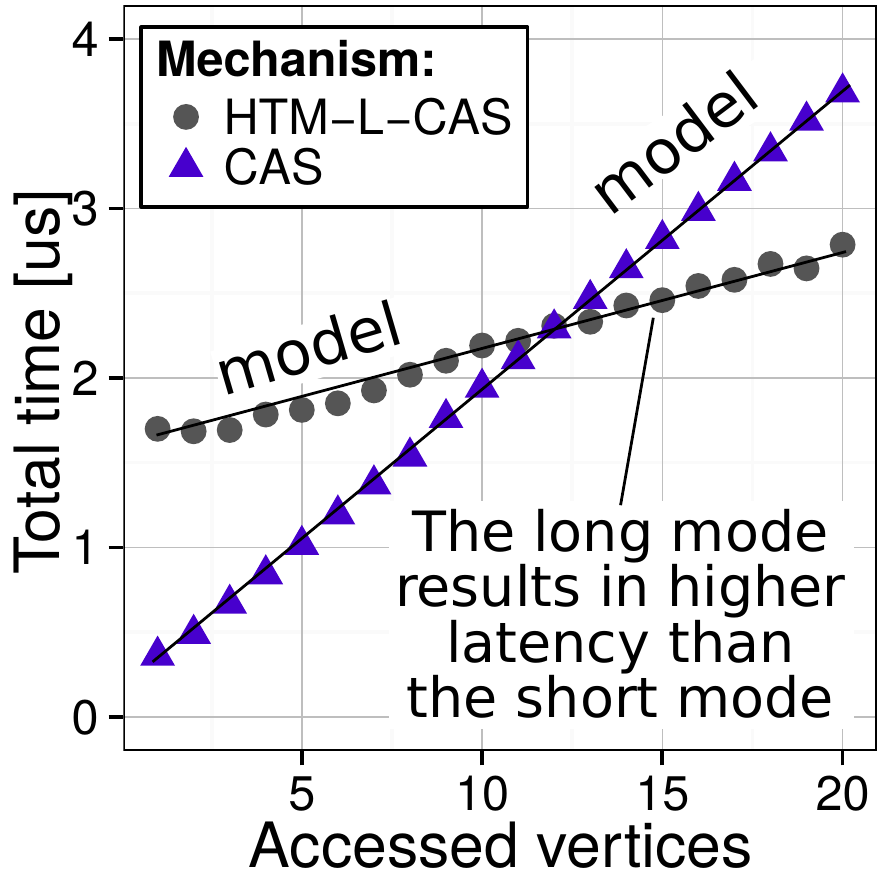}
  \label{fig:greina-cas-hle-model}
 }%\hfill
 \vspace{-0.8em}
 \caption{(\cref{sec:perf_model}) 
 The validation of the performance model.}
 
%\vspace{-1.0em}
\label{fig:perf_model}
\end{figure*}

%\vspace{-0.3em}
\section{Performance Model \& Analysis}
\label{sec:microbenchmarks}
%\vspace{-0.1em}

We now introduce a simple performance model that shows the tradeoffs between
atomics and HTM.
Then, we analyze the performance of AAM and answer the following research questions:
(1) what are HTM's advantages over atomics for implementing AAM activities,
(2) what are performance tradeoffs related to various HTM parameters,
and (3) what are the optimum transaction sizes
for analyzed architectures that enable highest speedups in 
selected graph algorithms.

%%%%We now introduce a simple performance model that illustrates the tradeoffs between
%%%%atomics and HTM.
%%%%%
%%%%Then, we analyze the performance of AAM based on Haswell {\text{RTM/HLE}} and BG/Q HTM.
%%%%This study has three
%%%%goals: (1) illustrating HTM's advantages over atomics for implementing AAM activities,
%%%%(2) analyzing performance tradeoffs related to various HTM parameters,
%%%%and (3) deriving the optimum transaction size
%%%%for analyzed architectures to enable highest speedups in 
%%%%selected graph algorithms.

%\vspace{-0.2em}
\subsection{Experimental Setup}
%\vspace{-0.1em}

We compile the code with gcc-4.8 (on Haswell) and with IBM XLC v12.1 (on BG/Q).
We use the following machines:

\begin{description}[leftmargin=0em]
\item[\textsf{ALCF BG/Q Vesta (BGQ)}] is a supercomputing machine where each compute node contains 16 1.6 GHz PowerPC A2 4-way multi-threaded cores, giving the total of 64 hardware threads per node. Each core has 16 kB of L1 cache. Every node has 32 MB of shared L2 cache and 16 GB of RAM. Nodes are connected with a 5D proprietary torus network~\cite{besta2014slim}.
This machine represents massively parallel supercomputers with HTM implemented in the shared last-level cache.
%\vspace{-0.5em}
\item[\textsf{Trivium V70.05 (Has-C)}] is a commodity off-the-shelf server where the processor (Intel Core i7-4770) contains 4 3.4 GHz Haswell 2-way multi-threaded cores, giving the total of 8 hardware threads. Each core has 32 KB of L1 and 256 KB of L2 cache. The CPU has 8 MB of shared L3 cache and 8 GB of RAM.
This option speaks for commodity computers with HTM operating in private caches.
%\vspace{-0.5em}
\item[\textsf{Greina (Has-P)}] is a high-performance cluster that contains two nodes connected with InfiniBand FDR fabric. Each node hosts 
an Intel Xeon CPU E5-2680 CPU with 12 2-way 2.50GHz multi-threaded cores; the total of 24 hardware threads. Each core contains 
64 KB of L1 and 256 KB of L2 cache. The CPU has 30 MB of shared L3 cache and 66 GB of RAM.
This machine represents high-performance clusters deploying HTM in private caches.
\end{description}	

%When the results for \textsf{Haswell-C} and \textsf{Haswell-P} follow similar performance trends, we
%refer to them collectively as \textsf{Haswell}.

\subsection{Considered Hardware Mechanisms}
\label{sec_analyzed_mechanisms}

For Haswell we compare the following mechanisms: RTM (\smallsf{Has-RTM}), HLE (\smallsf{Has-HLE}), GCC \realsmallsf{\_\_sync\_bool\_compare\_and\_swap} (\smallsf{Has-CAS}), and GCC \realsmallsf{\_\_sync\_add\_and\_fetch}
(\smallsf{Has-ACC}). We select CAS and ACC because they
can be used in miscellaneous graph codes such as BFS (a FF\&MF algorithm), PR (a FF\&AS algorithm), and ST Connectivity (a FR\&AS algorithm)~\cite{murphy2010introducing}.
For BG/Q we analyze: IBM XLC \realsmallsf{\_\_compare\_and\_swap} (\smallsf{BGQ-CAS}) and GCC \realsmallsf{\_\_sync\_add\_and\_fetch}
(\smallsf{BGQ-ACC}). We compare two modes of HTM in BG/Q: the \emph{short running mode}~\cite{Wang:2012:EBG:2370816.2370836}
(\smallsf{BGQ-HTM-S}) that bypasses L1 cache and performs better for shorter transactions, and the \emph{long running mode}~\cite{Wang:2012:EBG:2370816.2370836} (\smallsf{BGQ-HTM-L}) that keeps speculative states in L1 and is better suited for longer transactions~\cite{Wang:2012:EBG:2370816.2370836}.
%
%We also evaluate pthread locks on each architecture. As their performance is consistently worse than other mechanisms
%by $\approx$1-2 orders of magnitude we exclude them from the analysis and we omit the results for clarity of plots.

\subsection{Performance Model}
\label{sec:perf_model}

Our performance model targets graph processing and we argue in terms of activities and accessed vertices.
We predict that an activity implemented as a transaction that modifies one vertex is
more computationally expensive than an equivalent single atomic.
Yet, the transactional overheads (starting and committing) may be amortized with coarser transactions and respective activities
would outperform a series of atomics for a certain number of accessed vertices.

We now model the performance to determine the existence of crossing points; out model includes both the execution of the operations and fetching the operands from the memory.
The total time to execute an activity that modifies $N$ vertices (using either atomics or HTM) can be modeled with a simple linear function with $N$ as the argument.
We denote the slope and the intercept parameters of a function that targets atomics as $\mathcal{A}_{AT}$ and $\mathcal{B}_{AT}$;
the respective parameters for HTM are $\mathcal{A}_{HTM}$ and $\mathcal{B}_{HTM}$.
We predict that $\mathcal{B}_{HTM} > \mathcal{B}_{AT}$ due to high transactional overheads.
On the contrary, we conjecture that
$\mathcal{A}_{HTM} < \mathcal{A}_{AT}$ because HTM overheads will grow at a significantly
lower rate (determined by accesses to the memory subsystem) than that of atomics.

%We validate the model on \textsf{Has-C}, \textsf{Has-P}, and \textsf{BGQ}; the predictions and the data for \textsf{CAS} are illustrated in Figure~\ref{fig:perf_model}
%
We illustrate the model validation for \textsf{CAS} in Figure~\ref{fig:perf_model};
we plot only the results for RTM on \textsf{Has-C} and the long mode HTM on \textsf{BGQ} because all the other results differ marginally and follow similar performance patterns.
We use linear regression to calculate $\mathcal{A}_{AT}$, $\mathcal{B}_{AT}$, $\mathcal{A}_{HTM}$, and $\mathcal{B}_{HTM}$.
%$\mathcal{T}_{H,A}$ we use the first data point (total time for a transaction that accesses one vertex)
%for each $H$.
%
The analysis indicates that the model matches the data.
While a more extended model is beyond the scope of this paper, our analysis
illustrates that it is possible to amortize the transactional overhead
with coarser activities.
We now proceed to a performance analysis that illustrates
various tradeoffs between respective HTM parameters.

\begin{figure*}
%\vspace{-2em}
%\centering
 \begin{minipage}[c]{0.31\textwidth}
 \subfloat[Marking a vertex as visited 10 times (\cref{sec:intra-node-act-marking}).]{
  \includegraphics[width=1\textwidth]{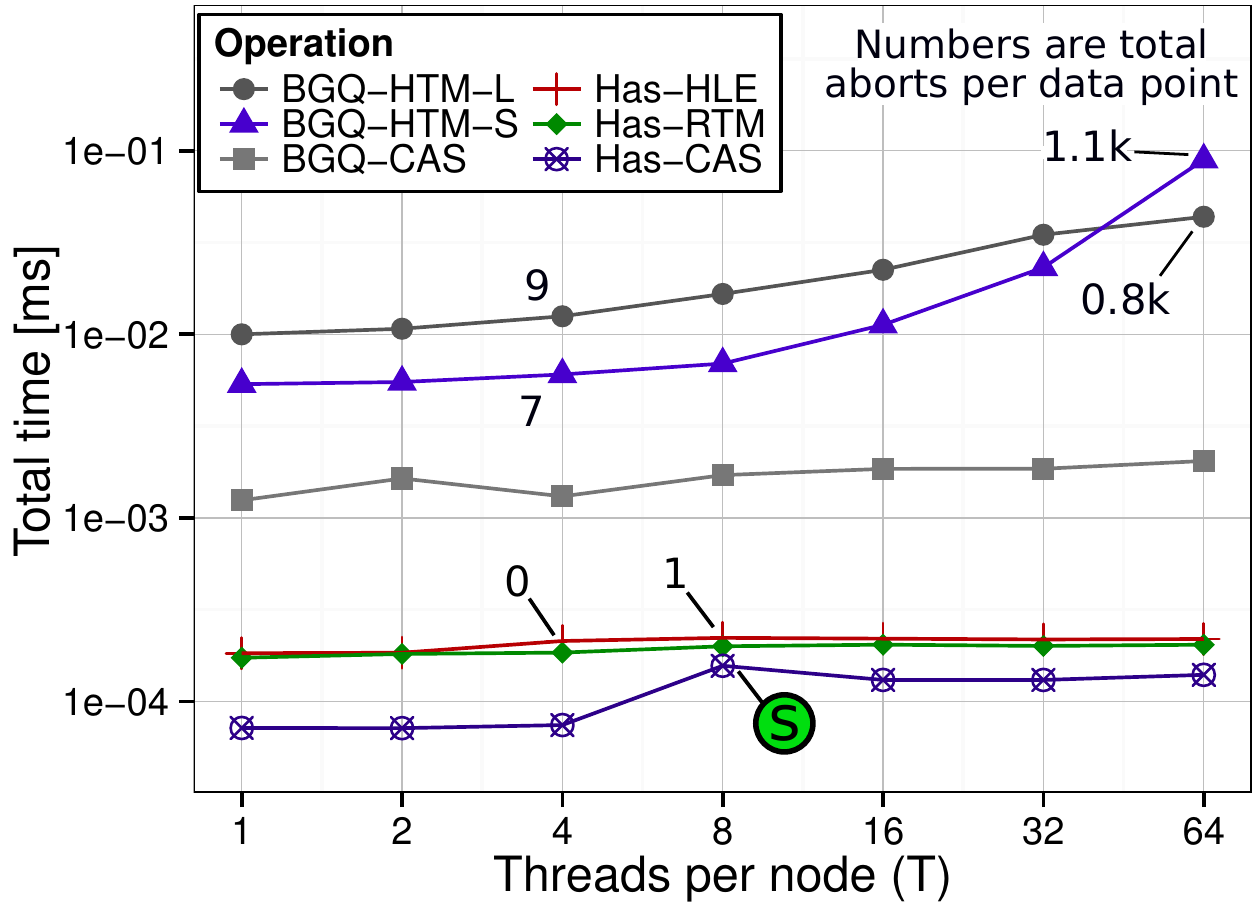}
  \label{fig:onnode_cas_no-cont}
 }
\end{minipage} \
% \endminipage\hfill
% \minipage{0.32\textwidth}
 \begin{minipage}[c]{0.31\textwidth}
   \subfloat[Marking a vertex as visited 100 times (\cref{sec:intra-node-act-marking}).]{
  \includegraphics[width=1\textwidth]{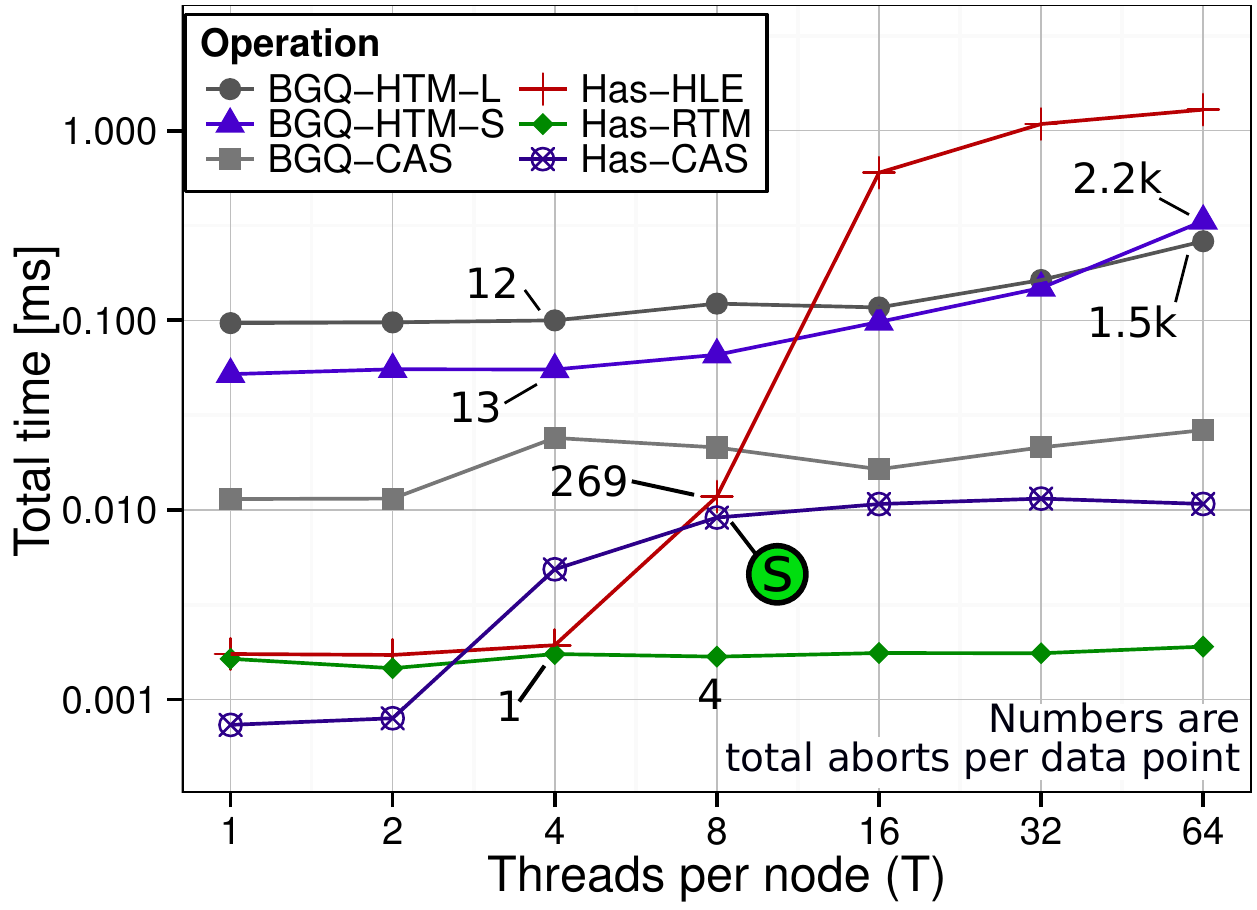}
  \label{fig:onnode_cas_cont}
 }
 \end{minipage}\quad
 \begin{minipage}[c]{0.3\textwidth}
 \scriptsize
\subfloat[Marking a vertex as visited: details of the aborts in the HTM
  implementations (\cref{sec:intra-node-act-marking}).]{\label{tab:onnode_cas_dets}
 
%\begin{minipage}[c]{0.3\textwidth}
%\vspace{-13em}
\begin{tabular}{@{}lllll@{}}
\multicolumn{2}{l}{}       & \multicolumn{3}{c}{Aborts due to:}                                                                                                                                                            \\ \toprule
\multicolumn{2}{l}{}       & \multicolumn{1}{l}{\begin{tabular}[l]{@{}l@{}}Memory\\ conflicts\end{tabular}} & \multicolumn{1}{l}{\begin{tabular}[l]{@{}l@{}}Buffer\\ overflows\end{tabular}} & \multicolumn{1}{l}{\begin{tabular}[l]{@{}l@{}}Other\\ reasons\end{tabular}} \\ \midrule
\parbox{0.06cm}{\multirow{3}{*}{\scriptsize\rotatebox[origin=c]{90}{\textsf{10 ops}}}} & \scriptsf{Has-RTM}    &                                                                                2 & 2                                                                               &  0                          \\
                  & \scriptsf{BGQ-HTM-L} &  802                                                                               &                                                                               3 &  1                          \\
                  & \scriptsf{BGQ-HTM-S} &  1,118                                                                               &                                                                               46 & 180                           \\ \midrule
\parbox{0.06cm}{\multirow{3}{*}{\rotatebox[origin=c]{90}{\textsf{100 ops}}}} & \scriptsf{Has-RTM}    &                                                                                2 &  2                                                                              & 0                           \\
                  & \scriptsf{BGQ-HTM-L} &  1,539                                                                               &                                                                               5 &  1                          \\
                  & \scriptsf{BGQ-HTM-S} &  2,242                                                                               &                                                                               13 & 2                           \\ \bottomrule
                  &        &                                                                                 &                                                                                &                            \\
                  &        &                                                                                 &                                                                                &                            \\
                  &        &                                                                                 &                                                                                &                            \\
                  &        &                                                                                 &                                                                                &     
%\vspace{-2em}                      
\end{tabular}
}
\end{minipage}

%\vspace{+1em}  

 \begin{minipage}[c]{0.31\textwidth}
 \subfloat[Incrementing a vertex' rank 10 times (\cref{sec:intra-node-act-increment}).]{
  \includegraphics[width=1\textwidth]{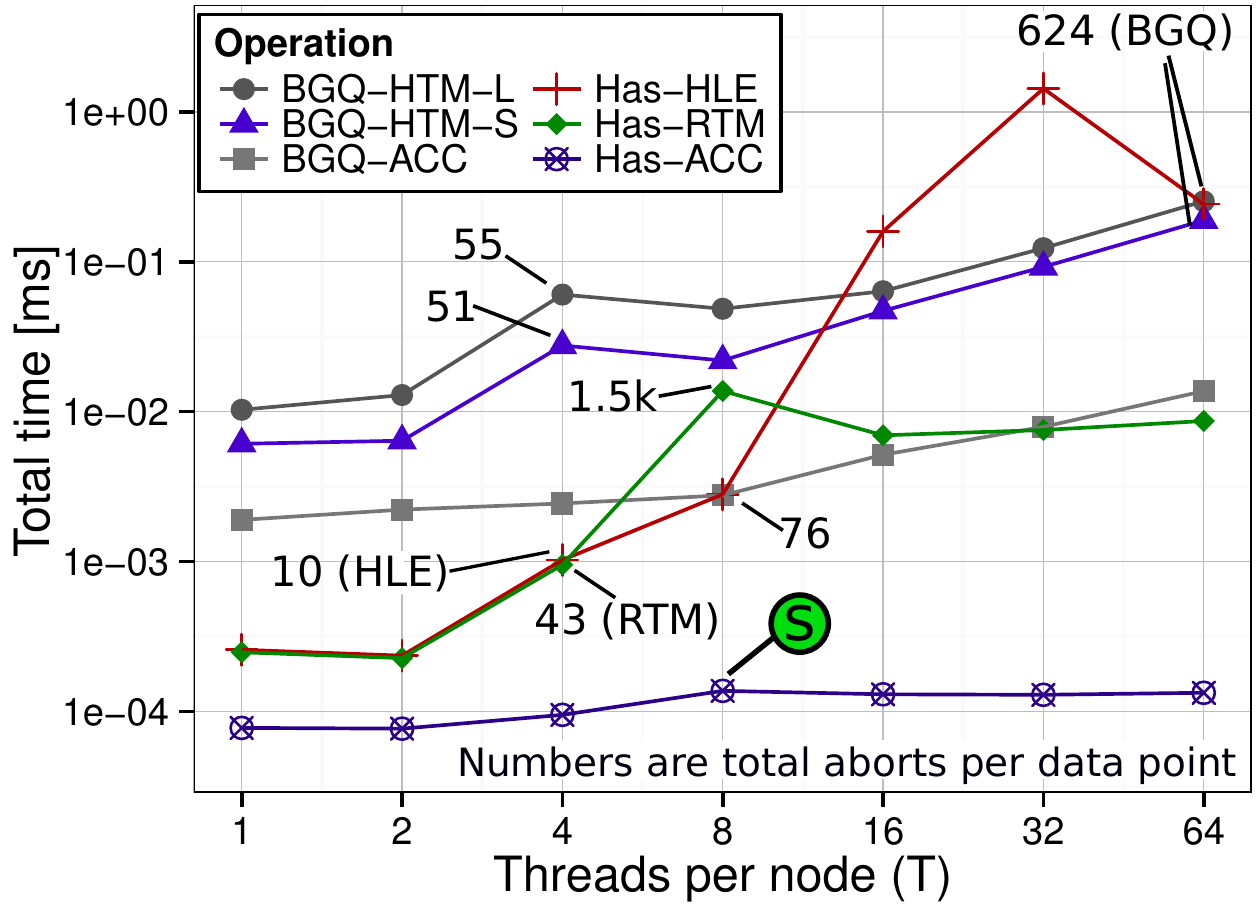}
  \label{fig:onnode_acc_no-cont}
 }
\end{minipage} \
% \endminipage\hfill
% \minipage{0.32\textwidth}
 \begin{minipage}[c]{0.31\textwidth}
 \subfloat[Incrementing a vertex' rank 100 times (\cref{sec:intra-node-act-increment}).]{
  \includegraphics[width=1\textwidth]{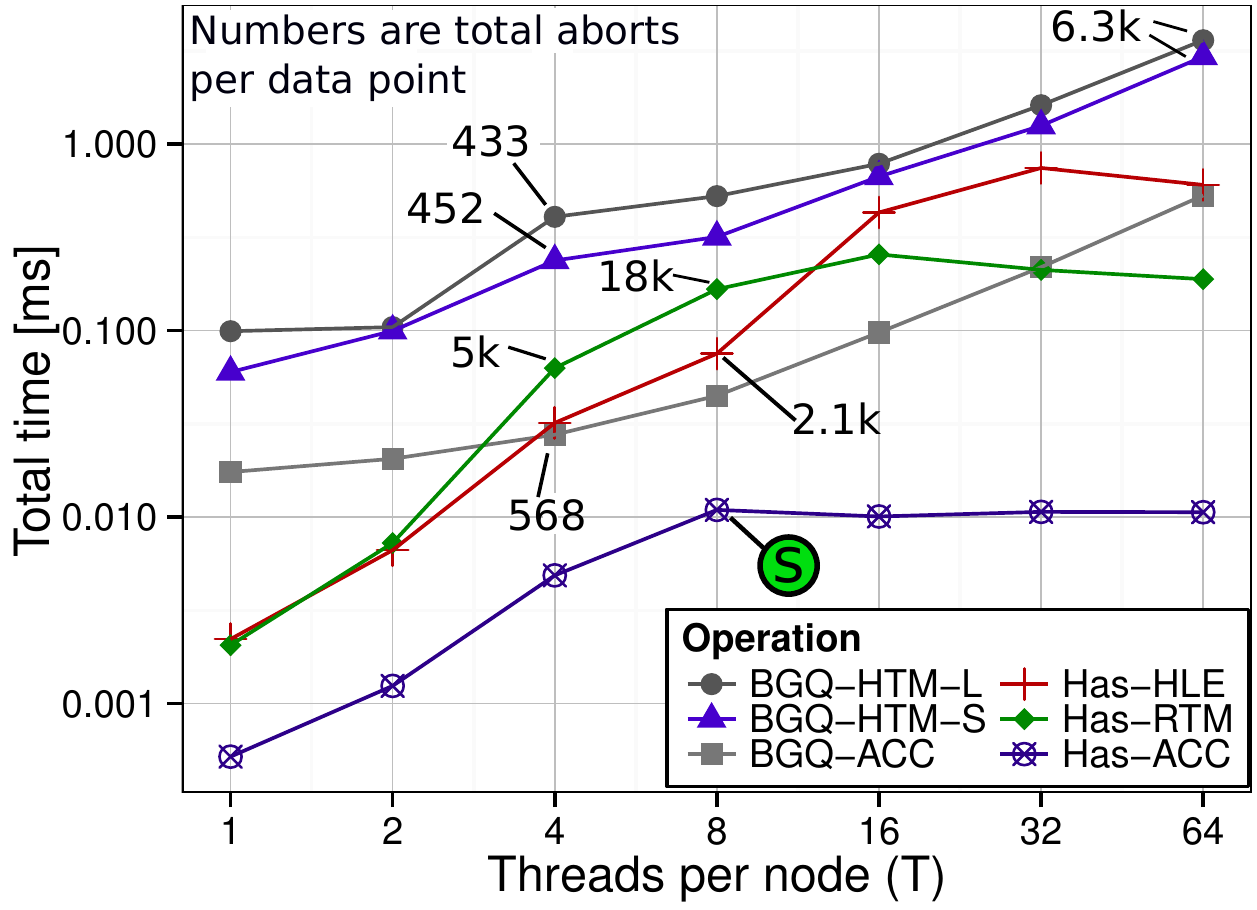}
  \label{fig:onnode_acc_cont}
 }
 \end{minipage}\quad
 \begin{minipage}[c]{0.3\textwidth}
 \scriptsize
\subfloat[Incrementing a vertex' rank: details of the aborts in the HTM implementations (\cref{sec:intra-node-act-increment}).]{\label{tab:onnode_acc_dets}

%\begin{minipage}[c]{0.3\textwidth}
%\vspace{-13em}
\begin{tabular}{@{}lllll@{}}
\multicolumn{2}{l}{}       & \multicolumn{3}{c}{Aborts due to:}                                                                                                                                                            \\ \toprule
\multicolumn{2}{l}{}       & \multicolumn{1}{c}{\begin{tabular}[c]{@{}l@{}}Memory\\ conflicts\end{tabular}} & \multicolumn{1}{c}{\begin{tabular}[c]{@{}l@{}}Buffer\\ overflows\end{tabular}} & \multicolumn{1}{c}{\begin{tabular}[c]{@{}l@{}}Other\\ reasons\end{tabular}} \\ \midrule
\parbox{0.06cm}{\multirow{3}{*}{\rotatebox[origin=c]{90}{10 ops}}} & \scriptsf{Has-RTM}    &                                                                                1,520 & 1                                                                               &  0                          \\
                  & \scriptsf{BGQ-HTM-L} &  624                                                                               &                                                                               62 &  614                          \\
                  & \scriptsf{BGQ-HTM-S} &  623                                                                               &                                                                               62 & 613                           \\ \midrule
\parbox{0.06cm}{\multirow{3}{*}{\rotatebox[origin=c]{90}{100 ops}}} & \scriptsf{Has-RTM}    &                                                                                18,952 &  33                                                                              & 0                           \\
                  & \scriptsf{BGQ-HTM-L} &  6,374                                                                               &                                                                               637 &  6,360                          \\
                  & \scriptsf{BGQ-HTM-S} &  6,392                                                                               &                                                                               639 & 6,380                           \\ \bottomrule
                  &        &                                                                                 &                                                                                &                            \\
                  &        &                                                                                 &                                                                                &                            \\
                  &        &                                                                                 &                                                                                &                            \\
                  &        &                                                                                 &                                                                                &                 
                  \vspace{-3em}                    
\end{tabular}
}
\end{minipage}

 \vspace{-0.5em}
 \caption{The analysis of the performance of intra-node activities implemented with atomics and HTMs (\cref{sec:intra-node-act}). Figures~\ref{fig:onnode_cas_no-cont} and~\ref{fig:onnode_cas_cont} illustrate the time it takes to mark a vertex as visited. Numbers in figures are sums
 of HTM aborts for a given datapoint (we report values for $T=4$ for \textsf{Has}/\textsf{BGQ}; we also plot numbers for $T=8$ (\textsf{Has}) and $T=64$ (\textsf{BGQ}) to illustrate the numbers of aborts generated by all the supported hardware threads). \protect\includegraphics[scale=0.35,trim=0 4 0 0]{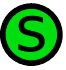} indicates the point where the latency of Haswell atomics stops to grow.
 Table~\ref{tab:onnode_cas_dets} shows
the distribution of the reasons of aborts for $T=64$ (BGQ) and $T=8$ (Haswell). We skip \smallsf{Has-HLE} as it does not provide functions to gather such statistics. A similar performance analysis for incrementing the rank of a vertex is presented in Figures~\ref{fig:onnode_acc_no-cont}-\ref{fig:onnode_acc_cont} and Table~\ref{tab:onnode_acc_dets}.}
 
%Performance of different implementations of intra-node and inter-node atomic operations. We analyze the performance of Compare-and-Swap (CAS) and Fetch-and-Op (FAO). We skip Accumulate due to space constraints and very similar performance patterns. For intra-node communication we compare: standard gcc atomics ($atomic$), implementations of these atomics based on HTM ($htm$) and pthread locks ($lock$). For inter-node we compare: remote atomics implemented over IBM PAMI library ($atomic$) and two variants of the implementation of these atomics built on top of Atomic Active Messages: \emph{htm-10} in which we perform the coalescing of 10 messages, and \emph{htm-100} where the number of coalesced messages is 100. 

%\vspace{-1em}
\label{fig:intra-node-analysis}
\end{figure*}

\begin{figure*}[!t]
%\vspace{-1em}
\centering
 %\endminipage\hfill
 \subfloat[BFS runtime ($T=1$) (\cref{sec:bgq_multivertex}).]{
  \includegraphics[width=0.235\textwidth]{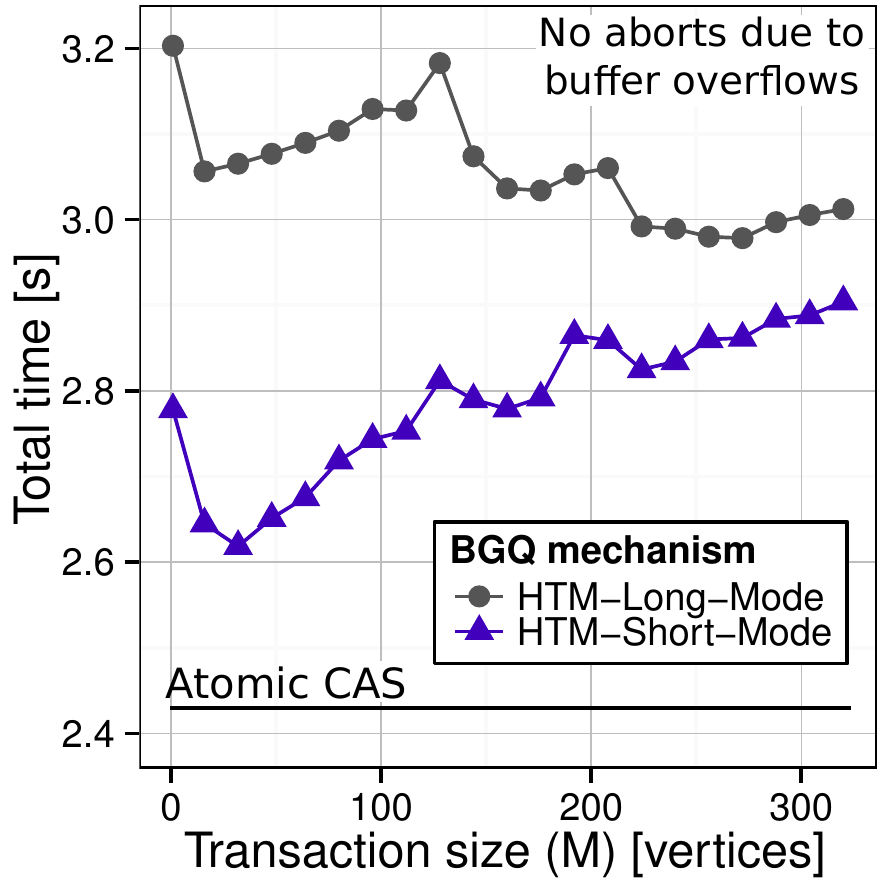}
  \label{fig:bfs_coal_study_t1_bgq}
 }\vspace{-0.5em}
 \subfloat[BFS runtime ($T=16$) (\cref{sec:bgq_multivertex}).]{
  \includegraphics[width=0.235\textwidth]{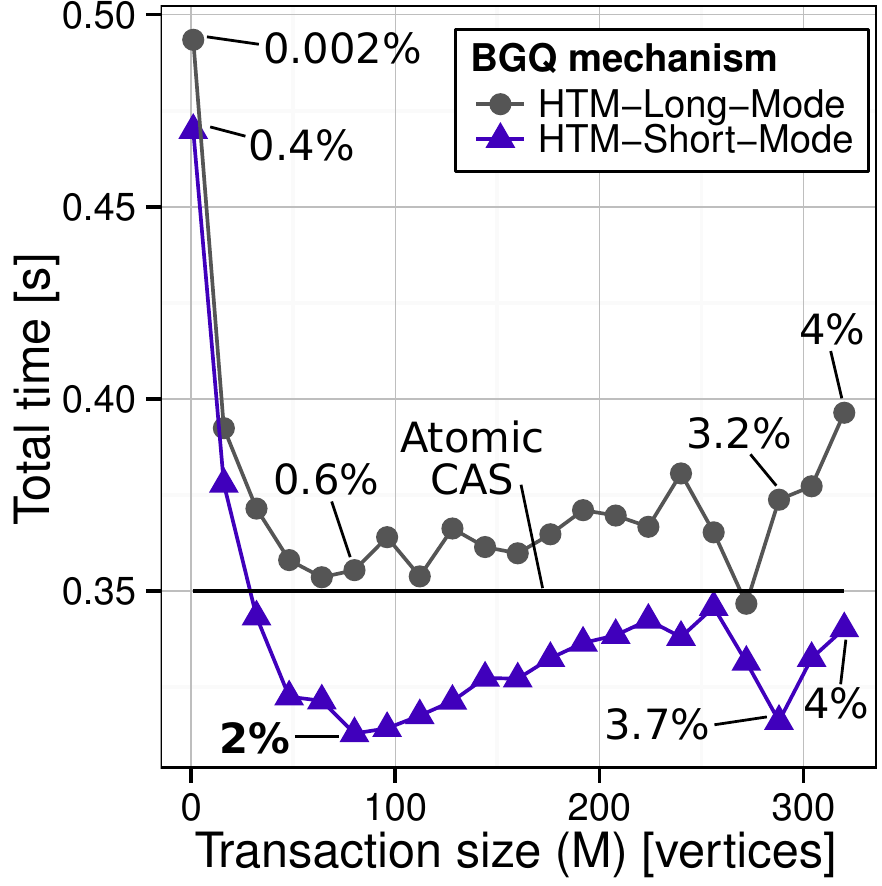}
  \label{fig:bfs_coal_study_t16_bgq}
 }
  \subfloat[BFS runtime ($T=64$) (\cref{sec:bgq_multivertex}).]{
  \includegraphics[width=0.235\textwidth]{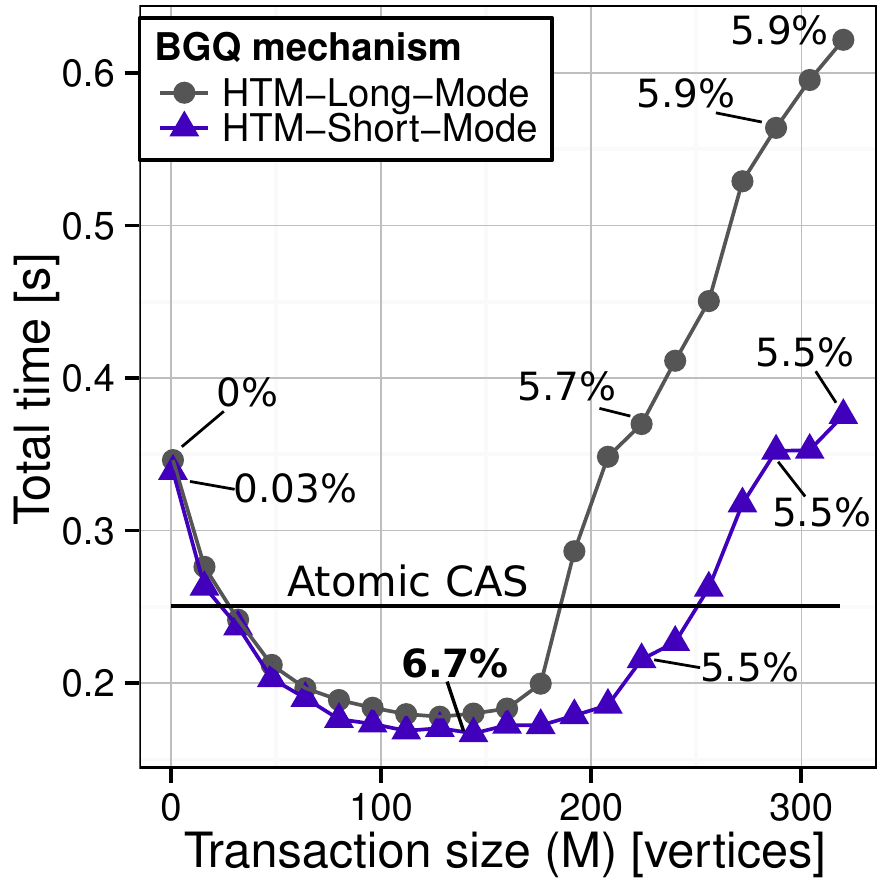}
  \label{fig:bfs_coal_study_t64_bgq}
 }
   \subfloat[BG/Q events ($T=64$) (\cref{sec:bgq_multivertex}).]{
  \includegraphics[width=0.235\textwidth]{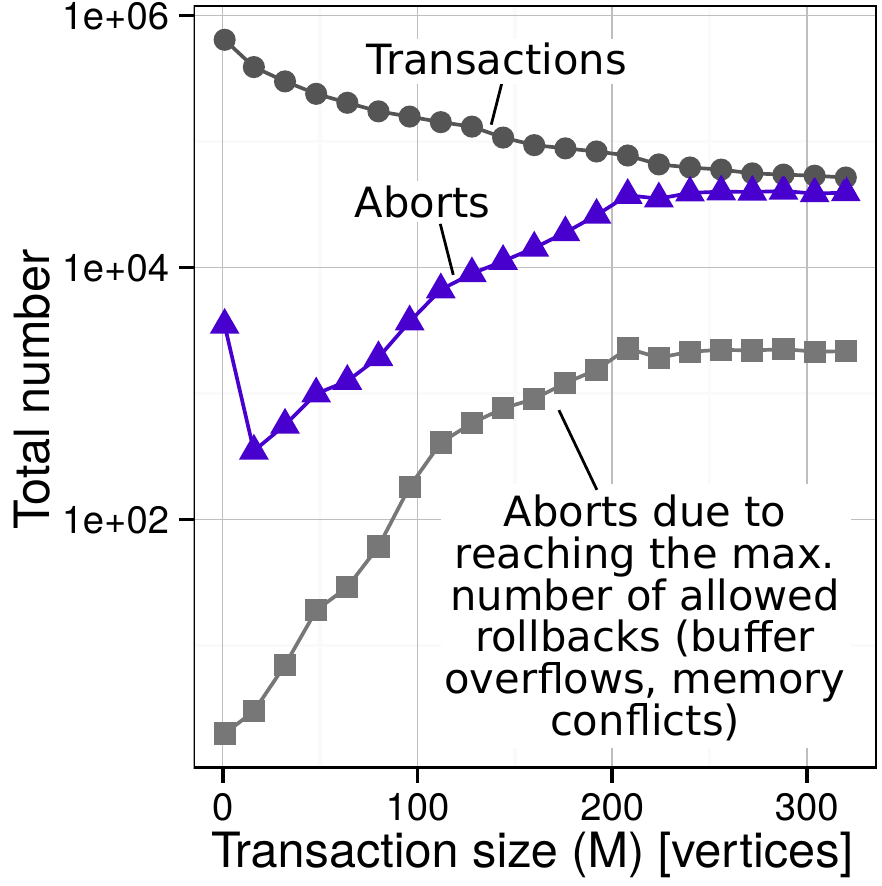}
  \label{fig:bfs_coal_study_t64_bgq_details}
 }\\
  \subfloat[BFS runtime ($T=1$) (\cref{sec:haswell-c_multivertex}).]{
  \includegraphics[width=0.235\textwidth]{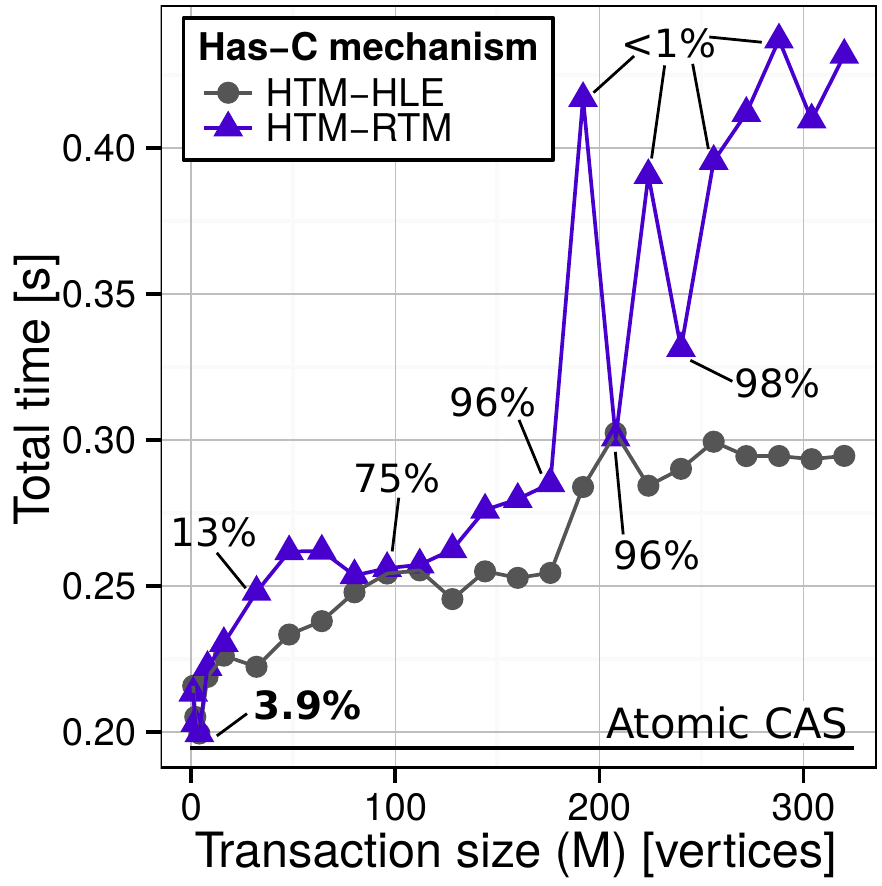}
  \label{fig:bfs_coal_study_t1_haswell}
 }
 \subfloat[BFS runtime ($T=4$) (\cref{sec:haswell-c_multivertex}).]{
  \includegraphics[width=0.235\textwidth]{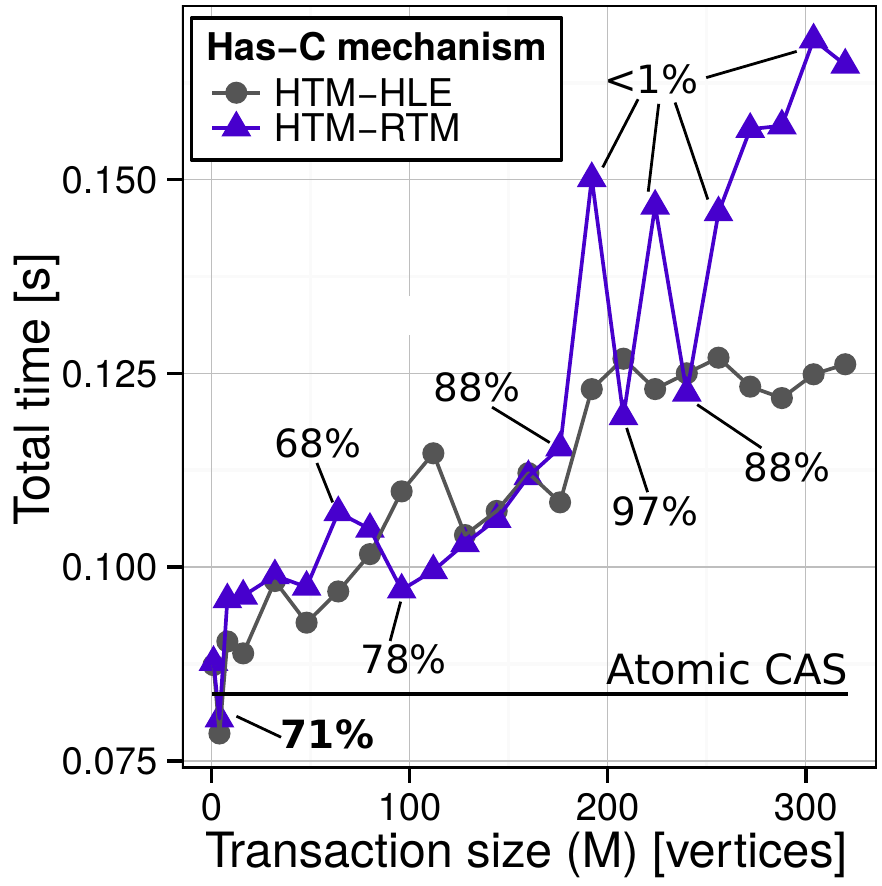}
  \label{fig:bfs_coal_study_t4_haswell}
 }
  \subfloat[BFS runtime ($T=8$) (\cref{sec:haswell-c_multivertex}).]{
  \includegraphics[width=0.235\textwidth]{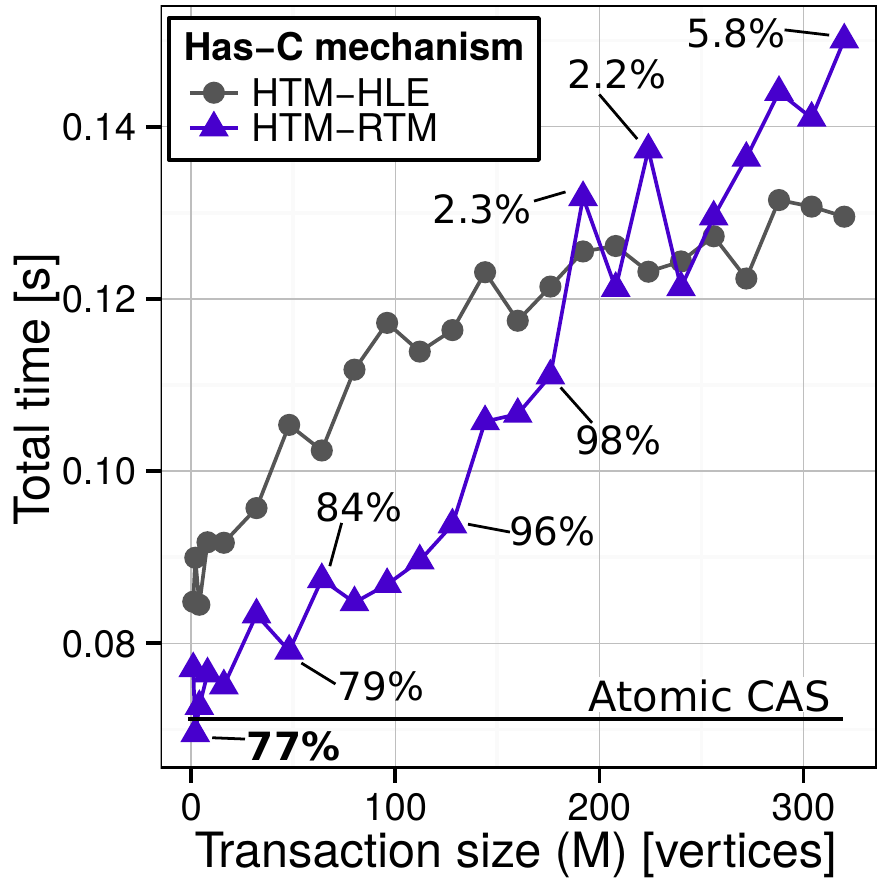}
  \label{fig:bfs_coal_study_t8_haswell}
 }
    \subfloat[Has-C events ($T=8$) (\cref{sec:haswell-c_multivertex}).]{
  \includegraphics[width=0.235\textwidth]{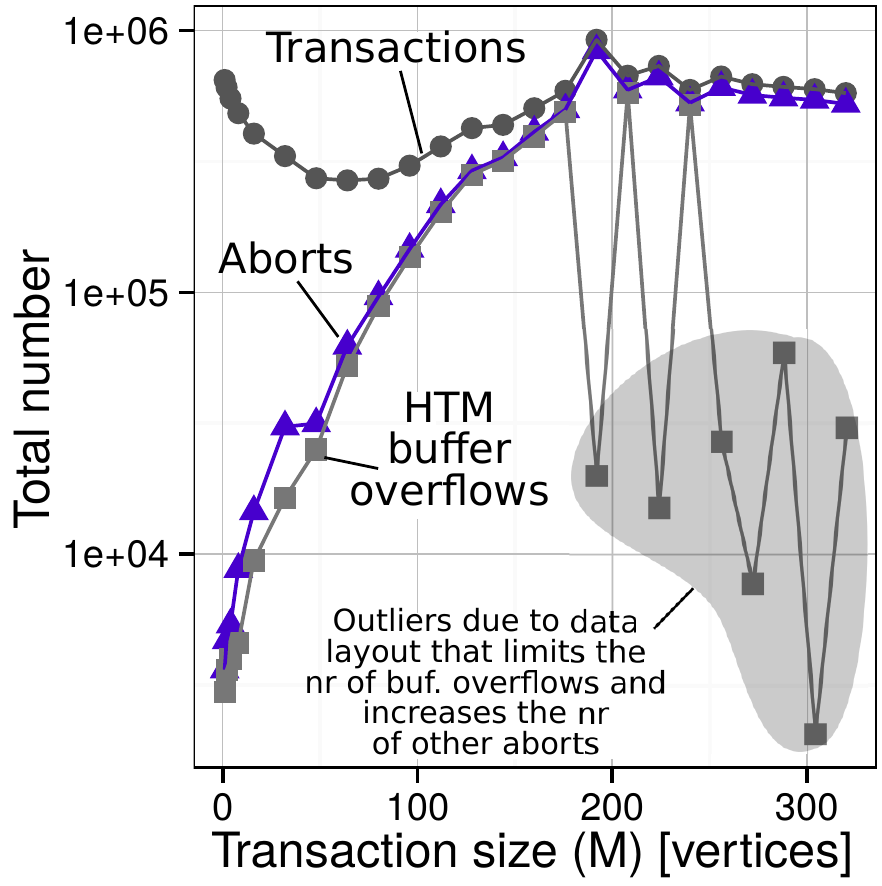}
  \label{fig:bfs_coal_study_t8_haswell_details}
 }\\
  \subfloat[BFS runtime ($T=1$) (\cref{sec:haswell-p_multivertex}).]{
  \includegraphics[width=0.235\textwidth]{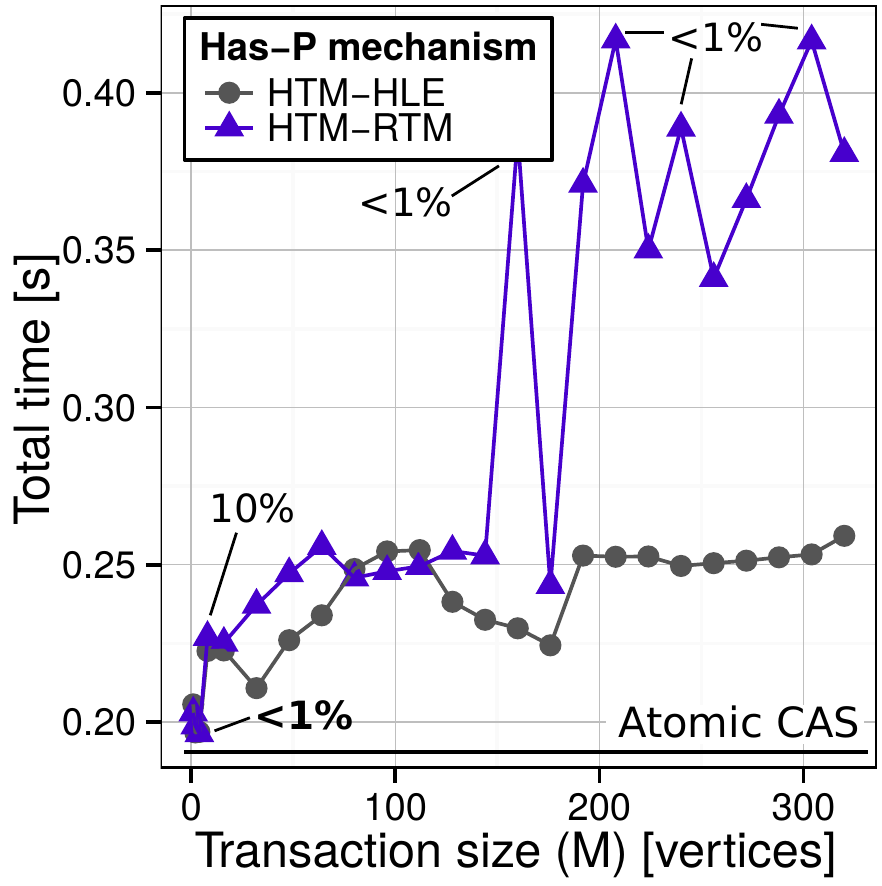}
  \label{fig:bfs_coal_study_t1_haswell_greina}
 }
 \subfloat[BFS runtime ($T=12$) (\cref{sec:haswell-p_multivertex}).]{
  \includegraphics[width=0.235\textwidth]{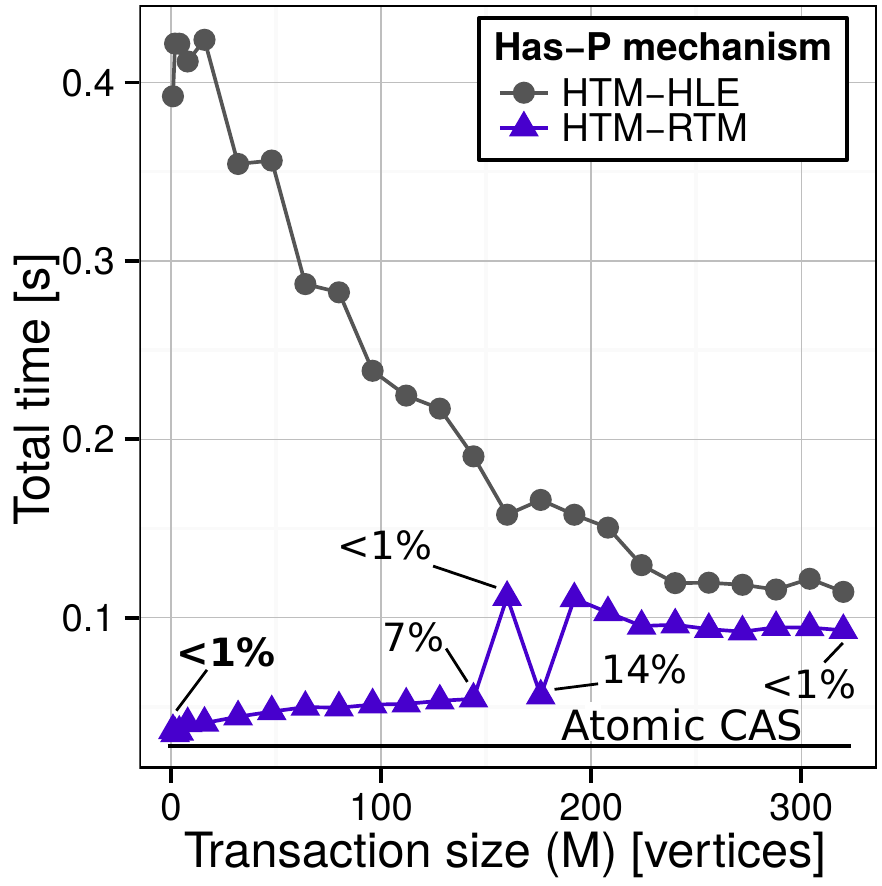}
  \label{fig:bfs_coal_study_t12_haswell_greina}
 }
  \subfloat[BFS runtime ($T=24$) (\cref{sec:haswell-p_multivertex}).]{
  \includegraphics[width=0.235\textwidth]{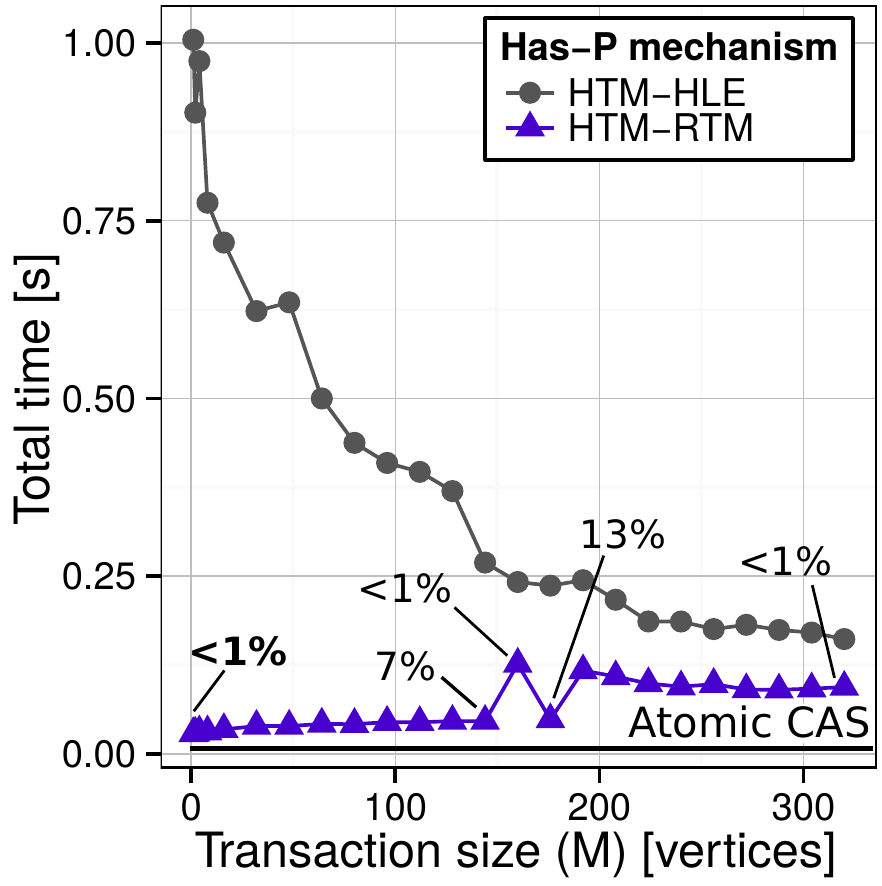}
  \label{fig:bfs_coal_study_t24_haswell_greina}
 }
    \subfloat[Has-P events ($T=24$) (\cref{sec:haswell-p_multivertex}).]{
  \includegraphics[width=0.235\textwidth]{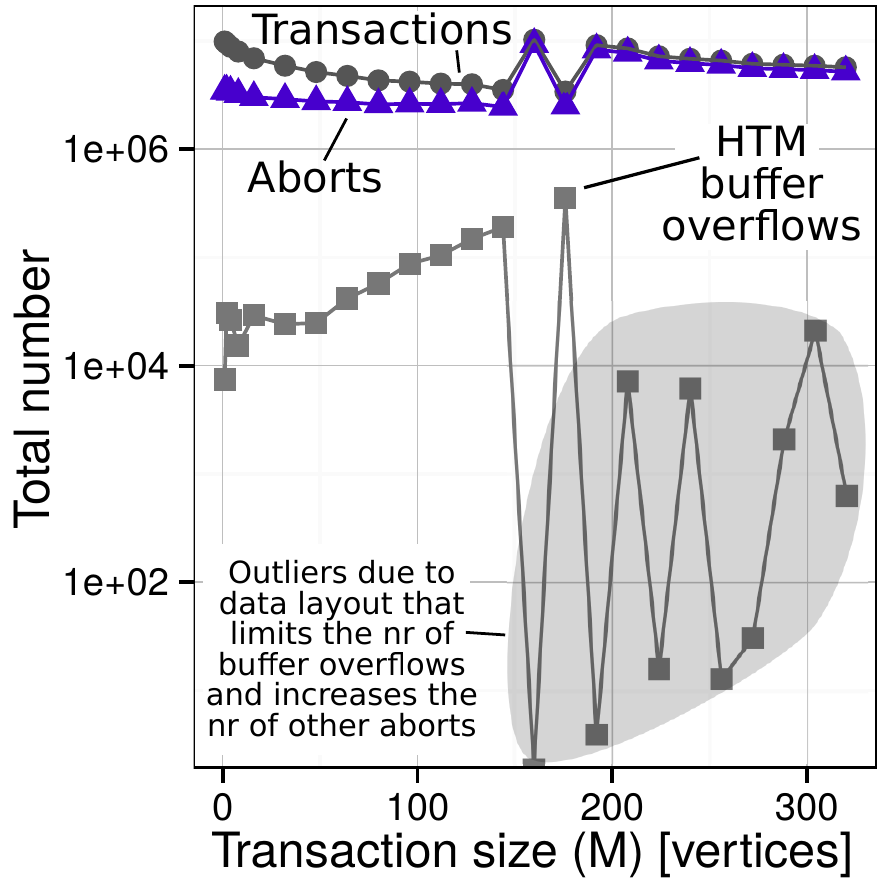}
  \label{fig:bfs_coal_study_t24_haswell_details_greina}
 }
 \vspace{-0.8em}
 \caption{(\cref{sec:multivertex-analysis-bfs}) The analysis of the performance of Graph500 OpenMP BFS implemented with hardware transactions on \textsf{BGQ} (Figures~\ref{fig:bfs_coal_study_t1_bgq}-\ref{fig:bfs_coal_study_t64_bgq_details}), \textsf{Has-C} (Figures~\ref{fig:bfs_coal_study_t1_haswell}-\ref{fig:bfs_coal_study_t8_haswell_details}), and \textsf{Has-P} (Figures~\ref{fig:bfs_coal_study_t1_haswell_greina}-\ref{fig:bfs_coal_study_t24_haswell_details_greina}). In each figure we vary the size of the transactions \smalltt{M} (i.e., the number of vertices visited). We also present the results for BFS implemented with atomics (horizontal lines). For BGQ, the percentages indicate the ratios of the numbers of serializations caused by reaching the maximum possible number of rollbacks to the numbers of all the aborts.
 For Haswell, the percentages are the ratios of the
  aborts due to HTM buffer overflows to all the aborts. Bolded numbers indicate the points with the minimum runtime per figure. We do not include the numbers for Haswell HLE because it does not enable gathering more detailed statistics~\cite{tsx-sc}. Figures~\ref{fig:bfs_coal_study_t64_bgq_details},~\ref{fig:bfs_coal_study_t8_haswell_details}, and~\ref{fig:bfs_coal_study_t24_haswell_details_greina} present the total number of HTM events (transactions, aborts, buffer overflows) for every analyzed $M$.}
 
%\vspace{-1em}
\label{fig:bfs_coal_study}
\end{figure*}

\subsection{Single-vertex Activities}
\label{sec:intra-node-act}
%\vspace{-0.1em}

First, we analyze the performance of single-vertex activities.
The results are illustrated in Figure~\ref{fig:intra-node-analysis}.
\textsf{Has-C} and \textsf{Has-P} follow similar performance trends and we show only the former (denoted as \textsf{Has}); we thus illustrate the results for both a multicore off-the-shelf system and a manycore high performance machine (\textsf{BGQ}).
\subsubsection{Activity 1: Marking a Vertex as Visited}
\label{sec:intra-node-act-marking}
%\vspace{-0.1em}

Here, each thread
uses a CAS or
an equivalent HTM code to atomically mark a single vertex; see Fig.~\ref{fig:onnode_cas_no-cont}-\ref{fig:onnode_cas_cont},
Table~\ref{tab:onnode_cas_dets}. This activity may be used in BFS or
any other related algorithm such as Single Source Shortest Path (SSSP). 
We analyze a negligibly contended scenario that addresses sparse graphs (Fig.~\ref{fig:onnode_cas_no-cont}; a vertex is marked 10 times to simulate low contention) and a more contended case for dense graphs with high $\bar{d}$ (Fig.~\ref{fig:onnode_cas_cont}; a vertex is marked 100 times).
We repeat the benchmark 1000 times and derive the average total time to finish the operations.

Figure~\ref{fig:onnode_cas_no-cont} shows that \smallsf{Has-CAS} finishes fastest and is slightly impacted by the
increasing $T$ ($\approx$50\% of difference between the results for $T=4$ and $T=8$). This is because \smallsf{Has-CAS} locks the respective cache line, causing contention in the memory system.
Both \smallsf{Has-RTM} and \smallsf{Has-HLE} have 1.5-3x higher latency than \smallsf{Has-CAS}, with \smallsf{Has-RTM} being 5-15\% faster than \smallsf{Has-HLE}. Their performance is not influenced by the increasing $T$ as they rarely abort.
Then, \smallsf{BGQ-HTM-S} and \smallsf{BGQ-HTM-L} are more sensitive to the
growing $T$ and their performance drops 11x when switching from
$T=1$ to $T=64$ due to expensive aborts.
As expected, \smallsf{BGQ-HTM-S} is faster than \smallsf{BGQ-HTM-L}, but as $T$ increases it also aborts more frequently, and becomes $\approx$2x less efficient ($T=64$) with $37.5\%$ more aborts.
\smallsf{BGQ-CAS} is least affected by the increasing $T$.

%%%\smallsf{BGQ-CAS}, \smallsf{BGQ-HTM-S}, and \smallsf{BGQ-HTM-L} are 7.2x, 30.1x, and 57.8x slower than \smallsf{Has-HLE} for $T=1$. HTM in BG/Q has much higher latency that HTM in Haswell because the former is implemented in L2 cache while the latter operates in very fast L1 cache~\cite{Wang:2012:EBG:2370816.2370836,tsx-sc}.
%%%%
%%%As expected, \smallsf{BGQ-HTM-S} is faster than \smallsf{BGQ-HTM-L}, but as $T$ increases it also aborts more frequently, and becomes $\approx$2 times less efficient for $T=64$ with $37.5\%$ more aborts.

Figure~\ref{fig:onnode_cas_cont} shows that \smallsf{Has-RTM}, \smallsf{BGQ-CAS}, \smallsf{BGQ-HTM-S}, and \smallsf{BGQ-HTM-L} follow similar performance patterns when threads access the vertex 100 times. The performance of \smallsf{Has-HLE} drops rapidly as it always performs the costly serialization after the first abort and thus forces all other transactions to abort. The latency of \smallsf{Has-CAS} grows proportionally to the contention in the memory system. It stabilizes at $T=8$ as for $T>8$ no more operations can be
issued in parallel.

\begin{figure*}
%\vspace{-1.7em}
\centering
 %\endminipage\hfill
 \subfloat[Aborts; Has-C.]{
  \includegraphics[width=0.19\textwidth]{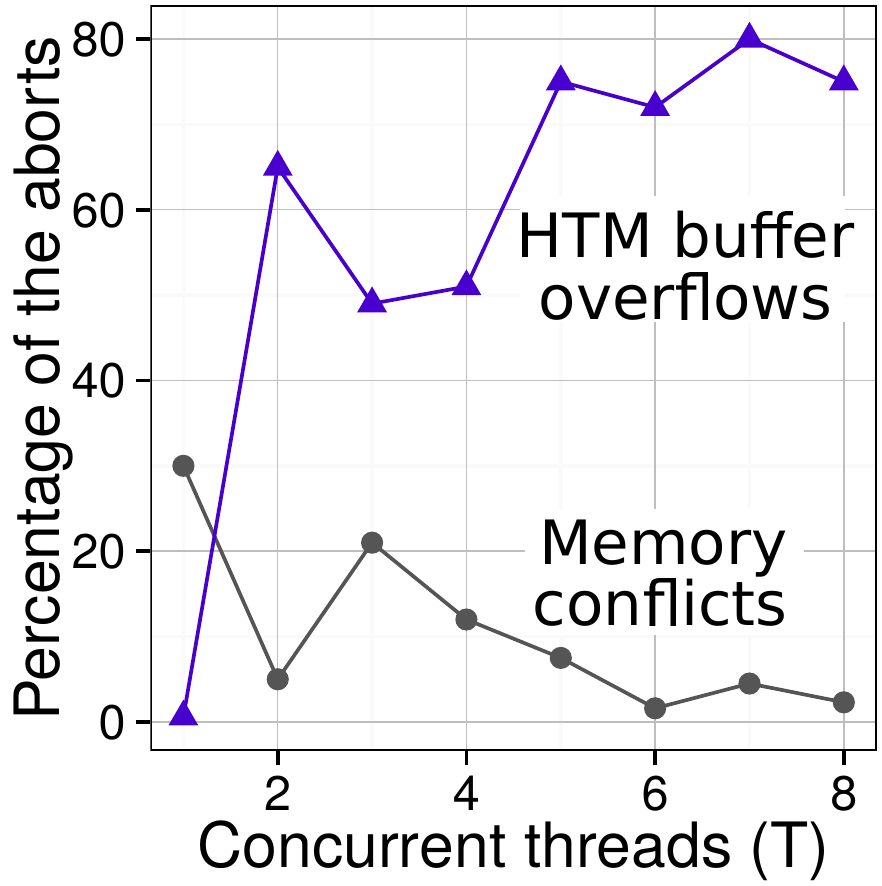}
  \label{fig:aborts-study-galilei}
 }%\hfill
  \subfloat[Aborts; Has-P.]{
  \includegraphics[width=0.19\textwidth]{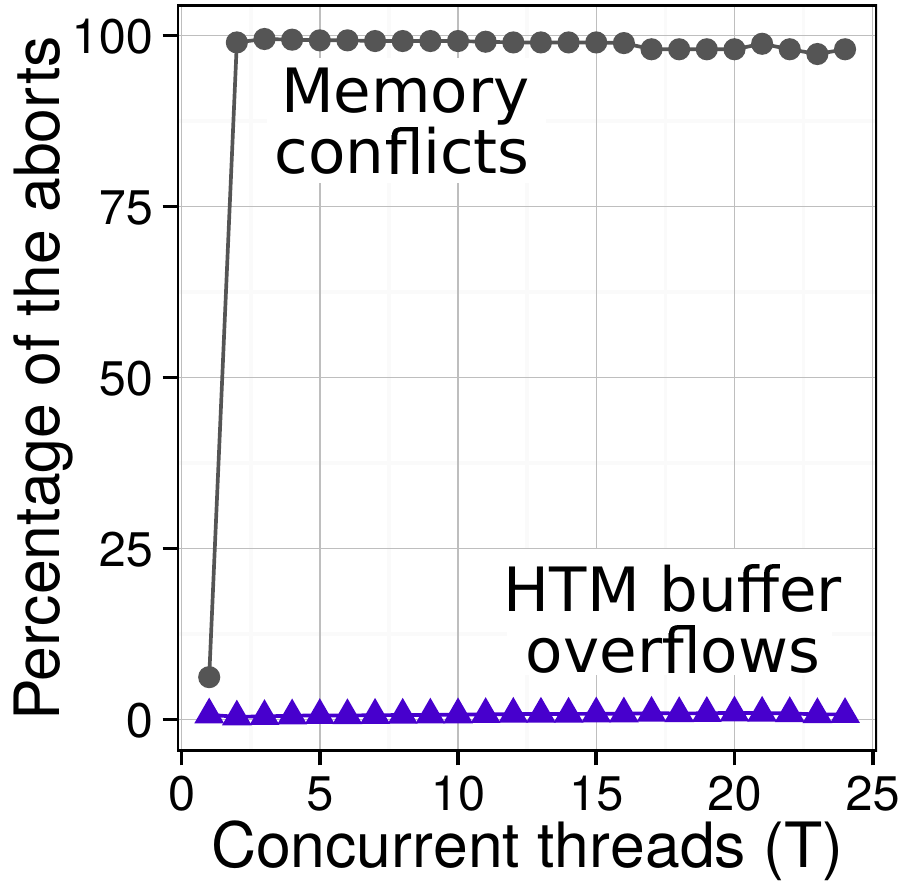}
  \label{fig:aborts-study-greina}
 }%\hfill 
 \subfloat[Remote CAS; BGQ.]{
  \includegraphics[width=0.19\textwidth]{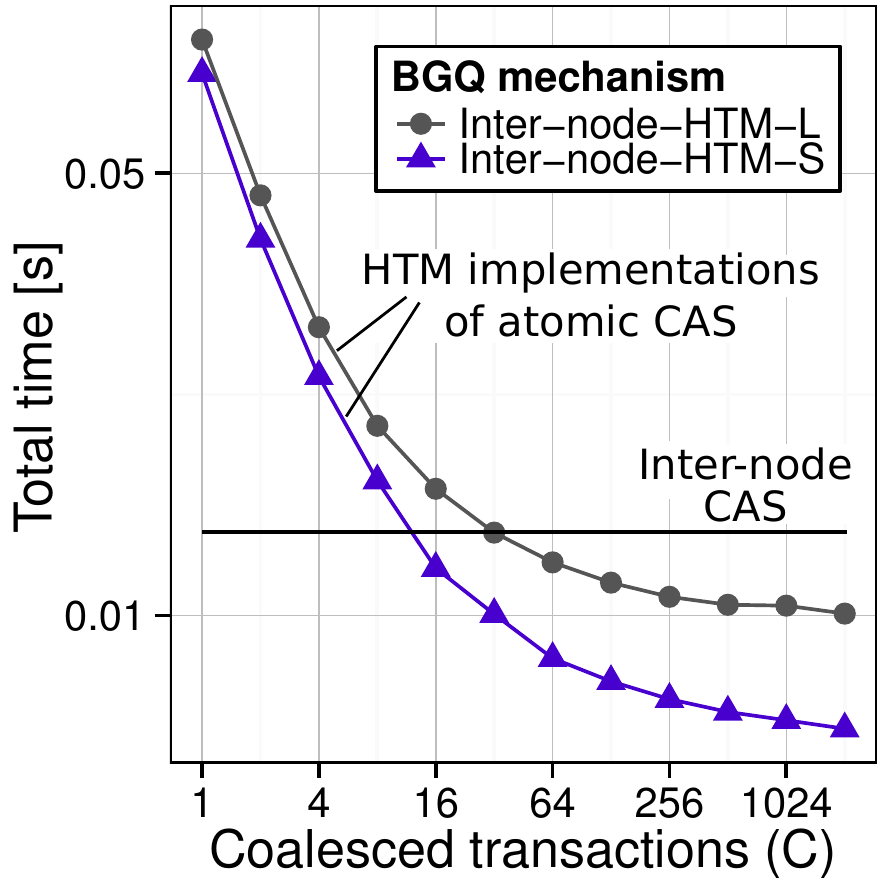}
  \label{fig:inter_cas_coal}
 }%\hfill
    \subfloat[Remote CAS; BGQ.]{
  \includegraphics[width=0.19\textwidth]{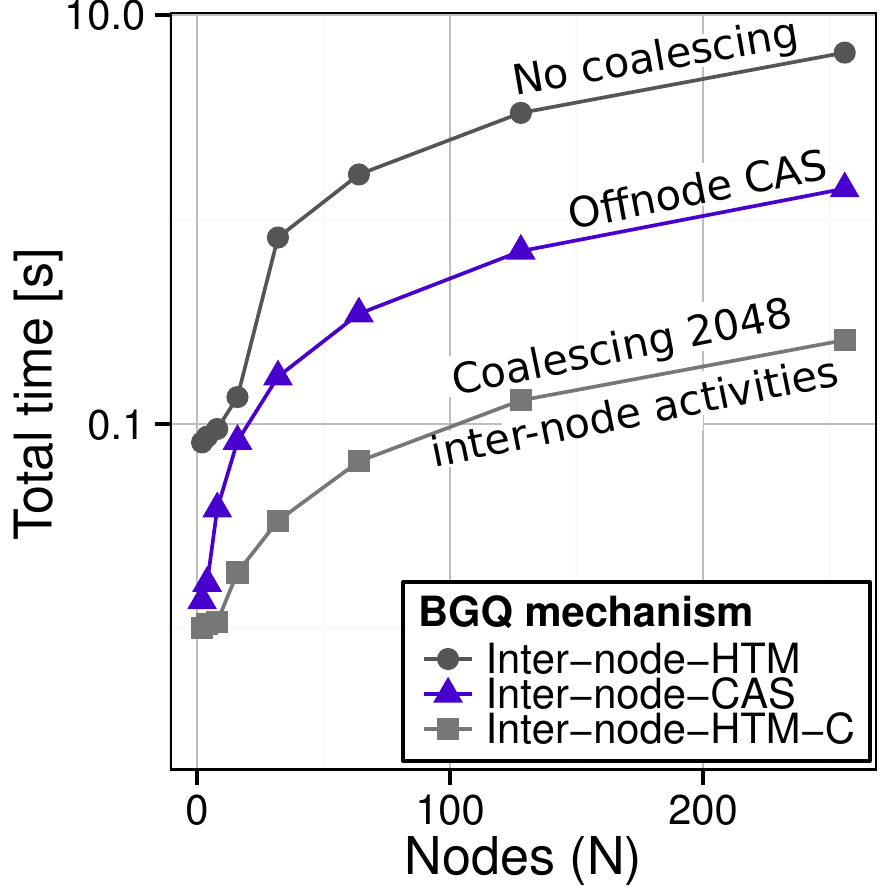}
  \label{fig:inter_cas_coal_nodes}
 }\\%\hfill
  \subfloat[Remote ACC; BGQ.]{
  \includegraphics[width=0.19\textwidth]{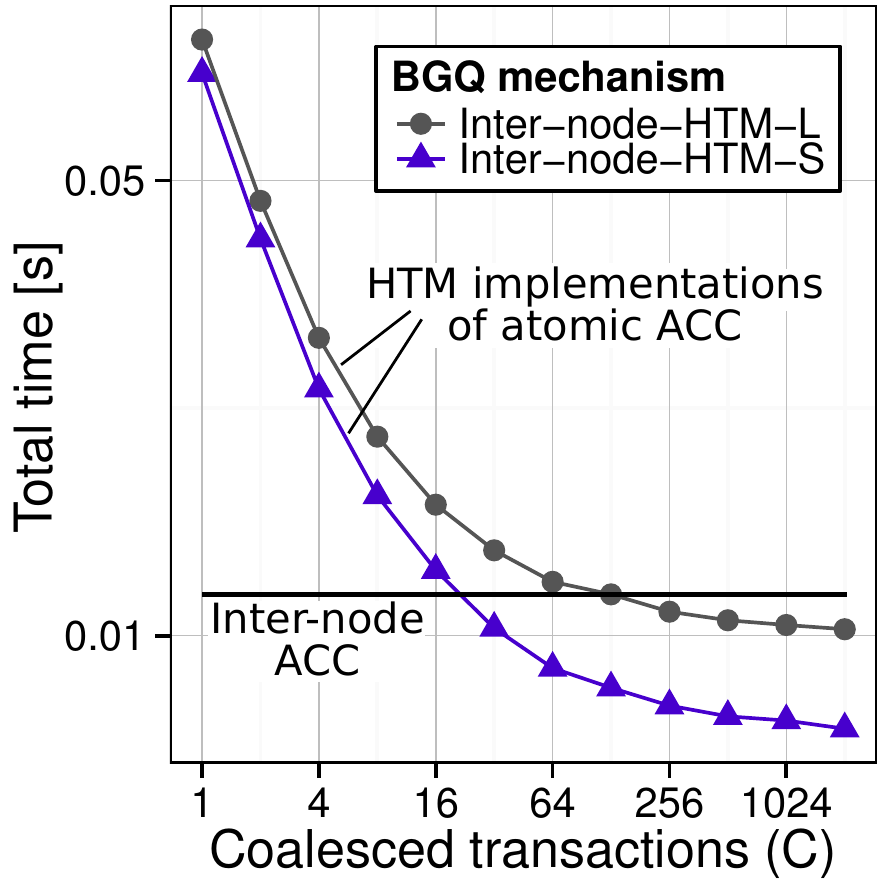}
  \label{fig:inter_fao_coal}
 }%\hfill
 \subfloat[Remote ACC; BGQ.]{
  \includegraphics[width=0.19\textwidth]{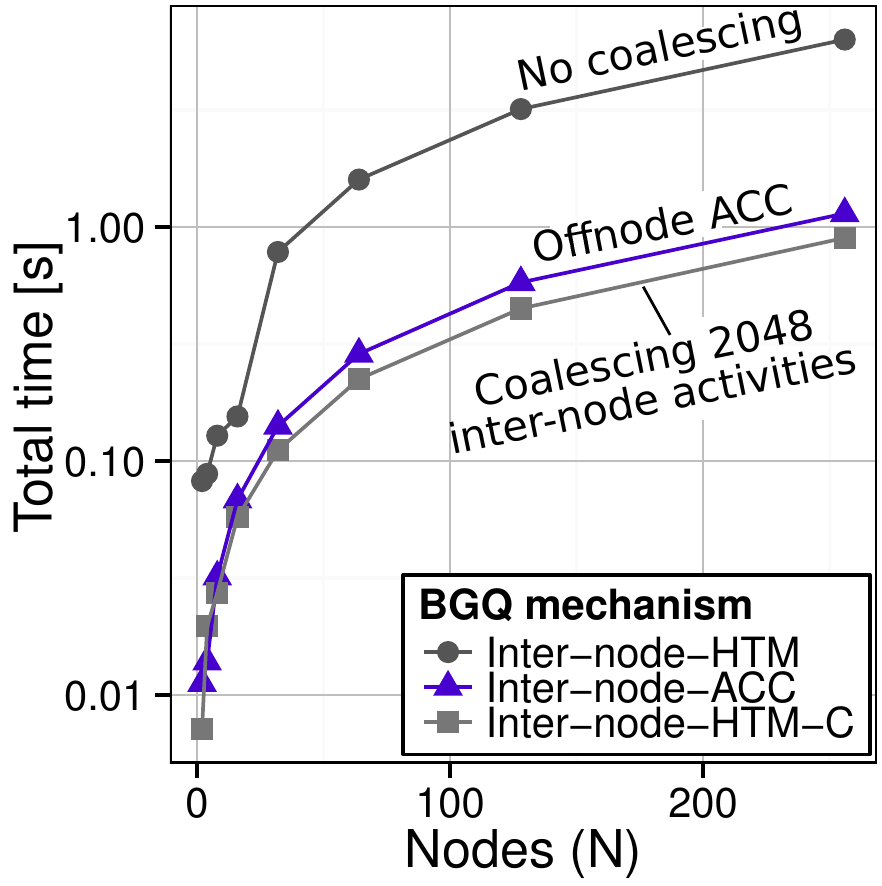}
  \label{fig:inter_fao_coal_nodes}
 }
  \subfloat[Remote CAS; Has-P]{
  \includegraphics[width=0.19\textwidth]{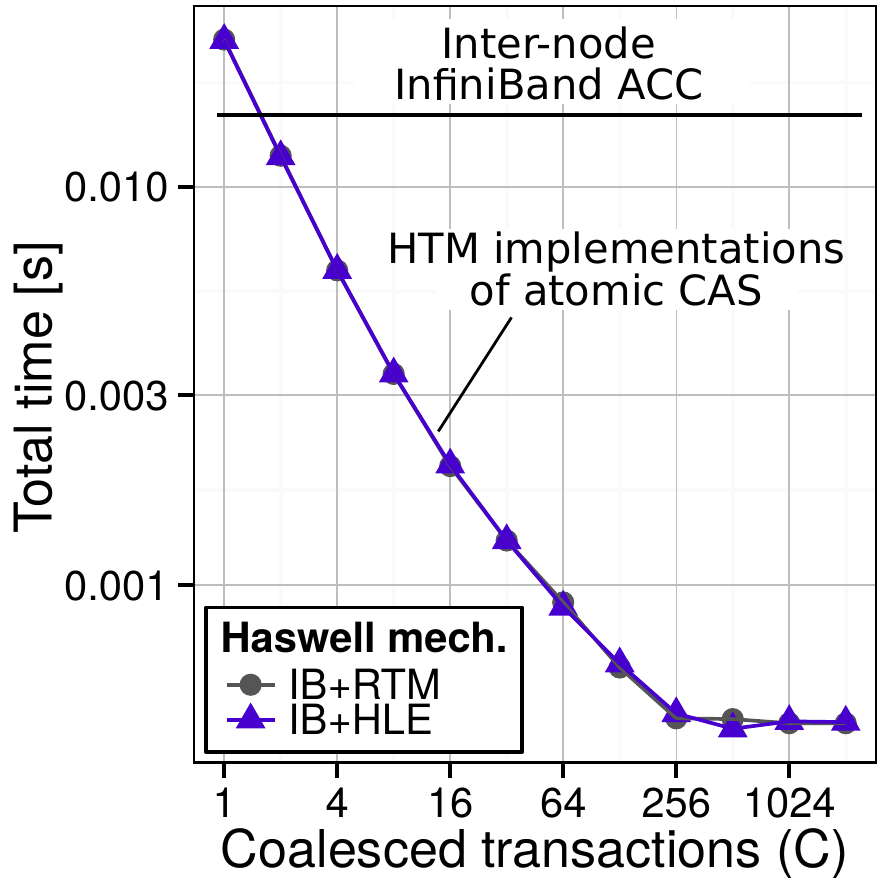}
  \label{fig:inter_cas_coal_greina}
 }%\hfill
  \subfloat[Remote ACC; Has-P.]{
  \includegraphics[width=0.19\textwidth]{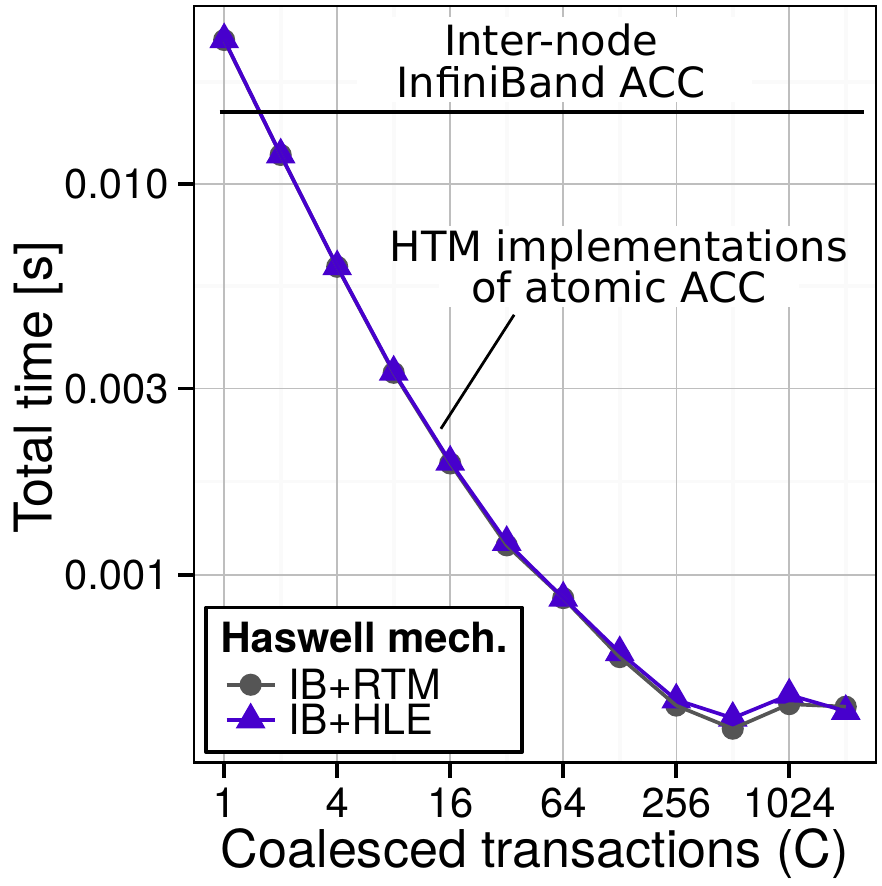}
  \label{fig:inter_fao_coal_greina}
 }%\hfill
 % \hfill
 \subfloat[Dist. transactions; BGQ.]{
  \includegraphics[width=0.19\textwidth]{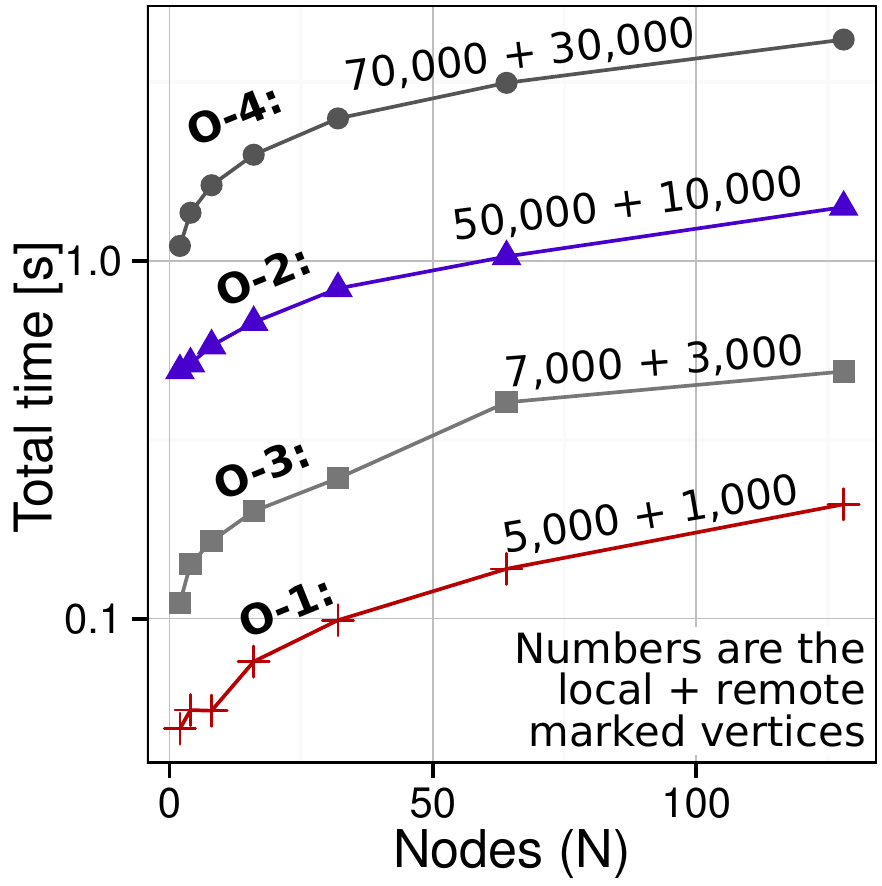}
  \label{fig:ownership}
 }

 \vspace{-0.8em}
 \caption{(\cref{sec:multivertex-analysis-bfs} \& \cref{sec:inter-node-bench}) 
The comparison of the percentage of the reasons of aborts on Has-C and Has-P (Figures~\ref{fig:aborts-study-galilei} and~\ref{fig:aborts-study-greina})
 and
the analysis of the performance of inter-node activities on BG/Q and Has-P. The results for BG/Q are following: marking a remote vertex as visited (Figures~\ref{fig:inter_cas_coal} and~\ref{fig:inter_cas_coal_nodes}), incrementing a vertex' rank (Figures~\ref{fig:inter_fao_coal} and~\ref{fig:inter_fao_coal_nodes}), and executing distributed transactions (Figure~\ref{fig:ownership}).
The results for Haswell are following: marking a remote vertex as visited (Figure~\ref{fig:inter_cas_coal_greina}) and incrementing a vertex' rank (Figure~\ref{fig:inter_fao_coal_greina}).
 }
 
%\vspace{-1em}
\label{fig:inter_coal_study}
\end{figure*}

\subsubsection{Activity 2: Incrementing Vertex Rank}
\label{sec:intra-node-act-increment}
%\vspace{-0.1em}

This activity can be used to implement PR.
Here, each thread
increments the rank of a single vertex 10 times (Figure~\ref{fig:onnode_acc_no-cont}) and 100 times (Figure~\ref{fig:onnode_acc_cont}) with an ACC or an equivalent HTM code;
see Table~\ref{tab:onnode_cas_dets} for details.
The most significant difference between the previous and the current benchmark is that the total time and the number of aborts of \smallsf{Has-RTM} and \smallsf{Has-HLE}
grow very rapidly in both scenarios as $T$ scales. This is because in the HTM implementation of ACC, the rank of the vertex
is modified by each transaction, generating a considerable number of conflicts and thus aborts. On the contrary, the HTM implementation of CAS generates few memory conflicts: once the vertex id is swapped, other threads only read it and do not modify it. \smallsf{BGQ-HTM-S} and \smallsf{BGQ-HTM-L} follow
a similar trend, with $\approx$3x more aborts than in the previous CAS benchmark.

%\vspace{0.5em}

%\vspace{-0.2em}
%\subsubsection{Discussion}
%\label{sec:single-vertex-act-discussion}
%\vspace{-0.3em}

\noindent
\textbf{\textsf{Discussion }}
%In all analyzed cases HTM in BG/Q has in general much ($\approx$41-57x) higher latency than HTM in Haswell because it is implemented in slower L2 cache~\cite{tsx-sc,Wang:2012:EBG:2370816.2370836}. 
We present the details of the above analysis in Tables~\ref{tab:onnode_cas_dets} and~\ref{tab:onnode_acc_dets}.
%; they contain the distribution
%of the reasons of aborts in analyzed HTM implementations.
%
We show that the considered single-vertex activities are in most cases best implemented with atomics. 
HTM is faster only in processing dense graphs with algorithms that use CAS (e.g., BFS) on Haswell.
We also conclude that while atomic CAS is more expensive than ACC, HTM implementation of single ACC is slower ($\approx$100x for RTM and $\approx$10x for BG/Q HTM)
than that of CAS as it generates more memory conflicts and thus costly aborts.

%\vspace{-0.2em}
\subsection{Multi-vertex Activities}
\label{sec:multivertex-analysis-bfs}
%\vspace{-0.1em}

The performance analysis of single-vertex intra-node activities illustrates that in
most cases a transaction modifying a single vertex is slower than an atomic
operation. We now analyze if it is possible to amortize
the cost of starting and aborting transactions by enlarging their size, i.e., \emph{coarsening}.
This section extends the model analysis (\cref{sec:perf_model}) by introducing effects such as
memory conflicts or HTM buffer overflows.
%
%%%The performance analysis of single-vertex intra-node activities illustrates that in
%%%most cases a transaction modifying a single vertex is slower than an atomic
%%%operation. We now analyze if it is possible to amortize
%%%the cost of starting and aborting transactions by enlarging their size, i.e., \emph{coarsening}.
%
We perform the analysis for 
the highly-optimized OpenMP BFS Graph500 code~\cite{murphy2010introducing}.
We modify the code so that
a single transaction atomically visits $M$ vertices and we evaluate
the modified code for
$M$ between $1$ and $320$ with the interval of 16.
%
%For both BG/Q and Haswell we test the performance of BFS for $T=1,2,4,8,16,32,64$.
We present the results for three scenarios: a single-threaded execution ($T=1$ for \textsf{BGQ}, \textsf{Has-C}, and \textsf{Has-P}),
a single thread per core ($T=16$ for BG/Q, $T=4$ for \textsf{Has-C}, and $T=12$ for \textsf{Has-P}), and a single thread per SMT hardware resource
($T=64$ for BG/Q, $T=8$ for \textsf{Has-C}, and $T=24$ for \textsf{Has-P}).
We use Kronecker
graphs~\cite{Leskovec:2010:KGA:1756006.1756039} with the power-law vertex degree distribution and $|V|=2^{20}, |E|=2^{24}$. The results are
shown in Figure~\ref{fig:bfs_coal_study}.

%We use R-MAT
%graphs~\cite{Chakrabarti04r-mat:a} with the parameters $A=0.57, B=0.19, C=0.19$ and $|V|=2^{20}, |E|=2^{24}$. The results of the analysis are
%presented in Figure~\ref{fig:bfs_coal_study}.

%\vspace{-0.5em}
\subsubsection{BG/Q (Supercomputer)}
\label{sec:bgq_multivertex}
%\vspace{-0.1em}

Figures~\ref{fig:bfs_coal_study_t1_bgq}-\ref{fig:bfs_coal_study_t64_bgq_details} present the analysis for BG/Q.
For $T=1$ the runtime of both \smalltt{HTM-Long-Mode} and \smalltt{HTM-Short-Mode} is always higher than that of \smalltt{Atomic-CAS} and it decreases initially with the increasing $M$ because higher $M$ reduces the number of transactions required to process the whole graph and thus amortizes the overhead of starting and committing transactions.
The runtime of \smalltt{HTM-Short-Mode} becomes higher with the increasing $M>32$ because this mode is better suited for short transactions. The runtime of \smalltt{HTM-Long-Mode}
decreases as expected and it stabilizes at $M \approx 240$.
%
%For $T=16$ the results become influenced by the overhead of aborts.
%
For $T=16$, initially the runtime drops rapidly for both HTM modes to reach the minimum (obtained for $M_{min}=80$ in \smalltt{HTM-Short-Mode}). Again, this effect is caused by amortizing the overheads of commits/aborts with coarser transactions. Beyond $M_{min}$ the runtime slowly increases with $M$ due to more frequent serializations caused by reaching the maximum number of allowed rollbacks (BGQ does not enable gathering more detailed statistics but we predict that these serializations are due to the higher number of HTM buffer overflows and memory conflicts). %Both modes have several additional local minimums due to advantaga
\smalltt{HTM-Long-Mode} is never more efficient than \smalltt{Atomic-CAS}. \smalltt{HTM-Short-Mode} becomes more efficient than \smalltt{Atomic-CAS} for $M=32$ and achieves the speedup of 1.11 at $M_{min}=80$.
A similar performance pattern can be observed for $T=64$; this time $M_{min}=144$ in \smalltt{HTM-Short-Mode} with the speedup of 1.49 over \smalltt{Atomic-CAS}. The runtime becomes dominated by aborts for $M>144$; cf.~Figure~\ref{fig:bfs_coal_study_t64_bgq_details} with more detailed numbers of aborts.

%\begin{figure}
%\vspace{-1.5em}
%\centering
%  \includegraphics[width=0.45\textwidth]{bfs_bgq_edges_vertices_labels-eps-converted-to.pdf}
%\label{fig:bgq-graph-study}
%\end{figure}

%\vspace{-0.1em}
\subsubsection{Has-C (Commodity Machine)}
\label{sec:haswell-c_multivertex}
%\vspace{-0.1em}

The results of the analysis for \textsf{Has-C} are presented in Figures~\ref{fig:bfs_coal_study_t1_haswell}-\ref{fig:bfs_coal_study_t8_haswell_details}.
In each scenario ($T=1,4,8$) the performance of both \smalltt{HTM-RTM} and \smalltt{HTM-HLE} decreases with increasing $M$. Several outliers are caused by disadvantageous graph data layouts that entail more aborts due to the limited associativity of L1 cache (8-way associative cache) that stores speculative states in Haswell~\cite{tsx-sc}. We perform a more detailed analysis for $M \in \{1,...,16\}$ to find out that $M_{min} = 2$. \smalltt{HTM-RTM} becomes less efficient than \smalltt{HTM-HLE} at M$\approx$200 because the cost of serializations due to the HTM buffer overflows dominates the runtime of \smalltt{HTM-RTM} beyond this point (serializations in \smalltt{HTM-HLE} are implemented in hardware~\cite{tsx-sc}, while in \smalltt{HTM-RTM} they have to be implemented in software).

\begin{figure*}
%\vspace{-1.5em}
\centering
 %\endminipage\hfill
 \subfloat[The performance of BFS on BG/Q for Kronecker graphs.]{
  \includegraphics[width=0.4\textwidth]{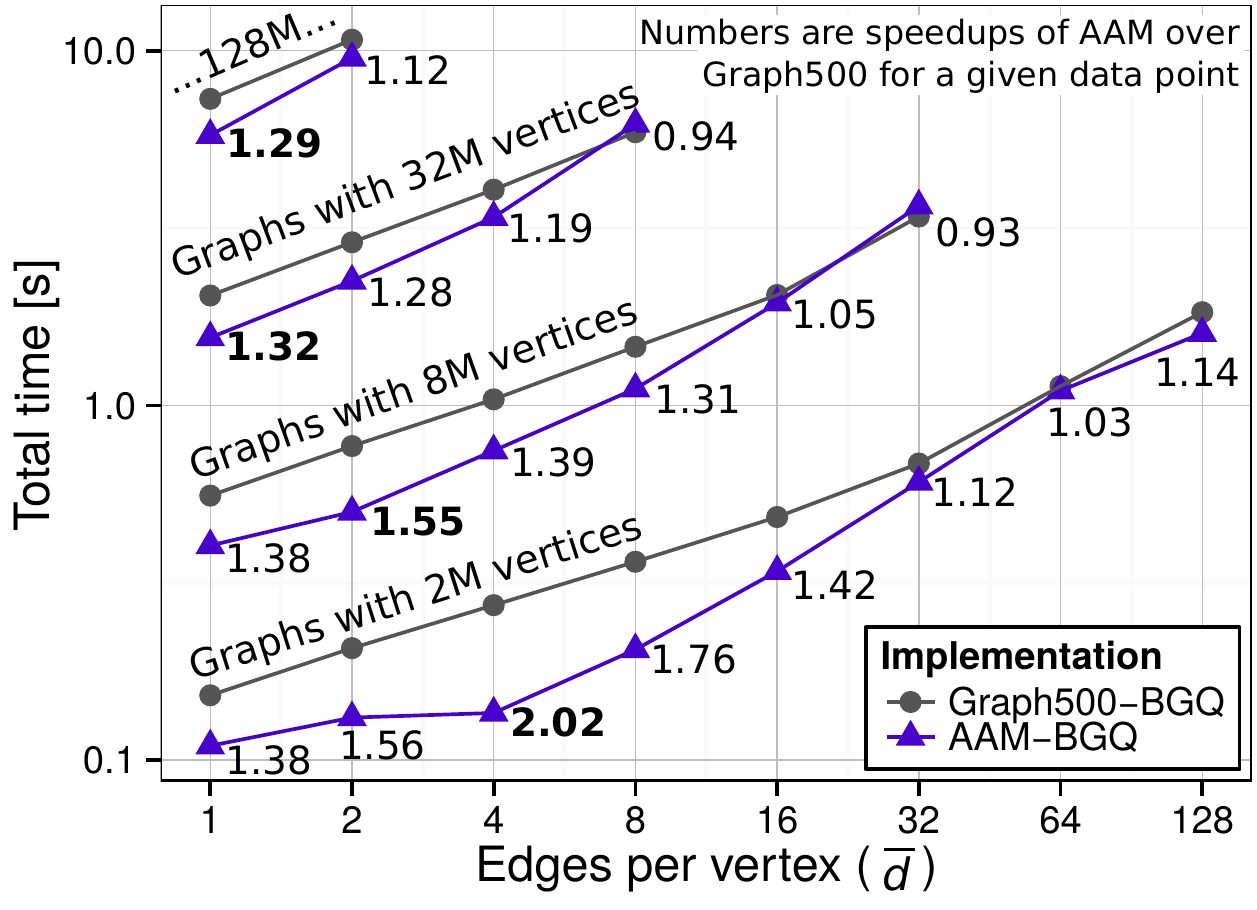}
  \label{fig:graphs_dets_bgq}
 }
 \subfloat[The performance of BFS on Haswell for Kronecker graphs.]{
  \includegraphics[width=0.4\textwidth]{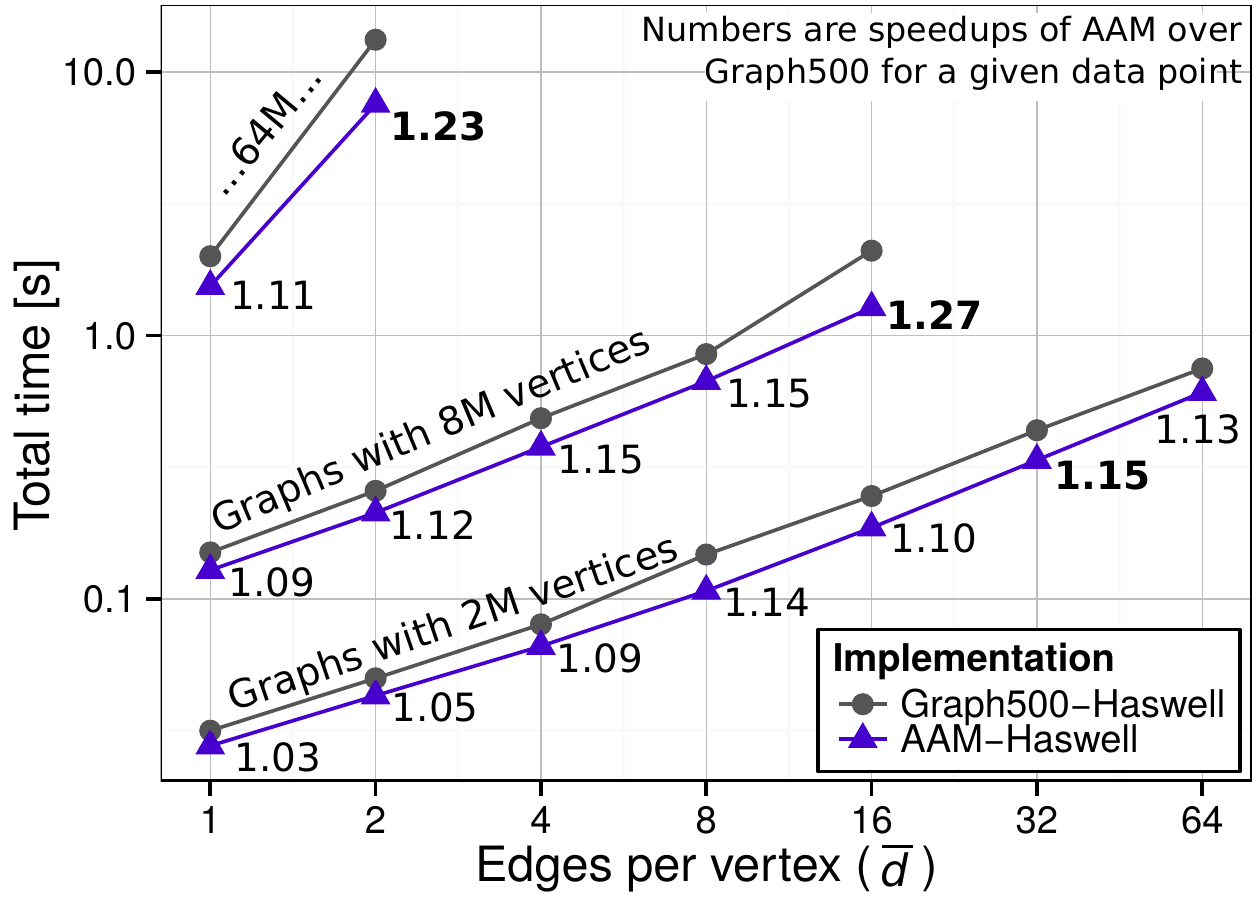}
  \label{fig:graphs_dets_haswell}
 }
%  \subfloat[The performance of BFS on BG/Q for real-world SNAP~\cite{snap} graphs  (\cref{sec:bfs_onnode_eval_bgq_real}).]{
%  \includegraphics[width=0.2\textwidth]{bgq_real_graphs_rev-eps-converted-to.pdf}
%  \label{fig:graphs_real_bgq}
% }
 \vspace{-1.0em}
  \caption{(\cref{sec:bfs_onnode_eval}) The overview of the performance of intra-node Graph500 BFS implemented with atomics (\smalltt{Graph500-BGQ}, \smalltt{Graph500-Haswell}), AAM RTM (\smalltt{AAM-Haswell}), and the short mode BG/Q HTM (\smalltt{AAM-BGQ}). We vary $|V|$ and $\bar{d}$; $T=64$ (on BG/Q) and $T=8$ (on Has-C).
 %We use 8 threads (on Haswell) and 64 threads (on BG/Q).
 }
% \caption{(\cref{sec:bfs_onnode_eval}) The overview of the performance of intra-node Graph500 BFS implemented with atomics (\smalltt{Graph500-BGQ}, \smalltt{Graph500-Haswell}), AAM RTM (\smalltt{AAM-Haswell}), and the short mode BG/Q HTM (\smalltt{AAM-BGQ}). We vary $|V|$ and $\bar{d}$.
% %We use 8 threads (on Haswell) and 64 threads (on BG/Q).
% }
%\vspace{-1em}
\label{fig:graphs_dets_perf}
\end{figure*}

%\vspace{-0.1em}
\subsubsection{Has-P (High-Performance Server)}
\label{sec:haswell-p_multivertex}
%\vspace{-0.1em}

The analysis of \textsf{Has-P} is presented in Figures~\ref{fig:bfs_coal_study_t1_haswell_greina}-\ref{fig:bfs_coal_study_t24_haswell_details_greina}.
The performance trends are partially similar to the observations for \textsf{Has-C}; especially for lower thread counts ($T \leq 4$).
A distinctive feature is a significantly lower number of HTM buffer overflows than in \textsf{Has-C}. To gain more insight we performed an additional analysis to compare the number of memory conflicts and HTM buffer overflows with varying $T$ for fixed $M=2$. We present the results in Figures~\ref{fig:aborts-study-galilei}-\ref{fig:aborts-study-greina}.
Surprisingly, we observe \textsf{Has-C} has significantly more buffer overflows than memory conflicts for the increasing $T$; a reverse trend is observed on \textsf{Has-P}.
This interesting insight may help improve the design of future HTM architectures.
%
%%%Surprisingly, \textsf{Has-C} has significantly more buffer overflows than memory conflicts for the increasing $T$; a reverse trend is observed on \textsf{Has-P}.
%%%%
%%%We conjecture that the complicated manycore design of \textsf{Has-P} generates significantly more memory conflicts than a simple multicore architecture of \textsf{Has-C}.
%%%%
%%%On the contrary, \textsf{Has-P} offers more hardware resources limiting the number of buffer overflows in comparison with \textsf{Has-C}.

%\begin{figure}[h!]
%\vspace{-1.5em}
%\centering
% %\endminipage\hfill
% \subfloat[Section~\cref{sec:haswell-p_multivertex} Has-C.]{
%%  \includegraphics[width=0.19\textwidth]{cas_offnode_coal_study_labels-eps-converted-to.pdf}
%  \includegraphics[width=0.19\textwidth]{results/aborts-study/galilei_aborts_e-eps-converted-to.pdf}
%  \label{fig:aborts-study-galilei}
% }%\hfill
%  \subfloat[Section~\cref{sec:haswell-p_multivertex} Has-P.]{
%  \includegraphics[width=0.19\textwidth]{results/aborts-study/greina_aborts_e-eps-converted-to.pdf}
%  \label{fig:aborts-study-greina}
% }%\hfill
%
% \vspace{-0.8em}
% \caption{(\cref{sec:haswell-p_multivertex}) 
% The comparison of the percentage of the reasons of aborts on Has-C and Has-P.}
% 
%%\vspace{-1.5em}
%\label{fig:aborts-threads-study}
%\end{figure}

%\vspace{0.3em}

\noindent
\textbf{\textsf{Discussion }}
Our analysis shows that RTM is more vulnerable to aborts than BG/Q HTM. The difference between the number of transactions and aborts never drops below 25\% for HTM in BG/Q for any analyzed $M$ (cf.~Figure~\ref{fig:bfs_coal_study_t64_bgq_details}), while for RTM this threshold is achieved for $M=144$ (\textsf{Has-C}).
%
%Another discovery is that HTM in BG/Q is negligibly influenced by HTM buffer overflows and memory conflicts: for $T=64$ and $M=320$ they constitute only $\approx$5\% of all the aborts.
%
Another discovery is that \textsf{Has-P} is only marginally impacted by buffer overflows (<1\% of all the aborts for $T=24$ and $M=320$). On the contrary, aborts in \textsf{Has-C} are dominated by HTM buffer overflows that constitute more than 90\% of all the aborts for $M>64$.
The only exception are the data points where the number of overflows drops rapidly as aborts become dominated by the limited L1 cache associativity (a similar effect is visible for \textsf{Has-P}). This effect is not visible in BG/Q because it stores its
speculative states in its L2 16-way associative cache~\cite{Wang:2012:EBG:2370816.2370836}, while both \textsf{Has-P} and \textsf{Has-C} have 8-way associative L1s.
%
%We conjecture that manycore high-performance design in \textsf{BGQ} and \textsf{Has-P} 

We conclude that the coarsening of transactions provides significant speedups (up to 1.51) over the \smalltt{Atomic-CAS} baseline on \textsf{BGQ} and \textsf{Has-C}; \textsf{Has-P} does not offer any speedups due to the overheads generated by memory conflicts. We find the following optimum transaction sizes for PowerPC in BG/Q: $M_{min} = 80$ ($T=16$), $M_{min} = 144$ ($T=64$). For \smallsf{x86} (\textsf{Has-C}) $M_{min} = 2$ for $T \in \{4,8\}$. We will use these values in Section~\ref{sec:evaluation} to accelerate Graph500~\cite{murphy2010introducing} for different types of graphs.
%We now proceed to analyze the performance of protocols for executing distributed transactions.
%
% We thus exclude it from the further analysis and we select \textsf{Has-C} to represent \textsf{x86}

%\vspace{-0.1em}
\subsection{Activities Spawned on a Remote Node}
\label{sec:inter-node-bench}
%\vspace{-0.1em}

We now analyze the performance of activities
spawned on a remote node.
We implement such activities as hardware transactions triggered upon
receiving an atomic active message. We again test both the long and the short running
mode (on BG/Q) and RTM/HLE (on Haswell).
To reduce the overhead of sending and receiving an
atomic active message and save bandwidth, we use \emph{activity coalescing}:
activities flowing to the same target are sent in a single message.

We run the benchmarks on BG/Q and Greina (\textsf{Has-P}); we skip \textsf{Has-C} because the Trivium server is not a distributed memory machine.
On BG/Q, we compare inter-node activities to optimized remote one-sided
CAS and ACC atomics provided by the generic function \smallsf{PAMI\_Rmw} in the IBM PAMI communication library~\cite{pami}.
On \textsf{Has-P} we compare activities to remote atomic operations provided by
MPI-3 RMA~\cite{mpi3, besta2015active, gerstenberger2013enabling, besta2014fault} implemented
over the InfiniBand fabric.
We evaluate the performance of marking a remote vertex as visited (addressing
distributed BFS computations) and incrementing the rank of a remote vertex (addressing
distributed PageRank).
%

%\vspace{-0.5em}
%\subsubsection{Marking a Vertex as Visited}
%\label{sec:marking-inter-bench}
%\vspace{-0.3em}

%\vspace{-0.1em}
\subsubsection{BG/Q (Supercomputer)}
\label{sec:inter-bgq}
%\vspace{-0.1em}

We first measure the time it takes a process $p_i$ to mark $2^{13}$ vertices stored
on a node $n_j$ as visited (targeting distributed BFS).
The results are presented in Figure~\ref{fig:inter_cas_coal}.
Without coalescing, HTM activities (\smalltt{Inter-node-HTM-L} for the long and \smalltt{Inter-node-HTM-S} for the short mode) are $\approx$5x slower than
PAMI atomics (\smalltt{Inter-node-CAS}). Still, for $C_{cross} = 16$ \smalltt{Inter-node-HTM-S} becomes more efficient.
Second, we scale the number of nodes $N$. Figure~\ref{fig:inter_cas_coal_nodes}
shows the time to mark a vertex stored in process $p_N$'s memory
by
$N-1$ other processes. We use the short HTM mode.
Coalesced AAMs (\smalltt{Inter-node-HTM-C}) outperform \smalltt{Inter-node-CAS} $\approx$5-7 times.

We also evaluate an inter-node activity that increments the rank of a vertex (targeting distributed PR).
We perform analogous benchmarks as for the remote \textsf{CAS}; % in~\cref{sec:marking-inter-bench}
we present the results in Figures~\ref{fig:inter_fao_coal}-\ref{fig:inter_fao_coal_nodes}.
Implementing ACC using HTM again generates costly aborts that dominate the runtime;
however coalescing enables a speedup of $\approx$20\% (for the short HTM running mode) over highly optimized
PAMI atomics.

%\vspace{-0.4em}
%\subsubsection{Incrementing Vertex Rank}
%\label{sec:inc-inter-bench}
%\vspace{-0.3em}

%\begin{figure}[h!]
%\vspace{-1.5em}
%\centering
% %\endminipage\hfill
% \subfloat[Section~\cref{sec:marking-inter-bench}.]{
%%  \includegraphics[width=0.19\textwidth]{cas_offnode_coal_study_labels-eps-converted-to.pdf}
%  \includegraphics[width=0.19\textwidth]{cas_offnode_coal_study_greina_e-eps-converted-to.pdf}
%  \label{fig:inter_cas_coal_greina}
% }%\hfill
%  \subfloat[Section~\cref{sec:inc-inter-bench}.]{
%  \includegraphics[width=0.19\textwidth]{fao_offnode_coal_study_greina_e-eps-converted-to.pdf}
%  \label{fig:inter_fao_coal_greina}
% }%\hfill
%
% \vspace{-0.8em}
% \caption{(\cref{sec:inter-node-bench}) 
% The analysis of the performance of inter-node activities on two nodes with Haswell CPUs and an InfiniBand network: marking a remote vertex as visited (Figure~\ref{fig:inter_cas_coal_greina} and incrementing a vertex' rank (Figure~\ref{fig:inter_fao_coal_greina}).}
% 
%\vspace{-1.5em}
%\label{fig:inter_coal_study_greina}
%\end{figure}

%\vspace{-0.4em}
%\subsection{Activities that Span Multiple Nodes}
%\label{sec:ownership-bench}
%\vspace{-0.3em}

%\vspace{-0.1em}
\subsubsection{Has-P (High-Performance Server)}
\label{sec:inter-has-p}
%\vspace{-0.1em}

Here, we test the performance of inter-node activities implemented on \textsf{Has-P}.
Our testbed has two nodes, thus we only vary $C$.
We present the results in Figures~\ref{fig:inter_cas_coal_greina} (\textsf{CAS}) and~\ref{fig:inter_fao_coal_greina} (\textsf{ACC}).
Setting $C=2$ enables AAM to outperform remote atomics provided by MPI-3 RMA.

%\vspace{-0.1em}
\subsection{Distributed Activities}
\label{sec:dist_activities_eval}
%\vspace{-0.1em}

Finally, we
test the ownership protocol for executing
activities that span multiple nodes (see Figure~\ref{fig:ownership} for \textsf{BGQ} results).
Each process issues $x$ transactions; each transaction
marks $a$ local and $b$ remote randomly selected vertices. We compare
four scenarios: \smalltt{O-1} ($x=10^3, a=5, b=1$), \smalltt{O-2} ($x=10^4,
a=5, b=1$), \smalltt{O-3} ($x=10^3, a=7, b=3$), and \smalltt{O-4} ($x=10^4,
a=7, b=3$).
We measure the total time to execute transactions. 
\smalltt{O-1} finishes fastest, \smalltt{O-3} is slower as more
remote vertices have to be acquired. \smalltt{O-2} and \smalltt{O-4} follow
the same performance patterns; additional overheads are due to the backoff scheme.
If no time is spent on backoff, then the
protocol may livelock and may make no progress.

We conclude that AAM can be used in various environments (e.g., IBM networks or InfiniBand)
to enable remote transactions and to accelerate distributed processing.

\begin{table*}
%\vspace{-0.5em}
\centering
\scriptsize
\setlength{\tabcolsep}{3pt}
\begin{tabular}{@{}l|llll|c|cc|cc|cccc@{}}
\toprule
\multicolumn{5}{c|}{\textbf{\textsf{Input graph properties}}}                  & \multicolumn{3}{c|}{\textbf{\textsf{BG/Q analysis}}} & \multicolumn{6}{c}{\textbf{\textsf{Haswell analysis}}} \\ \midrule
\multicolumn{1}{c|}{Type}        & ID & Name & $|V|$ & $|E|$ & \parbox{1.6cm}{\centering $S$ over g500\\($M=24$)}  & $M$     & \multicolumn{1}{c|}{\parbox{1.0cm}{\centering $S$ over\\g500}}  & \parbox{1.5cm}{\centering $S$ over g500\\($M=2$)} & \parbox{1.7cm}{\centering $S$ over Galois\\($M=2$)}  & $M$    &  \parbox{1.0cm}{\centering $S$ over\\g500}   &  \parbox{1.0cm}{\centering $S$ over\\Galois}   &  \parbox{1.0cm}{\centering $S$ over\\HAMA}   \\ \midrule
\multirow{2}{*}{\parbox{1.9cm}{\centering Comm. networks (\texttt{CN}s)}}  & cWT & wiki-Talk & 2.4M & 5M  &  2.82   &  48       &  3.35                     & 0.91 & 1.22  & 6    &  0.96   &  1.28       & 344     \\
                                 & cEU & email-EuAll & 265k & 420k & 3.67    &   32      &   4.36                    & 0.76 & 0.88 & 4      & 0.97    &  1.12       & 1448      \\ \midrule
\multirow{6}{*}{\parbox{1.5cm}{\vspace{-0.1em}\centering Social networks\\(\texttt{SN}s)}} & sLV  & soc-LiveJ. & 4.8M & 69M & 1.44     &   12      &  1.56                     & 1.05  & 1.1  & 3     &  1.07   &  1.12       &  $>10^4$     \\
                                 & sOR & com-orkut & 3M & 117M & 1.22     &   20      &   1.27                     & 1.06  & 0.69 & 4      & 1.13    & 0.74        &  $>10^4$   \\
                                 & sLJ & com-lj & 4M & 34M  &  1.44   &    12     &  1.54                     & 1.03  & 1.03 & 4      &  1.04   &  1.04     &  603    \\
                                 & sYT & com-youtube & 1.1M & 2.9M & 1.67     &  8       &  1.84                     & 0.96  & 1.1 & 5      & 0.98    &  1.11      &  670       \\
                                 & sDB & com-dblp & 317k & 1M &  1.33    &   8      &  1.80                     & $\approx$1  & 2.5  & 2     &  $\approx$1   & 2.53    &    2160       \\
                                 & sAM & com-amazon & 334k & 925k & 1.14    &    8     &   1.62                    &  1.04   & 1.64  & 2     & 1.04    &  1.64   &  1426 \\ \midrule
\parbox{1.7cm}{\centering Purchase network (\texttt{PN}s)}                 & pAM & amazon0601  & 403k & 3.3M & 1.45     &  8       &   1.91                     & $\approx$1  & 1.25 & 3       & 1.03    & 1.30    &     618    \\ \midrule
\multirow{3}{*}{\parbox{1.7cm}{\vspace{-0.1em}\centering Road networks (\texttt{RN}s)}}   & rCA & roadNet-CA & 1.9M & 5.5M   & $\approx$1    &   2      &   1.59                   & 1.33   & 1.74 & 8      & 1.38    &  1.80   &   $>10^4$      \\
                                 & rTX & roadNet-TX  & 1.3M & 3.8M  &  $\approx$1    &   2      &  1.53                   &  1.29  & 1.89 & 6      & 1.42    & 2.08       & $>10^4$    \\
                                 & rPA & roadNet-PA  & 1M & 3M & $\approx$1      &      2   &      1.52               & $\approx$1  & 2.00 & 9     &  1.07   & 2.16   &  $>10^4$   \\ \midrule
\parbox{1.5cm}{\vspace{-0.2em}\centering Citation graphs  (\texttt{CG}s)}                 & ciP  &  cit-Patents  & 3.7M & 16.5M & 1.16   &   8      &  1.57  & 1.01  & 1.26 & 2      & 1.01    & 1.26   &1875    \\ \midrule
\multirow{3}{*}{\parbox{1.5cm}{\vspace{-0.1em}\centering Web graphs (\texttt{WG}s)}}      & wGL & web-Google  & 875k & 5.1M &  1.78    &  12       &    2.08                   & 0.98  & 1.26  & 6    & 1.06    & 1.35    &     365      \\
                                 & wBS & web-BerkStan & 685k & 7.6M & 1.91     &   24      &  1.91                     &  0.93 & 1.31 & 5     & 1.07    &  1.40   &      755      \\
                                 & wSF & web-Stanford & 281k & 2.3M &  1.89   &   24      &  1.89                   &   0.98  & 1.54  & 5     & 1.07    & 1.58    &      1077      \\ \bottomrule
\end{tabular}
%\vspace{-1.0em}
%\caption{(\cref{sec:bfs_onnode_eval_real-world}) The performance of AAM for real-world SNAP graphs. $S$ denotes speedup and g500 denotes Graph500. $\approx$1 indicates that the respective $S$ is in the range $(0.99,1.01)$.}
\caption{(\cref{sec:bfs_onnode_eval_real-world}) The performance of AAM for real-world graphs. $S$ and g500 denote speedup and Graph500. $\approx$1 indicates that the given $S \in (0.99;1.01)$.}
\label{tab:snap-bgq-haswell}
%\vspace{-1em}
\end{table*}

\section{Evaluation}
\label{sec:evaluation}
%\vspace{-0.3em}

%We now use the information obtained in~\cref{sec:microbenchmarks}
%to accelerate graph computations based on AAM. We use $M_{min} \in \{2, 80, 144\}$ for the most
%advantageous size of transactions on BG/Q and Haswell, and $C=2,048$ for activity coalescing.
%We use Kronecker~\cite{Leskovec:2010:KGA:1756006.1756039}, Erd\H{o}s-Renyi~\cite{Erdos60onthe} (ER), and real-world SNAP\footnote{available at \url{https://snap.stanford.edu/data/index.html}} graphs to evaluate computations
%on graphs with different vertex distributions (power-law, binomial, Poisson).

%We now use the information obtained in Section~\ref{sec:microbenchmarks}
%to accelerate graph computations based on AAM. We use $M_{min} \in \{2, 80, 144\}$ for the most
%advantageous size of transactions on BG/Q and Haswell, and $C=2,048$ for activity coalescing.
%We use Kronecker~\cite{Leskovec:2010:KGA:1756006.1756039}, Erd\H{o}s-Renyi~\cite{Erdos60onthe} (ER), and real-world SNAP\footnote{Available at \url{https://snap.stanford.edu/data/index.html}.} graphs with different vertex distributions (power-law, binomial, Poisson).

We now use AAM to accelerate the processing of large Kronecker~\cite{Leskovec:2010:KGA:1756006.1756039} and Erd\H{o}s-Renyi~\cite{Erdos60onthe} (ER) graphs with different vertex distributions (power-law, binomial, Poisson). We also evaluate real-world SNAP graphs\footnote{\scriptsize \textsf{Available at \url{https://snap.stanford.edu/data/index.html}.}}.
We evaluate BFS and PR because they are the basis of various data analytics benchmarks such as Graph500 and
because they are proxies of many algorithms such as Ford-Fulkerson.

%\vspace{-0.1em}
\subsection{BFS: Massively-Parallel Manycores}
\label{sec:bfs_onnode_eval}
%\vspace{-0.1em}

%We first evaluate the speedup that AAM delivers in highly-parallel multi- and manycore environments.
%We compare the OpenMP Graph500 reference code~\cite{murphy2010introducing} (\smalltt{Graph500-BGQ}, \smalltt{Graph500-Haswell}) based on atomics with
%the coarsened variants that use the short mode HTM in BG/Q (\smalltt{AAM-BGQ}) and RTM in Haswell (\smalltt{AAM-Haswell}).
%The parameters used are $M=144$, $T=64$ (for BG/Q) and $M=2$, $T=8$ (for Haswell).
%The long mode and HLE are omitted as they follow similar performance patterns and vary by up to 10\%.

We first evaluate the speedup that AAM delivers in highly-parallel multi- and manycore environments.

\textbf{\textsf{Comparison Baseline: }}
Here, we use the OpenMP Graph500 highly optimized reference code~\cite{murphy2010introducing} (\smalltt{Graph500-BGQ}, \smalltt{Graph500-Haswell}) based on atomics 
as the comparison baseline.
The baseline applies several optimizations; among others it reduces the amount of fine-grained synchronization
by checking if the vertex was visited before executing an atomic.

We compare the Graph500 baseline with
the coarsened variants that use the short mode HTM in BG/Q (\smalltt{AAM-BGQ}) and RTM in Haswell (\smalltt{AAM-Haswell}).
We only use \textsf{Has-C} (denoted as \textsf{Haswell}) because it provides higher speedups over atomics than \textsf{Has-P} as we show in Figure~\ref{fig:bfs_coal_study}.
%
%%Here, we only use \textsf{Has-C} (denoted as \textsf{Haswell}) because it provides higher speedups over atomics than \textsf{Has-P} as we show in Figure~\ref{fig:bfs_coal_study}.
%
The long mode and HLE are omitted as they follow similar performance patterns and vary by up to 10\%.
We set $T=64$ (for BG/Q) and $T=8$ (for Haswell) for full parallelism.

%\vspace{-0.1em}
\subsubsection{Processing Kronecker Power-Law Graphs}
\label{sec:bfs_onnode_eval_kronecker}
%\vspace{-0.1em}

%Consult the technical report for graphs with other values of $M$ and $T$.

%We scale $|V|$ from
%$2^{20}$ to $2^{28}$, and we use $\bar{d} \in \{1,2,...,256\}$;
%highest values generate graphs that fill the whole available memory. The results
%for other graphs and for the remaining values of $M$ and $T$ can be found in the technical report.

%Here, we set $M=144$, $T=64$ (for BG/Q) and $M=2$, $T=8$ (for Haswell).

Here, we use the results of the analysis in Section~\ref{sec:microbenchmarks}
and set $M_{min} \in \{2, 80, 144\}$ for the most
advantageous size of transactions on BG/Q and Haswell.
%
%The parameters used are $M=144$, $T=64$ (for BG/Q) and $M=2$, $T=8$ (for Haswell).
%
We present the results in Figure~\ref{fig:graphs_dets_perf}.
%We present the results for Kronecker graphs in Figure~\ref{fig:graphs_dets_perf}.
%
We scale $|V|$ from
$2^{20}$ to $2^{28}$, and we use $\bar{d} \in \{1,2,...,256\}$;
highest values generate graphs that fill the whole available memory.
For BG/Q, \smalltt{AAM-BGQ} outperforms \smalltt{Graph500-BGQ} by up to 102\% for a graph with $\approx$2 millions vertices and $\bar{d}=4$.
For higher $\bar{d}$ \smalltt{AAM-BGQ} becomes comparable to \smalltt{Graph500-BGQ}. This is because adding more edges for fixed $|V|$
generates more transactions that conflict and abort more often.
%
%For Haswell, \smalltt{AAM-Haswell} consistently outperforms \smalltt{Graph500-Haswell} by up to 27\%.
For Haswell, AAM consistently outperforms Graph500 by up to 27\%.
The speedup does not change significantly when increasing $\bar{d}$. This is because we use smaller transactions in \smalltt{AAM-Haswell}
($M=2$) than in \smalltt{AAM-BGQ} ($M=144$) and thus they do not incur considerably more memory conflicts when $\bar{d}$ is increased.
%The speedup is in general lower than for BG/Q due to the higher number of costly buffer overflows.

%\vspace{-0.4em}
\subsubsection{Processing Real-World Graphs}
\label{sec:bfs_onnode_eval_real-world}
%\vspace{-0.1em}

Next, we evaluate AAM for real-world graphs
(see Table~\ref{tab:snap-bgq-haswell}).
For this, we extend Graph500 
so that it can read graphs from a file.
%
%The tested graphs are presented in the first part of Table~\ref{tab:snap-bgq-haswell}.
We selected directed/undirected graphs with $|V|>250k$
that could fit in memory and we excluded graphs
that could not easily be loaded into Graph500 framework (e.g., amazon0505).

%The results for BG/Q are presented in the second part of Table~\ref{tab:snap-bgq-haswell}.

\textbf{\textsf{BlueGene/Q: }}
The tested graphs are generally sparser than the analyzed Kronecker graphs.
%Thus, we expect that the optimum $M$ will be smaller than 144 and we set it to 24.
We discovered that the optimum $M$ is smaller than 144 (we set it to 24).
This is because in dense graphs more data is contiguous in memory and thus
can be processed more efficiently by larger transactions.
The results show that graphs with similar structure entail similar performance gains. 
The highest $S$ (speedup) is achieved for \texttt{CN}s (up to 3.67) and \texttt{WG}s (up to 1.91).
\texttt{SN}s, \texttt{PN}s, and \texttt{CG}s offer moderate $S$ (1.14-1.67).
\texttt{RN}s entail no significant change in performance.
We also searched for optimum values of $M$ for specific graphs;
this improves $S$ across all the groups.
The results indicate that respective groups have similar optimum values of $M$.
The differences are due to the structures of the graphs that may either
facilitate coarsening and reduce the number of costly aborts
(\texttt{CN}s and \texttt{WG}s) or entail more significant
overheads (\texttt{RN}s).

%\begin{turn}{90}comm.\end{turn}

%\begin{table}[h]
%\centering
%\scriptsize
%\setlength{\tabcolsep}{5.0pt}
%\begin{tabular}{@{}l|llll|l@{}}
%\toprule
%Graph type        & Graph ID & Graph name & $|V|$  & $|E|$  & Speedup \\ \midrule
%\multirow{3}{*}{Comm. networks} & cWT & wiki-Talk & 2.4M & 5M & 2.94 \\
%					& cEU & email-EuAll & 265k & 420k & 2.58 \\
%					& cEN & email-Enron & 36k & 367k & 1.15 \\ \midrule
%\multirow{6}{*}{Social networks} & sLV  & soc-LiveJ. & 4.8M & 69M & 1.07 \\
%					 & sOR & com-orkut & 3M & 117M & 1.17 \\
%					 & sLJ & com-lj & 4M & 34M & 1.05 \\
%					 & sYT & com-youtube & 1.1M & 2.9M & 1.31 \\
%					 & sDB & com-dblp & 317k & 1M & $\approx$1 \\
%					 & sAM & com-amazon & 334k & 925k & $\approx$1 \\ \midrule
%                 Purchase network & pAM & com-amazon  & 403k & 3.3M & 1.07 \\ \midrule
%\multirow{3}{*}{Road networks} & rCA & roadNet-CA & 1.9M & 5.5M & $\approx$1 \\
% & rTX & roadNet-TX  & 1.3M & 3.8M & $\approx$1 \\
% & rPA & roadNet-PA  & 1M & 3M & $\approx$1 \\ \midrule
%                Citation graphs   & ciP  &  cit-Patents  & 3.7M & 16.5M & $\approx$1 \\ \midrule
%\multirow{3}{*}{Web graphs} & wGL & web-Google  & 875k & 5.1M & 1.44 \\
% & wBS & web-BerkStan & 685k & 7.6M & 1.89 \\
% & wSF & web-Stanford & 281k & 2.3M &  1.73 \\ \bottomrule
%\end{tabular}
%\vspace{-1.0em}
%\caption{(\cref{sec:bfs_onnode_eval_real-world}) The speedup of AAM over Graph500 on BG/Q.}
%\label{tab:snap-bgq}
%\end{table}

\begin{figure*}[t!]
%\vspace{-1.5em}
\centering

% $|V_i| = 2^{12}$, $P_{ER} = 0.0005$.
% $|V_i| = 2^{12}$, $P_{ER} = 0.0005$.
% $p=16$, $P_{ER} = 0.005$.

 \subfloat[Scalability of AAM; BG/Q.]{
  \includegraphics[width=0.185\textwidth]{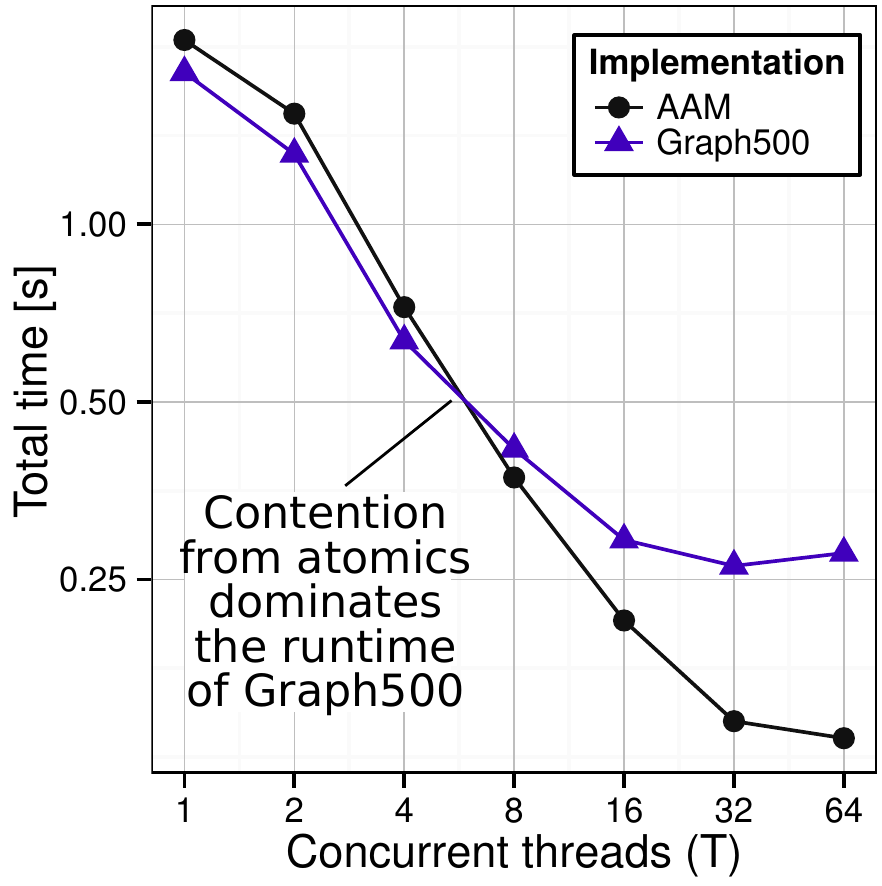}
  \label{fig:onnode-scal_bgq}
 }\hfill
 \subfloat[Scalability of AAM; Haswell.]{
  \includegraphics[width=0.185\textwidth]{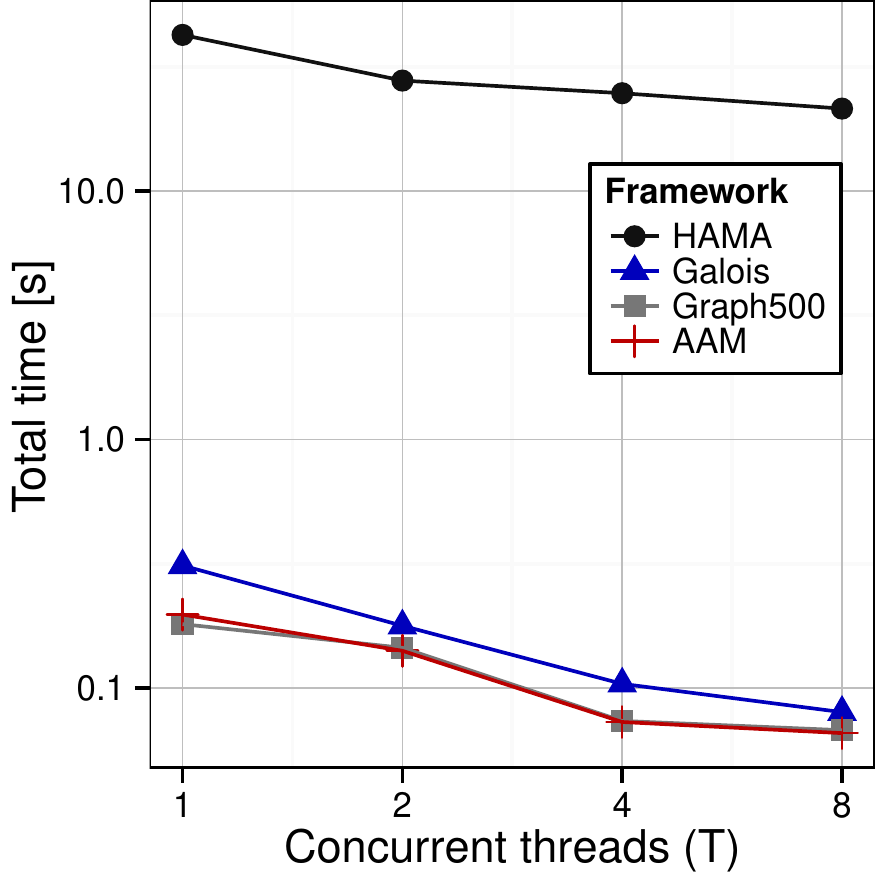}
  \label{fig:onnode-scal_haswell}
 }\hfill
\subfloat[PageRank, $ER = 0.0005$.]{
  \includegraphics[width=0.185\textwidth]{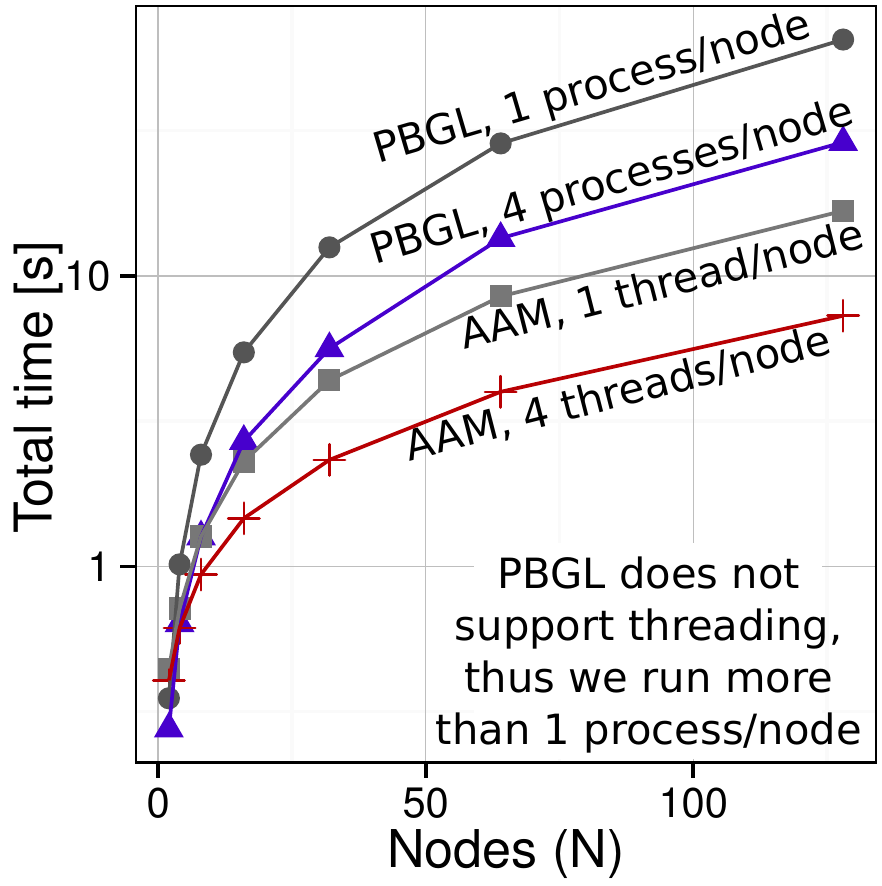}
  \label{fig:pr_N}
 }\hfill
 \subfloat[PageRank, $ER = 0.0005$.]{
  \includegraphics[width=0.185\textwidth]{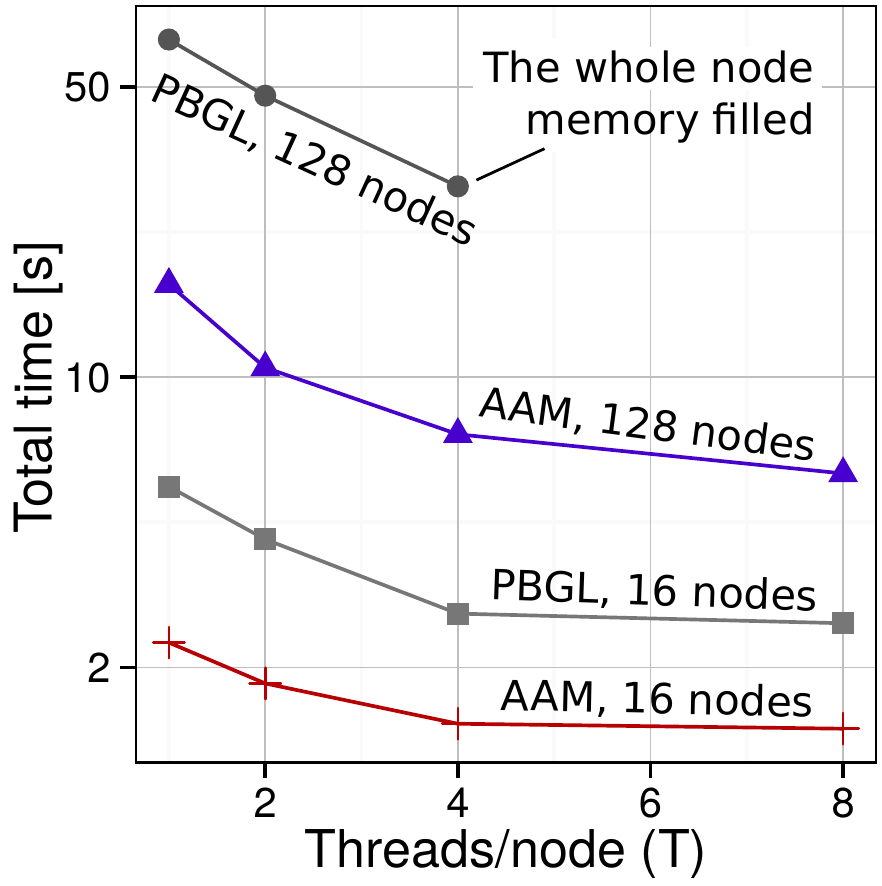}
  \label{fig:pr_T}
 }\hfill
  \subfloat[PageRank, $ER = 0.005$.]{
  \includegraphics[width=0.185\textwidth]{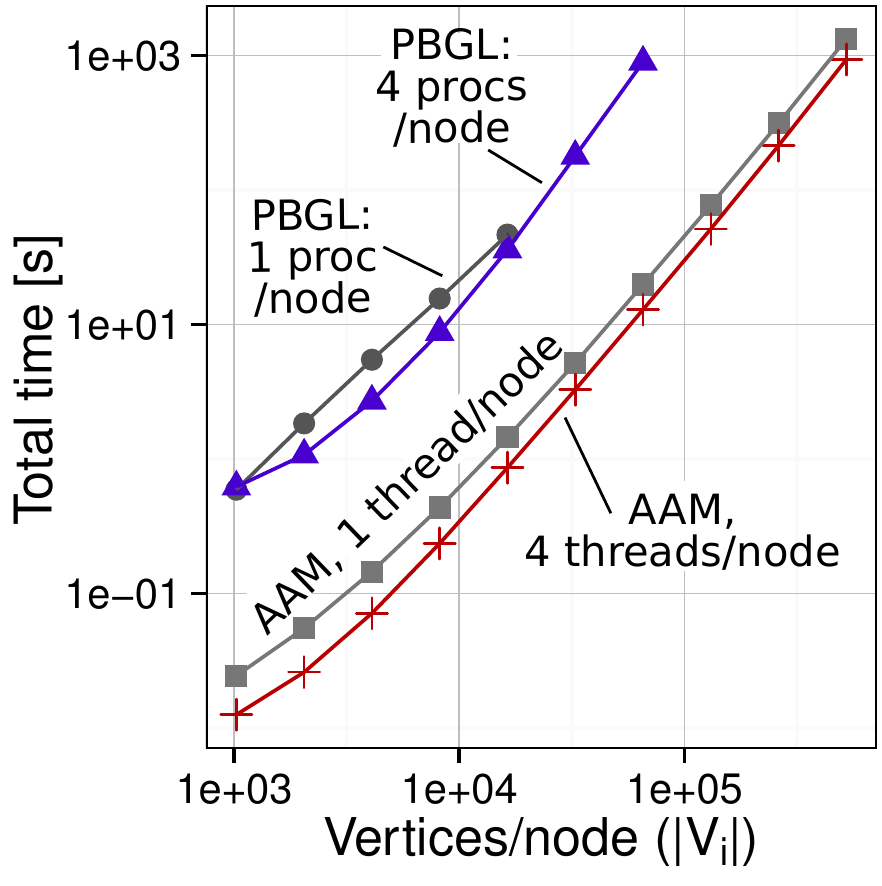}
  \label{fig:pr_S}
 }
 \vspace{-0.8em}
 \caption{The analysis of the performance of: BFS (when varying $T$; \cref{sec:bfs_onnode_eval_scalability} and Figures~\ref{fig:onnode-scal_bgq}-\ref{fig:onnode-scal_haswell}) and distributed PR (\cref{sec:pr_eval}, Figures~\ref{fig:pr_N}-\ref{fig:pr_S}).}
%\vspace{-1em}
\label{fig:pr_general}
\end{figure*}

\textbf{\textsf{Haswell: }}
%
%We perform a similar study on Haswell. 
%
We compare AAM to several state-of-the-art graph processing engines: Galois~\cite{Kulkarni:2008:OPB:1346281.1346311} (represents runtime systems that exploit amorphous data-parallelism), SNAP~\cite{leskovec2008dynamics} (represents network analysis and data mining libraries),
and HAMA~\cite{Seo:2010:HEM:1931470.1931872}\footnote{Update (31.07.15): we use HAMA 0.6.4} (an engine similar to Pregel~\cite{Malewicz:2010:PSL:1807167.1807184} that represents Hadoop-based BSP processing engines).
%
%Here, we compare to several state-of-the-art graph processing engines: SNAP~\cite{leskovec2008dynamics} and GraphLab~\cite{low2010graphlab} (represent network analysis and data mining libraries), Galois~\cite{Kulkarni:2008:OPB:1346281.1346311} (represents runtime systems that exploit amorphous data-parallelism),
%and HAMA~\cite{Seo:2010:HEM:1931470.1931872} (a clone of Pregel that represents Hadoop-based BSP processing engines).
%
We do not evaluate these engines on BG/Q due to various compatibility problems
(e.g., BG/Q does not support Java required by HAMA).
BFS in Galois only returns the diameter. We modified it (with fine locks) so that it constructs a full BFS tree, 
analogously to AAM and Graph500.
%
%The results are presented in Table~\ref{tab:snap-haswell}. 

First, we set $M=2$.
While AAM is in general faster than Graph500
(up to 33\% for \texttt{rCA}), several inputs
entail longer AAM BFS traversals. AAM is up to a factor of two faster than Galois
but is slower for two inputs (\texttt{cEU} and \texttt{sOR}). 
There is some diversity in the results because AAM on Haswell is significantly more sensitive to small changes of $M$
than on BG/Q. 
Thus, we again searched for the optimum $M$ for each
input separately which resulted in higher AAM's speedups. 
The performance of HAMA and SNAP is generally much lower than AAM (results for SNAP are consistently worse than for HAMA and we exclude them due to space constraints).
HAMA suffers from overheads caused by the underlying MapReduce architecture
and expensive synchronization.
The analyzed real-world graphs have usually high diameters (e.g., 33 for \texttt{sAM}) and thus require many
BSP steps that are expensive in HAMA. This is especially visible for \texttt{RN}s that have particularly big diameters (554 for \texttt{rCA}) and accordingly long runtimes. As we will show in the next section, processing Kronecker graphs with lower diameters
reduces these overheads.
We also investigated SNAP and we found out that it is particularly inefficient for undirected graphs and
it does not efficiently use threading.
Our final discovery is that, similarly to BG/Q, respective groups of graphs have similar optimum values of $M$.

\subsubsection{Evaluating the Scalability of AAM}
\label{sec:bfs_onnode_eval_scalability}
%\vspace{-0.1em}

Finally, we evaluate the scalability of AAM by varying $T$.
The results are presented in Figure~\ref{fig:onnode-scal_bgq}
(BG/Q) and~\ref{fig:onnode-scal_haswell} (Haswell).
We use a Kronecker graph with $2^{21}$ vertices and $2^{24}$ edges.
We vary $T$ between 1 and the
number of available hardware threads.
The BG/Q results indicate that AAM utilizes onnode parallelism
more efficiently than Graph500. For Haswell, the performance
patterns for AAM and Graph500 are similar; both
frameworks deliver positive speedups for any $T$ and outperform
other schemes by $\approx$20-50\% (Galois) and $\approx$2 orders of magnitude (HAMA).
We skip SNAP for clarity; it is consistently 2-3x slower
than HAMA.

%\begin{figure}[h!]
%\vspace{-1.5em}
%\centering
% %\endminipage\hfill
% \subfloat[Scalability of AAM on BG/Q.]{
%  \includegraphics[width=0.22\textwidth]{scal_bfs_bgq-eps-converted-to.pdf}
%  \label{fig:onnode-scal_bgq}
% }\hfill
% \subfloat[Scalability of AAM on Haswell.]{
%  \includegraphics[width=0.22\textwidth]{scal_bfs_haswell-eps-converted-to.pdf}
%  \label{fig:onnode-scal_haswell}
% }
% \vspace{-0.9em}
% \caption{(\cref{sec:bfs_onnode_eval_scalability}) The analysis of the performance of BFS when varying $T$\maciej{to be finished}.}
% 
%\vspace{-0.7em}
%\label{fig:scalability}
%\end{figure}

%\begin{figure*}
%\vspace{-1.5em}
%\centering
%
%% $|V_i| = 2^{12}$, $P_{ER} = 0.0005$.
%% $|V_i| = 2^{12}$, $P_{ER} = 0.0005$.
%% $p=16$, $P_{ER} = 0.005$.
%
% %\endminipage\hfill
%\subfloat[$ER = 0.0005$.]{
%  \includegraphics[width=0.31\textwidth]{PR_N_labels-eps-converted-to.pdf}
%  \label{fig:pr_N}
% }\hfill
% \subfloat[$ER = 0.0005$.]{
%  \includegraphics[width=0.31\textwidth]{PR_T_labels-eps-converted-to.pdf}
%  \label{fig:pr_T}
% }\hfill
%  \subfloat[$ER = 0.005$.]{
%  \includegraphics[width=0.31\textwidth]{PR_S_labels-eps-converted-to.pdf}
%  \label{fig:pr_S}
% }
% \vspace{-0.5em}
% \caption{(\cref{sec:pr_eval}) The overview of the performance of distributed PageRank.}
%\vspace{-2.0em}
%\label{fig:pr_general}
%\end{figure*}

%\htor{maybe say something about implementation complexity}

%\vspace{-0.1em}
\subsection{PR: Distributed Memory Machines}
\label{sec:pr_eval}
%\vspace{-0.1em}

As the last step, we provide an initial large-scale evaluation of AAM in a distributed environment.
We select PR
to illustrate that expensive and numerous aborts generated by the HTM implementation
of \textsf{ACC} (cf.~\cref{sec:intra-node-act-increment}) can be amortized with the coalescing of activities. We compare AAM to
a version of Parallel Boost Graph Library (PBGL)~\cite{Gregor05theparallel} based on active messages.
The utilized variant of PBGL applies various optimizations; for example it processes incoming edges to reduce the amount of synchronization and to limit
the performance overheads caused by atomics.
We run the benchmarks on BG/Q to enable large-scale evaluation.
%due to space constraints and the insufficient scale we exclude \textsf{Has-P} as it
%only hosts two nodes.
%
We use ER graphs with the probability parameter $ER \in \{0.005, 0.0005\}$ and the number of vertices up to $2^{23}$.
PBGL does not support threading, we thus spawn multiple
processes per node and an equal number of threads in AAM; we scale $T$ until PBGL fills in the whole memory.

The results of the analysis are presented in Figure~\ref{fig:pr_general}. 
We scale $N$ (Figure~\ref{fig:pr_N}), $T$ (Figure~\ref{fig:pr_T}), and $|V_i|$ (Figure~\ref{fig:pr_S}).
In each scenario AAM outperforms PBGL $\approx$3-10 times thanks to the coalescing of activities and more efficient utilization
of intra-node parallelism. 

\section{Related Work and Discussion}
%\vspace{-0.3em}

% \cite{Lublinerman:2011:DI:2048066.2048133} (Delegated Isolation),
% \cite{stapl} (STAPL),
% , and \cite{kdtsdm12}.

The challenges connected with the processing of graphs are presented by Lumsdaine et al.~\cite{DBLP:journals/ppl/LumsdaineGHB07}.
%
%Example abstractions and formalizations for graph computations are Tao~\cite{Pingali:2011:TPA:1993498.1993501}, active messages for graph computations~\cite{active-pebbles,Edmonds:2013:EGA:2464996.2465441}, and Galois~\cite{Kulkarni:2008:OPB:1346281.1346311}.
%
Example frameworks for parallel graph computations are Pregel~\cite{Malewicz:2010:PSL:1807167.1807184}, PBGL~\cite{Gregor05theparallel}, HAMA~\cite{Seo:2010:HEM:1931470.1931872}, GraphLab~\cite{low2010graphlab}, and Spark~\cite{Zaharia:2010:SCC:1863103.1863113}.
There exist several comparisons of various engines~\cite{Satish:2014:NMG:2588555.2610518,Guo:2014:WGP:2650283.2650530,Lu:2014:LDG:2735508.2735517,6691555}.
%
%%%%%A recent comparison of various engines was done by Satish et al.~\cite{Satish:2014:NMG:2588555.2610518}.
%
%Some interesting algorithms for solving popular graph problems are presented in, e.g., \cite{Bader:2006:FSA:1226705.1226708, Cong:2009:FPC:1809961.1809979, 5713180, McLendonIII:2005:FSC:1096268.1096270, Prountzos:2013:BCA:2442516.2442521}.
%
AAM differs from these designs as it is a mechanism that can be used to implement abstractions and to accelerate
processing engines. It uses HTM to reduce the amount of synchronization and thus to accelerate graph analytics.

GraphBLAS~\cite{DBLP:journals/corr/MattsonBBBDFFGGHKLLPPRSWY14} is an emerging standard for expressing
graph computations in terms of linear algebra operations. AAM can be used to implement the GraphBLAS
abstraction and to accelerate the performance of graph analytics based on sparse linear algebra computations.

The Galois runtime~\cite{Kulkarni:2007:OPR:1250734.1250759} optimizes graph processing by
coarsening fine graph updates.
AAM can be integrated with Galois.
In AAM, we focus on scalable techniques for implementing coarsening with HTM.
First, we provide a detailed performance analysis of HTM for graph computations, a core paper contribution. Instead, Galois mostly addresses locking~\cite{Kulkarni:2008:OPB:1346281.1346311}. Second, contrary to Galois, AAM targets both shared- and distributed-memory systems.
% it also proposes the ownership protocol for executing distributed hardware transactions.
Third, our work performs a holistic extensive performance analysis of coarsening. Instead, coarsening in Galois is not evaluated on its own. 
We conclude that AAM's techniques and analysis can be used to accelerate the Galois runtime.

Active Messages (AM) were introduced by Eicken et al.~\cite{von1992active}. Various AM implementations were proposed~\cite{Edmonds:2013:EGA:2464996.2465441, willcock-amplusplus, active-pebbles, pami, bonachea2002gasnet}.
Our work enhances these designs by combining AM with HTM. We illustrate how to program AAM and we conduct an extensive analysis 
to show how to tune AAM's performance on state-of-the-art manycore architectures.

Transactional memory was introduced by Herlihy et al.~\cite{Herlihy:1993:TMA:165123.165164}.
%There has been an extensive amount of research into STM~\cite{Shavit:1995:STM:224964.224987}. Distributed STM systems were also proposed and analyzed~\cite{Lesani:2011:CMT:1941553.1941577}.
Several implementations of HTM were introduced, but their performance was not extensively analyzed~\cite{tsx-sc, Wang:2012:EBG:2370816.2370836, Dice:2009:EEC:1508244.1508263,Chaudhry:2009:RHS:1550399.1550516}.
Yoo et al.~\cite{tsx-sc} present performance gains from using Haswell HTM in scientific workloads such as simulated annealing. Our analysis generalizes these findings, proposes a simple performance model, and provides a deep insight into the performance of both BG/Q and Haswell HTM for a broad range of transaction sizes and other parameters in the context of data analytics.

%\vspace{-0.4em}
%\subsection{Compiling Transactions into Atomics}
%\vspace{-0.3em}

We envision that the potential of AAM could be further expanded by
combining it with some ideas related to code analysis.
For example, 
one could envision a simple
compiler pass that pattern-matches each single-vertex transaction against the set of
atomic operations to transform it if possible to accelerate graph processing. However, such an analysis
is outside the scope of this paper.

Finally, AAM can be extended with algorithms for the online selection of $M$. Here, as our study exhaustively illustrates performance tradeoffs in the available HTM implementations, it would facilitate the runtime decisions on how to select $M$. For example, the runtime can prune the space of all the applicable values of HTM parameters depending on which HTM is utilized. In addition, our performance model can be further extended and combined with data mining techniques to enable effective online decisions based on graph sampling. We leave this study for future research.

\section{Conclusion}
%\vspace{-0.3em}

Designing efficient algorithms for massively parallel and distributed graph computations is becoming one
of the key challenges for the parallel programming community~\cite{Kulkarni:2008:OPB:1346281.1346311}.
Graph processing is fine-grained by nature and its traditional implementations
based on atomics or fine locks are error-prone and may entail significant overheads~\cite{Kulkarni:2008:OPB:1346281.1346311}.

%As Pingali et
%al. stated~\cite{Pingali:2011:TPA:1993498.1993501}, many inherent difficulties connected with designing such algorithms (data irregularity, data-driven computations, etc.) may be merely the instances of the lack of a proper abstraction.

We propose Atomic Active Messages (AAM), a mechanism
that reduces the amount of fine-grained synchronization in 
irregular graph computations. AAM
is motivated by recent advances towards implementing transactional memory
in hardware.
AAM provides several high performance techniques for
executing fine-grained graph modifications as coarse transactions, it
facilitates the utilization of state-of-the-art hardware mechanisms
and resources, and it can be used to accelerate highly
optimized codes such as Graph500 by more than 100\%.

%%%%We propose Atomic Active Messages (AAM), a mechanism
%%%%that reduces that amount of fine-grained synchronization in graph analytics. AAM
%%%%is motivated by recent advances towards implementing transactional memory
%%%%in hardware.
%%%%AAM provides several high performance techniques for
%%%%executing fine-grained graph modifications as coarse transactions; it
%%%%facilitates the utilization of state-of-the-art hardware mechanisms
%%%%and resources and enables significant speedups of more than 100\% over highly
%%%%optimized codes such as Graph500 or Galois.

%We propose Atomic Active Messages (AAM), a generic and intuitive abstraction and a mechanism
%that addresses the shortcomings of existing models. AAM
%is motivated by recent advances towards implementing transactional memory
%in hardware. It provides a novel classification of atomic active messages that can
%be used to express both shared- and distributed-memory graph computations.
%The AAM framework uses several high performance techniques for
%executing fine-grained graph modifications as coarse transactions; it
%facilitates the utilization of state-of-the-art hardware mechanisms
%and resources and enables significant speedups of more than 100\% over highly
%optimized codes such as Graph500~\cite{murphy2010introducing}.

AAM targets highly-parallel multi- and manycore architectures
and distributed-memory machines. It provides a novel classification of atomic active messages that can
be used to design and program both shared- and distributed-memory graph computations.
AAM enables different optimizations
from both of these worlds
such as coarsening intra-node transactions and coalescing inter-node activities.
We illustrate how to implement AAM with HTM;
however, other mechanisms such as distributed STM~\cite{Lesani:2011:CMT:1941553.1941577},
flat-combining~\cite{Hendler:2010:FCS:1810479.1810540}, or optimistic
locking~\cite{Kung:1981:OMC:319566.319567} could also be used.

Finally, to the best of our knowledge, our work is the first detailed performance analysis of hardware transactional
memory in the context of graph computations and the first to compare HTMs implemented
in Intel Haswell and IBM Blue Gene/Q. Among others, we conjecture that implementing HTM in the bigger L2 cache (BG/Q)
enables higher performance than in the smaller L1 cache (Haswell).
We believe our analysis and data can be used by architects
and engineers to develop a more efficient HTM that would offer even higher speedups for
irregular data analytics.
%
%We believe our study and data can be used by architects
%and engineers to develop a more performant HTM that would offer even higher speedups for
%irregular data analytics.

\maciej{TODO: more focus on Tao? Put in background?}

\maciej{TODO: our classification is >driven< by the AM and HTM semantics. Tao is driven by...}

\maciej{TODO: check future citation}

\maciej{TODO: pout more focus on the fact that we use classification to show that different algorithms can be implemented using the same protocol}

\maciej{TODO: group listings in one top figure (?)}

\maciej{TODO: more focus on the ownership protocol}

\maciej{TODO: when choosing a vertex to process, have some selection scheme that minimizes the probability of conflicts (random?)}

\maciej{TODO: put 'Activities and Atomic Operations' to the discussion?}

\maciej{TODO: a nice figure/table that shows both categories and protocols}

%\appendix
%\section{Appendix Title}
%
%This is the text of the appendix, if you need one.

%Acknowledgments.

% We recommend abbrvnat bibliography style.

{\small 
\vspace{0em}\subsubsection*{Acknowledgements}
We thank Hussein Harake and the CSCS and ALCF teams
granting access to the Greina, Monte Rosa, and Vesta machines,
and for their excellent technical support.}

%\vspace{-1em}

{
%\def\bibfont{\scriptsize}
%\scriptsize
%\setlength{\parskip}{0pt}
%\setlength{\bibsep}{2pt plus 1ex}
%\setlength{\bibsep}{-0.2pt plus 10ex}
\bibliographystyle{abbrv}
%\raggedright
%\textsf{\bibliography{aam_hpdc-2015}}
\bibliography{references}
}

%\vspace{-0.5em}
%%\def\bibfont{\scriptsize}
%%\setlength{\bibsep}{0.25pt plus -0.3ex}
%\bibliographystyle{abbrv}
%%\bibliographystyle{abbrv}
%% The bibliography should be embedded for final submission.
%
%\bibliography{aam_hpdc-2015}
%%\begin{thebibliography}{}
%%\softraggedright
%%
%%\end{thebibliography}

\end{document}